\documentstyle[12pt,epsfig]{article}
\begin{document}
\def\caja{\mathsurround=0pt}
\def\eqalign#1{\,\vcenter{\openup1\jot \caja
        \ialign{\strut \hfil$\displaystyle{##}$&$
        \displaystyle{{}##}$\hfil\crcr#1\crcr}}\,}

\newcommand{\QGindx}[1]{\mbox{\scriptsize #1}}
\newcommand{\QGbar}[1]{\overline{\mbox{#1}}}
\newenvironment{QGitemize}{\begin{list}{$\bullet$}%
{\setlength{\topsep}{-2.8mm}\setlength{\partopsep}{0.2mm}%
\setlength{\itemsep}{0.2mm}\setlength{\parsep}{0.2mm}}}%
{\end{list}}
\newcounter{QGenumct}
\newenvironment{QGenumerate}{\begin{list}{\arabic{QGenumct}.}%
{\usecounter{QGenumct}\setlength{\topsep}{-2.8mm}%
\setlength{\partopsep}{0.2mm}\setlength{\itemsep}{0.2mm}%
\setlength{\parsep}{0.2mm}}}{\end{list}}

\setcounter{tocdepth}{2}

\newcommand{\QGee}{\mbox{e}^+\mbox{e}^-}
\newcommand{\QGq}{\mbox{q}}
\newcommand{\QGqbar}{\overline{\mbox{q}}}
\newcommand{\QGg}{\mbox{g}}
\newcommand{\QGgaZ}{\gamma^*/\mbox{Z}^0}
\newcommand{\QGalphaem}{\alpha_{\mbox{\scriptsize em}}}
\newcommand{\QGalphas}{\alpha_{\mbox{\scriptsize s}}}
\newcommand {\QGecm} {$E_{\QGindx{cm}}$}
\newcommand {\QGnch} {$\langle n_{\QGindx{ch}} \rangle$}
\newcommand {\QGnchdist} {$n_{\QGindx{ch}}$}
\newcommand {\QGptin} {$p_{\perp \QGindx{in}}$}
\newcommand {\QGptout} {$p_{\perp \QGindx{out}}$}
\newcommand {\QGhjet} {$M_{\QGindx{heavy}}/E_{\QGindx{cm}}$}
\newcommand {\QGdjet} {$M_{\QGindx{diff}}/E_{\QGindx{cm}}$}
\newcommand {\QGdurham} {$k_\perp$}
\newcommand {\QGycut} {$y_{\QGindx{cut}}$}
\newcommand {\QGcosnr} {$\left|\cos{\theta^{*}_{\QGindx{NR}}}\right|$}
\newcommand {\QGleang} {$\cos{\alpha_{34}}$}
\newcommand {\QGtfamily} {$T,\;T_{\QGindx{major}}$ and $T_{\QGindx{minor}}$}
\newcommand {\QGtmajor} {$T_{\QGindx{major}}$}
\newcommand {\QGtminor} {$T_{\QGindx{minor}}$}
\newcommand {\QGthrustm} {$\langle T \rangle$}
\newcommand {\QGtmajorm} {$\langle T_{\QGindx{major}} \rangle$}
\newcommand {\QGtminorm} {$\langle T_{\QGindx{minor}} \rangle$}

\newcommand{\QGcpc}[3] {Comp.\ Phys.\ Comm.\ {\bf#1} (#2) #3\/}
\newcommand{\QGjetp}[3] {JETP Lett.\ {\bf#1} (#2) #3\/}
\newcommand{\QGnp}[3] {Nucl.\ Phys.\ {\bf#1} (#2) #3\/}
\newcommand{\QGpl}[3] {Phys.\ Lett.\ {\bf#1} (#2) #3\/}
\newcommand{\QGprep}[3] {Phys.\ Rep.\ {\bf#1} (#2) #3\/}
\newcommand{\QGprev}[3] {Phys.\ Rev.\ {\bf#1} (#2) #3\/}
\newcommand{\QGprl}[3] {Phys.\ Rev.\ Lett.\ {\bf#1} (#2) #3\/}
\newcommand{\QGrmp}[3] {Rev.\ Mod.\ Phys.\ {\bf#1} (#2) #3\/}
\newcommand{\QGzp}[3] {Zeit.\ Phys.\ {\bf#1} (#2) #3\/}

\begin{center}
{\LARGE \bf QCD EVENT GENERATORS}
\end{center}
\begin{center}
{\it Conveners}: I.G.~Knowles and T.~Sj\"ostrand
\end{center}
\begin{center}
{\it Working group}: A.~Blondel, A.~Boehrer, C.D.~Buchanan, 
D.G.~Charlton, S.-L. Chu, S.~Chun, G.~Dissertori, 
D.~Duchesneau, J.W.~Gary, M.~Gibbs, A.~Grefrath, 
G.~Gustafson, J.~H\"akkinen, K.~Hamacher, K.~Kato, 
L.~L\"onnblad, W.~Metzger, R.~M{\o}ller, T.~Munehisa, 
R.~Odorico, Y.~Pei, G.~Rudolph, S.~Sarkar, M.H.~Seymour,
J.C.~Thompson, \v{S}.~Todorova and B.R.~Webber.
\end{center}
\vspace*{1cm}
\tableofcontents
\newpage
\section{Introduction}
\label{QGsectintro}

This section is devoted to QCD generators, relevant for LEP~2 processes where
hadrons may be found in the final state: $\QGee \to \QGgaZ \to \QGq \QGqbar,\;
\QGee \to \mbox{W}^+\mbox{W}^- \to  \QGq \QGqbar \QGq' \QGqbar',\;\QGee \to
\mbox{Z}^0 \mbox{h}^0 \to \nu \overline{\nu} \mbox{b}\QGbar{b}$, {\it etc.}
In fact, almost all interesting processes at LEP~2 give hadronic final states,
ensuring that QCD generators will remain of vital importance. 

It is instructive to contrast the EW and QCD generator perspectives for LEP~2.
In the EW physics program, the main emphasis is on four-fermion final states.
This is different from LEP~1, where the Z$^0$ line shape was a major focus
of attention \cite{QGlep1}. Dedicated four-fermion generators are new 
creations, that have to stand on their own and cannot be tested at LEP~1. 
Therefore there
is little sense of continuity with respect to the LEP~1 workshop
\cite{QGLEPoneQED} and subsequent LEP~1 activities. QCD physics, by contrast,
extrapolates logically from LEP~1. New aspects may enter, such as colour
reconnection, but these are expected to be relatively small perturbations on
the basic picture (though of importance for precision physics). Therefore the
QCD generators write-up for the LEP~1 workshop \cite{QGLEPone} is still partly
relevant and subsequent LEP~1 experience very much so. The high Z$^0$
statistics will make LEP~1 a significant testing ground for many new QCD
physics ideas also in the LEP~2 era.

It is thus logical to begin this section with an assessment of experience from
LEP~1, with emphasis on areas where generators are known to have shortcomings.
Any improvements for LEP~1 will directly benefit LEP~2. This is followed by
a comparison of extrapolations to LEP~2 energies, from which the current range
of uncertainty can be estimated. Next comes a survey of existing generators,
ranging from major programs, with coverage of the full generation chain, to
shorter pieces of code for specific purposes. Finally, there is a section on
standardization efforts, to help ease life for users who rely on several
generators.
  
This report is not a complete description of the topic. However it should
provide a convenient starting point, with ample references to further relevant
literature.

\newpage

\section{Experience from LEP~1}
\label{QGsectexper}

\subsection{Event shapes and inclusive distributions}

A large quantitative improvement in the description of event shape and
inclusive distributions has been made at LEP~1 with respect to the era of
PETRA and PEP. This is due mainly to the vast amount of high quality data 
available and the need to achieve good agreement in model/data comparisons
so as to obtain small systematic errors for the high precision electroweak
measurements. To help facilitate this goal flexible fitting algorithms were
developed, based on previous work \cite{QGKH_Billsthesis,QGKH_Tasso_tune}.
In many cases the dependence of the model's response to its parameters is
analytically interpolated
\cite{QGKH_O_tune,QGKH_L_tune,QGKH_A_tune,QGKH_D_tune}.
This strategy is flexible, allows easy exchange of input distributions but
also the simultaneous fitting of very many, 10--15, parameters 
\cite{QGKH_D_tune,QGKH_A_afbbrussels}.

Evidently the choice of input distributions used to constrain the model
parameters is important. In general a distribution depends on very many
parameters, thus the parameters resulting from a fit are in general
correlated. A survey has been undertaken to determine which distributions
have the highest sensitivity to the individual model parameters
\cite{QGKH_D_tune,QGKH_MWdiss}. It turns out that semi-inclusive spectra
are most important, as has been observed before \cite{QGKH_A_tune}. The
charged particle momentum and transverse momentum spectra strongly
constrain the fragmentation function or, alternatively, the cluster
parameters. However their dependence on the fragmentation parameters is
not exclusive. Inclusive distributions may depend even more strongly on
$\Lambda_{\QGindx{QCD}}$ and/or the parton shower cut-off. In fact, the 
latter parameter strongly influences the high-momentum tail of the momentum
spectrum. The 3-jet rate as defined using the Durham or JADE algorithms
almost only depends on $\Lambda_{\QGindx{QCD}}$. This emphasizes the 
reliability
of $\QGalphas$ determinations using this quantity. In contrast, and
somewhat surprisingly, the AEEC depends strongly on very many model
parameters. Measures of the general event topology, e.g. thrust and
sphericity, depend mainly on $\Lambda_{\QGindx{QCD}}$ and only in the 
2-jet regime
on fragmentation parameters. Shape measures sensitive to radiation out of
the event plane, like minor or aplanarity, show strong dependence
both on fragmentation parameters and on $\Lambda_{\QGindx{QCD}}$. 
In summary,
model parameters are best determined by fitting the model to inclusive
distributions, jet rates and shape distributions simultaneously.

It appears that the ``partonic'' phase of the models is best tested by
studying the properties of jets defined using jet algorithms
\cite{QGKH_jetalgos}. At large resolution parameter $y_{\QGindx{cut}}$, when
dealing with few jets or the emission of the ``first'' hard gluons at
large angles, fragmentation effects are almost negligible. In contrast
at smaller $y_{\QGindx{cut}}$, where higher jet rates are sizable i.e. 
when the
subjet structure described by multiple emission of soft and collinear 
gluons is important, also fragmentation effects are of increasing
importance.

The parton shower models {\sc Ariadne} \cite{QGariadne}, {\small HERWIG}
\cite{QGherwig} and {\sc Jetset} \cite{QGjetset} describe well the
general evolution of the individual jet rates with $y_{\QGindx{cut}}$, 
especially the 3-jet rate \cite{QGKH_D_tune} (see Fig.~\ref{QGKHnjet}
\cite{QGKH_A_jetrates} and Fig.~\ref{QGKHD2delphi} \cite{QGKH_D_tune}).
A more detailed 3-jet Dalitz plot study using the ordered normalized jet 
energies $x_i~(i=1,2,3)$ and ${\cal Z}=(x_2-x_3)/\sqrt{3}$ unveiled slight
differences among the models \cite{QGKH_A_dalitz}. {\sc Ariadne} is in
perfect agreement with the data, {\sc Jetset} is slightly below the data in
the almost 2-jet like case (${\cal Z} \rightarrow 1/\sqrt{3}$) and slightly
above when the lower energetic jets have similar energies (${\cal Z}
\rightarrow 0$). {\small HERWIG} shows a somewhat bigger deviation along
the diagonal (i.e. the $x_3$ direction) of the Dalitz plot. Also the ${\cal
O}(\QGalphas^2)$ ME option of {\sc Jetset} is in good agreement with the data.
It is interesting to observe that the agreement is less good when optimized
scales are used to achieve a better agreement for the 4-jet rate.

\begin{figure}[tb]
\begin{center}
\vspace{-1.cm}
\begin{minipage}[t]{7.5cm}
\mbox{\epsfig{file=qgnjet.eps,height=9cm}}
\caption{\label{QGKHnjet} 
Differential $n$-jet rates compared to {\sc Ariadne}, {\small HERWIG}
and {\sc Jetset PS}.}
\end{minipage}
\hfill
\begin{minipage}[t]{7.5cm}
\mbox{\epsfig{file=qgd2delphi.eps,width=10.5cm}}
\caption{\label{QGKHD2delphi}
Differential 2-jet rate compared to {\sc Ariadne}, {\small HERWIG}
and {\sc Jetset PS \& ME}.}
\end{minipage}
\end{center}
\end{figure}

The discrepancies observed are due either to the different shower evolution 
strategies used or can be traced back to way in which the PS models perform
the matching of the first splitting to the first order matrix element. In
{\sc Ariadne} this matching is performed naturally, since the splitting
function is just the lowest order matrix element expression. If no matching
is performed (a possible option in {\small HERWIG} and {\sc Jetset}) the
agreement with the data is poor.

The 4- and 5-jet rates predicted by {\sc Jetset PS} ({\small HERWIG}) decrease
more (less) rapidly with $y_{\QGindx{cut}}$ than the data 
(see Fig.~\ref{QGKHnjet}). At large $y_{\QGindx{cut}}$ the 
discrepancy is up to 20\% \cite{QGKH_D_tune}. {\sc Ariadne} 
however is in perfect agreement with the data.

Clear discrepancies have been observed at PEP and PETRA comparing the 
{\sc Jetset ME} model to the 4-jet rate. This discrepancy has been 
resolved by
introducing optimized scales \cite{QGKH_scales}. Today using optimized scales
the 4-jet rate is perfectly described by the {\sc Jetset ME} model
\cite{QGKH_A_tune,QGKH_D_tune}. However the 5-jet rate predicted by the ME
model, as is to be expected, decreases far too rapidly and is one order of 
magnitude below the data at large $y_{\QGindx{cut}}$. Recently it has been 
shown that
the 5-jet rate is also correctly described \cite{QGKH_O_5jet} when the ${\cal
O}(\QGalphas^3)$ tree-level graphs are included in the model
\cite{QGKH_karlsruhe_5jet}. The scale in this case can be chosen similarly to
that for the standard ${\cal O}(\QGalphas^2)$ case.

The observations made for the jet rates consistently lead to the following
picture if the models are compared to event shape distributions: general event
shape measures, mainly sensitive to hard gluon radiation, like thrust,
sphericity, $M^2_{\QGindx{high/sum}}/\sqrt{s}$ or $B_{\QGindx{max/sum}}$, are
reproduced extremely well by all PS models \cite{QGKH_D_tune}. The only
significant discrepancy is a slight overestimation of very spherical events by
{\small HERWIG}. Observables sensitive to higher order radiation like minor,
aplanarity, $M^2_{\QGindx{low}}/\sqrt{s}$ and $B_{\QGindx{min}}$ are 
consistently
overestimated (underestimated) by {\small HERWIG} ({\sc Jetset}) for large
values of the observables. Due to the normalization of the distributions this
must also lead to (in general smaller) deviations at intermediate or small
values of these observables. For example the minor distribution in the
case of {\small HERWIG} is predicted to be too wide. {\sc Ariadne} is in
perfect agreement for most distributions. As {\sc Jetset} and {\sc Ariadne}
both use the {\sc Jetset} string fragmentation model, it is evident that the
discrepancies observed for {\sc Jetset} are due to the parton shower part of
the model.

The general fragmentation part of the models are best tested using inclusive 
charged particle distributions which depend strongly on the interplay between
the partonic and fragmentation phases of the models. The average charged
multiplicity \QGnch\ is the integral of the scaled momentum 
($x$) distribution. Both quantities have to be described simultaneously by the
models. When fitting only to the scaled momentum spectrum, {\small HERWIG}
predicts $\langle n_{\QGindx{ch}} \rangle \approx 20.8$ close to the the very 
precisely known LEP~1
average $\langle n_{\QGindx{ch}} \rangle = 20.92\pm0.24$ 
\cite{QGKH_allesandro_brussels}. {\sc Ariadne}
and {\sc Jetset PS} give values that are too small ($\approx20.3$) and 
{\sc Jetset ME} gives too high a multiplicity 
($\approx22.7$) \cite{QGKH_MWdiss}.

The {\small HERWIG} $x$ distribution oscillates slightly around the data
distribution. For small $x$ it is below, for $0.3\leq x\leq 0.7$ it is above
(max. 10\%) and for larger $x$ again below the data. If the multiplicity is
constrained to the measured value, the $x$ spectrum is well described by the
{\sc Jetset PS} and {\sc Ariadne} for $x\leq 0.5 $ but drops 20\%--30\% below
the data for large $x$. This should not to be overinterpreted because
experimental smearing is important in this momentum range and systematic
errors increase. The data so far available from ALEPH and DELPHI
\cite{QGKH_D_tune,QGKH_A_xdistr} agree here only within the full experimental
error. The {\sc Jetset ME} result also oscillates slightly around the data
curve ($\pm5\%$).

Thus the multiplicity distribution is described well by {\sc Ariadne} and 
{\sc Jetset} (compare Fig.~\ref{QGKH_multiplicity}). {\small HERWIG}  
predicts a
slightly too wide distribution thus overestimating the dispersion of the
number of hadrons; in {\small HERWIG} this is strongly coupled to the number
of primary partons.

The transverse momentum in the event plane, \QGptin, is strongly sensitive
to hard gluon radiation and almost correctly described by all models. Only
the large \QGptin\ tail is slightly underestimated. The predicted
\QGptout\ distribution for $p_{\perp\QGindx{out}}>0.8$ GeV falls off 
more rapidly than
the data in all models and at large \QGptout\ is $\approx$ 30\% below the
data! This fact is shown in figure \ref{bildptout} which also compares the
data of ALEPH~\cite{QGKHnewal} and DELPHI~\cite{QGKH_D_tune} to depict the
precision of the experimental data.
The large \QGptout\ tail is mainly due to gluon radiation. This
failure of the shower models is presumably due to missing large angle 
contributions in the basic LLA used by the models. A matching of the second
order matrix element and the LLA shower formalism should lead to an improved
description similar to that of the matched NLLA and ${\cal O}(\QGalphas^2)$
calculations used in $\QGalphas$ determinations \cite{QGKH_alphaspapers}.
For the ME model the situation can be improved by including higher order
terms as has been shown recently by OPAL \cite{QGKH_O_5jet}.
\begin{figure}[t]
\begin{center}
\vspace{-1.cm}
  \begin{minipage}[t]{7.5cm}
     \mbox{\epsfig{file=qgpt_out.eps,width=10.5cm}}
\caption[$p^{out}_{t}$ with respect to the sphericity axis]
{\label{bildptout}
Distribution of \QGptout\ with respect to the sphericity axis 
compared to {\sc Ariadne}, {\small HERWIG} and {\sc Jetset PS \& ME}
\cite{QGKH_D_tune}.}
  \end{minipage}
\hfill
  \begin{minipage}[t]{7.5cm}
     \mbox{\epsfig{file=qgmulti.eps,width=10.5cm}}
\caption[charged multiplicity]
{\label{QGKH_multiplicity}
Multiplicity distribution \cite{QGKH_A_multiplicity} compared to the
DELPHI tuning of {\sc Ariadne}, {\small HERWIG} and {\sc Jetset PS \& ME}
\cite{QGKH_D_tune}.}
  \end{minipage}
\end{center}
\end{figure}

\subsection{Particle composition and spectra}
Experimental studies of the spectra and composition of particles in hadronic
jets provide an unique way to understand the fragmentation of quarks and
gluons into hadrons. Thanks to the excellent performance of the detectors
and high statistics available, very careful work by all four LEP experiments
has given us a very complete picture of the production of identified
particles from e$^+\mbox{e}^-$ annihilation. All states in the $SU(3)$
pseudoscalar and vector meson nonets, except the $\rho^+$, and at least one
state per isospin multiplet in the baryon octet and decuplet, plus the scalar
f$_0(980)$ and the tensors f$_2(1270),\;\mbox{K}^*_2(1430)$ and f$_2^\prime
(1525)$ \cite{QGKH_D_f2prime} have been measured. The average production
rates per hadronic Z event, together with the predictions from the tuned
\cite{QGKH_D_tune} {\sc Jetset} 7.4 and {\small HERWIG} 5.8, are listed in
table \ref{QGYPrate} \cite{QGKH_allesandro_brussels}. The measurements are in
good agreement between experiments for all mesons and octet baryons. However
for the decuplet baryons there are still discrepancies between experiments,
reflecting difficulties in the measurements. In particular, the $\Delta^{++}$
signal is difficult to measure because of its large width and the large
combinatorial background in combinations of $\pi^+$p. The $\Omega^{-}$ rate
seems to be established around the value expected from {\sc Jetset}, contrary
to the old claims of an anomalously high production rate.
\begin{table}[tbp]
\begin{center}
\begin{tabular}{lllll} \hline
 Particle    &  Rate         &  Experiments &  Rate        &   Rate     \\   
             &  Measured     &             & {\sc Jetset 7.4}   &
{\small HERWIG 5.8}  \\
\hline   
All charged    & $20.92\pm 0.24$ & ADLO &    20.81      &     20.94     \\
\hline 
$\pi^0$        & $9.19\pm 0.73$  & DL   &     9.83      &      9.81     \\ 
$\pi^+$        & $8.53\pm 0.22$  & O    &     8.55      &      8.83     \\ 
K$^0$          & $1.006\pm 0.017$ & ADLO &    1.09      &      1.04     \\ 
K$^+$          & $1.185\pm 0.065$ & DO   &    1.12      &      1.06     \\ 
$\eta$         & $0.95\pm 0.11$   & AL   &   1.10       &      1.02     \\ 
$\eta^\prime$  & $0.22\pm 0.07$   & AL   &   0.09       &      0.14     \\ 
\hline 
f$_0(980)$     & $0.140\pm 0.034$ & DL   &   0.16       &      ---      \\
\hline 
$\rho^0$       & $1.29\pm 0.13$   & AD   &   1.27       &     1.43       \\ 
K$^{*0}$       & $0.380\pm 0.021$ & ADO  &   0.39       &     0.37       \\ 
K$^{*+}$       & $0.358\pm 0.034$ & DO   &   0.39       &     0.37       \\ 
$\omega$       & $1.11\pm 0.14$   & AL   &   1.32       &     0.91       \\ 
$\phi$         & $0.107\pm 0.009$ & ADO  &  0.107       &     0.099      \\
\hline 
f$_2(1270)$   & $0.25 \pm 0.08$  & DL   &  0.29        &     0.26       \\ 
K$^*_2(1430)^0$ & $0.095\pm 0.035$ & O  &  0.075       &     0.0785     \\
f$_2^\prime(1525)$ & $0.0224 \pm 0.0062$ & D &  0.026   &     0.03       \\
\hline 
p              & $0.49\pm 0.05$   & DO   &   0.485      &     0.39       \\ 
$\Lambda$      & $0.186\pm 0.008$ & ADLO &   0.175      &     0.184      \\ 
$\Sigma^0$     & $0.0355\pm 0.0065$ & DO &   0.036      &     0.0265     \\ 
$\Sigma^+$     & $0.044\pm 0.006$ & DO   &   0.0343     &     0.0298     \\ 
$\Xi^-$        & $0.0129\pm 0.0007$ & ADO &  0.015      &     0.0247*    \\
\hline 
$\Delta^{++}$  & $0.064\pm 0.033$ & DO   &     0.080    &     0.077      \\ 
$\Sigma(1385)^+$ & $0.011\pm 0.002$ & ADO &    0.009    &     0.0163     \\ 
$\Xi(1530)^0$  & $0.0031\pm 0.0006$ & ADO &    0.00345  &     0.0125*    \\ 
$\Omega^-$     & $0.00080\pm 0.00025$ & ADO &  0.00095  &     0.00385*   \\
\hline 
$\Lambda\bar{\Lambda}$ & $0.089\pm 0.007$ & ADO & 0.085 &     0.134*     \\ 
$\Lambda\Lambda+\bar{\Lambda}\bar{\Lambda}$
               & $0.0249\pm 0.0022$ & ADO &    0.023    &     0.029      \\
\hline 
\end{tabular}
\end{center}
\caption{
Average particle production rates in hadronic Z decays (excluding charge
conjugates and antiparticles), compared to the predictions of {\sc Jetset}
and {\small HERWIG}. A *~indicates that the predicted rate differs from
measurement by more than three standard deviations.}
\label{QGYPrate}
\end{table}

Particle rates could depend on many things, such as flavor content, spin,
mass, phase space, hadron wave functions, Bose-Einstein interference and
other collective effects. The two most frequently used models {\small
HERWIG} and {\sc Jetset} use different ways to account for the particle
production rates. In the Lund/{\sc Jetset} approach (similarly to the old
Field \& Feynman model \cite{QGKH_FF}), the production rate of a specific
hadron type depends principally on its flavor content and spin. One can
also use essentially pure phase space as in the case of the {\small
HERWIG} cluster fragmentation approach. 

Studies of general features of particle production, such as the strangeness
suppression factor s/u or the fraction of mesons produced in spin-1 states,
$V/(V+P)$, or in orbitally excited states provide useful information about
the main production mechanisms. The (one dimensional) string model suggests
the production of orbitally excited states is small $\approx$10\%
\cite{QGlund} whilst $V/(V+P)=3/4$ is expected from simple spin counting.

Measurements of the f$_0(980),\;\mbox{f}_2(1270),\;\mbox{K}_2^*(1430)^{0}$
\cite{QGKH_allesandro_brussels} and f$_2^\prime(1525)$ \cite{QGKH_D_f2prime}
as well as of D$^{**}$ and B$^{**}$ mesons indicate that
orbitally excited states, most of which so far were not included in {\small
HERWIG}, {\sc Jetset} and other models, are copiously produced ($\approx$
30\% of the primary hadrons). Thus a quite large fraction of the observed
stable particles come from decays of these numerous states. As a result, the
$V/(V+P)$ ratio can differ significantly from that when no orbitally excited
states are considered. From a global tuning, where the orbitally excited
meson states are included, a value of $V/(V+P)\approx 0.4--0.6$ is obtained
for light mesons \cite{QGKH_D_tune,QGYPhemi}. This low value of $V/(V+P)$
could be explained by mass differences between the vector and pseudoscalar 
mesons, i.e.
by the relatively larger binding energy of pseudoscalar mesons\cite{QGlund}.
The measured ratio of $V/(V+P)=0.75\pm0.04$ \cite{QGYPbbst} for B mesons
agrees well with the expected value of 3/4. However for D mesons the much
lower value $0.46 \pm0.06$ \cite{QGYPddst} is still not understood.

In the string fragmentation model, one  expects the strangeness suppression
factor s/u to be around 0.3 using the typical values of (constituent) quark
masses. This parameter can be measured from the production rates of strange
compared with non-strange mesons and from the momentum spectrum of strange
mesons. The results, which are summarized in table~\ref{QGYPs2u}, are very
consistent with the expectation\footnote{However, neutrino experiments at
lower energies \cite{QGYPs2u1} and recently both ZEUS \cite{QGYPs2u2} and H1
\cite{QGYPs2u3} require a lower value of about 0.2 for s/u. More careful 
studies in this area are needed in the future.}. It is interesting to see
that s/u determined from heavy mesons agrees well with the values obtained
from light mesons. This suggests that the strangeness suppression occurs at
the quark level. 
\begin{table}[htbp]
\begin{center}
\begin{tabular}{lll} \hline   
Technique                         & Results        & References \\ \hline
$\frac{\mbox{K}^{*0}}{\mbox{$\rho^0+\omega$}},\;
\frac{\mbox{K}^{*\pm}}{\mbox{$\rho^0+\omega$}},\;
\frac{\mbox{$2\phi$}}{\mbox{K}^{*0}},\;
\frac{\mbox{$2\phi$}}{\mbox{K}^{*\pm}},\;
\sqrt{\frac{\mbox{$2\phi$}}{\mbox{$\rho^0+\omega$}}}$ & 
$0.29\pm 0.03$ & table~\ref{QGYPrate} ~\cite{QGYPypei}
\\
$\frac{\mbox{K}^+}{\mbox{$\pi^+$}}$ at high momentum  & 
$0.35\pm 0.07$(stat) & ~\cite{QGYPypei,QGYPokp} \\
$\frac{\mbox{K}^+}{\mbox{$\pi^+$}}$ at high momentum  & 
$0.25\pm 0.03$(stat) & ~\cite{QGYPypei,QGYPakp} \\
K$^0$ momentum spectrum  & $0.285\pm 0.035$  & ~\cite{QGYPok0p} \\
K$^0$ momentum spectrum  & $0.30 \pm 0.02$(stat) & ~\cite{QGYPdk0p} \\
\hline
Ratio $ \frac{\mbox{$2f(\mbox{c}\rightarrow\mbox{D}^+_{\QGindx{s}})$}}%
{\mbox{$f(\mbox{c}\rightarrow\mbox{D}^+
+f(\mbox{c}\rightarrow\mbox{D}^0)$}} $ & $0.31\pm 0.07$ & ~\cite{QGYPlepd} \\
Ratio $ \frac{\mbox{2B}_{\QGindx{s}}}%
{\mbox{B}_{\QGindx{u,d}}}$ & 
$0.32\pm 0.08$ & ~\cite{QGYPaleb} \\
Ratio $ \frac{\mbox{2B}^{**}_{\QGindx{s}}}%
{\mbox{B}^{**}_{\QGindx{u,d}}}$ & $0.28\pm 0.11$ &
~\cite{QGYPdelb} \\
B$^0\bar{\mbox{B}}^0$ mixing ($\chi_{\QGindx{B}},
\chi_{\QGindx{B}_{\QGindx{d}}}),\;
f_{\Lambda_{\QGindx{b}}}$ & $\sim 0.3$ & ~\cite{QGYPbmix} \\
\hline
\end{tabular}
\end{center}
\caption{Measurements of s/u at LEP}
\label{QGYPs2u}
\end{table}

The relative rates of the decuplet baryons as well as those of the
$\Sigma^{0,+}$ and $\Xi^-$ provide a clean test of models, since they
are less affected by resonance decays. From table~\ref{QGYPrate}, we
obtain the following ratios:
\begin{eqnarray*} 
\mbox{Ratio~~~~~~~~~}         & \mbox{Measured} &
\mbox{\sc Jetset~~~~}\mbox{\small HERWIG} \\
\Xi^-/\Sigma^+     \mbox{~~~~~~~}    & 0.29 \pm 0.04 &
\mbox{~~~0.44~~~~~~~~~~0.84} \\ 
\Xi^-/\Sigma^0     \mbox{~~~~~~~~}   & 0.36 \pm 0.07 &
\mbox{~~~0.42~~~~~~~~~~0.93} \\
\Sigma(1385)^+/\Delta^{++}~~         & 0.17 \pm 0.09 &
\mbox{~~~0.11~~~~~~~~~~0.21} \\
\Xi(1530)^0/\Sigma(1385)^+           & 0.28 \pm 0.07 &
\mbox{~~~0.38~~~~~~~~~~0.77} \\
\Omega^-/\Xi(1530)^0     \mbox{~~~~} & 0.26 \pm 0.09 &
\mbox{~~~0.28~~~~~~~~~~0.31}
\end{eqnarray*} 
One obtains from the above ratios a suppression factor of about 0.28 per
s quark for baryons (0.24 if only the decuplet baryons are considered).
This is similar to the value obtained for mesons, suggesting that the
additional suppression for diquarks might be small.

After being tuned to LEP~1 data, {\small HERWIG} and {\sc Jetset}\footnote{New
parameters have to be introduced, as attempted in \cite{QGKH_D_tune}, to treat
the quark type dependent production probabilities for pseudoscalar, vector and
orbitally excited mesons.} describe well the measured rates in the meson 
sector.
There is a fairly good agreement in the baryon octet, except that the proton
rate is slightly underestimated and the $\Xi^-$ rate is overestimated by about
a factor of two by {\small HERWIG}. In the baryon decuplet {\sc Jetset}
predictions are consistent with the data while the predictions of {\small
HERWIG} differ from the data in most of the cases. Differences in the ratios 
of
the baryon rates between {\small HERWIG} and data, as shown above, can not be
solved simply by tuning the cluster fragmentation parameters, indicating the
need for real dynamics beyond phase space and spin counting.

Although in general {\sc Jetset} describes the measured rates better than
{\small HERWIG}, it contains a large number of free parameters. As a result,
it has little predictive power. The {\sc UCLA} model \cite{QGucla}, 
a variant of {\sc Jetset} with less parameters, does a good job in many
cases but has problems in the baryon decuplet. Also the copiously produced
orbitally excited mesons so far are not included in the {\sc UCLA} model. In
\cite{QGYPchli} an interesting regularity in production rates is shown 
for all particles (except pions) belonging to the pseudoscalar and vector
meson nonets and the baryon octet and decuplet. The particle multiplicity
can be described by a simple exponential fall off in mass squared and $2J+1$
spin counting factors. This regularity seems to be energy independent and
has recently been established similarly also in $\QGbar{p}$p interactions
\cite{QGpbarp}. However it is necessary to use generalized isospin multiplets
and to not seperate the contributions from resonance decays. Recently a new
approach~\cite{QGYPbeca} has been proposed which uses only three free
parameters but reproduces the measured rates quite well. The basic assumption
used is that hadrons reach complete thermal and chemical equilibrium, in
contrast to the general belief that e$^+\mbox{e}^-\rightarrow$hadrons is a
rapidly expanding process and during fragmentation the hadronic density is
rather low. More tests are needed to check this thermodynamic approach.  

Since all fragmentation models contain a number of parameters which can be
tuned according to data (more dramatic in the case of {\sc Jetset}),
measurements of production rates do not provide a high discriminating power
among different models. A more effective method is to look at baryon
correlations. In {\sc Jetset} the major source of baryon production is the
creation of a diquark-antidiquark pair within the fragmentation. The
baryon-antibaryon (${\cal B}\bar{{\cal B}}$) rate is much higher than the
${\cal BB}$ and $\bar{{\cal B}}\bar{{\cal B}}$ rate (see table~\ref{QGYPrate})
and ${\cal B}\bar{{\cal B}}$ are more likely to occur close in phase space
than ${\cal BB}$ or $\bar{{\cal B}}\bar{{\cal B}}$. Correlations between
${\cal B}\bar{{\cal B}}$ can be reduced by the {\it popcorn} mechanism,
allowing a meson to be created in between a ${\cal B}\bar{{\cal B}}$ pair.  
As can be seen from table~\ref{QGYPrate}, {\small HERWIG} overestimates the 
$\Lambda\bar{\Lambda}$ rate (note that the prediction for the $\Lambda$ is
quite good), while {\sc Jetset} with popcorn describes the data well. It has
been shown in \cite{QGYPpop} that ${\cal B}\bar{{\cal B}}$ correlations, for
example in rapidity, are overestimated by {\small HERWIG}, whilst {\sc
Jetset} with a high probability of the popcorn occurrence ($\sim 80\%$) 
reproduces the data well. A more impressive test is to study the angle
between the baryon and the event axis in the ${\cal B}\bar{{\cal B}}$ rest
frame. The string model predicts that baryon production is preferentially 
lined up along the event axis, while the cluster model predicts an isotropic
distribution. Data~\cite{QGYPpop} clearly favor the string model. Also
measurements of baryon and antibaryon production in quark and antiquark jets
with polarized beams by SLD~\cite{QGYPsld} and jet charge
studies~\cite{QGKH_A_afbbrussels} support the string model but disfavor the
idea of isotropic cluster decays. 

Identified particle spectra have been studied  as function of both the scaled
momentum $x_p=p/p_{\QGindx{beam}}$ and the variable 
$\xi_p=\log{\frac{1}{x_p}}$. 
In general all models describe the data fairly well, with few discrepancies
remaining:

\begin{QGitemize} 
\item
Data show a harder momentum spectrum for the $\eta$ produced in gluon jets
\cite{QGYPl3et}.
\item
K$^\pm$ momentum spectra predicted by the models are too
soft~\cite{QGYPokp,QGYPakp,QGlepk}. This might be caused by wrong branching
fractions of b hadrons in the models.
\item
Momentum spectra of light quark baryons predicted by the models are too
hard~\cite{QGYPokp,QGYPakp,QGYPpop,QGlepk,QGYPlepb}. This indicates a
different production mechanism for baryons than for mesons. Partly it may
also be due to missing orbitally excited baryon states in the models.
However, so far these states have not been observed in e$^+\mbox{e}^-$
annihilation.
\end{QGitemize}  

The heavy quark fragmentation function has been measured at LEP~1 mainly
using D$^{(*)}$ reconstruction in the c-quark case \cite{QGKH_cfrag} and
using high-$p_{\perp}$ lepton spectra \cite{QGKH_bfrag_leptons}, D$^*-$lepton
combinations \cite{QGKH_bfrag_Dstarlepton}, and exclusive
\cite{QGKH_bfrag_exclusive} and inclusive b-reconstruction
\cite{QGKH_bfrag_inclusive} in the case of b-quark fragmentation. The
D-meson distributions are obscured by contributions from b-hadron decays.
Today the (experimentally involved) inclusive b-hadron reconstruction yields
the best statistical precision. It allows for the first time (besides a
precise determination of the average b-hadron energy $\langle x_E \rangle$) 
a decisive
comparison to different fragmentation models. This, so far incomplete
comparison, gives best agreement for LLA based parton shower models 
({\sc Ariadne, Jetset} and {\small HERWIG}) combined with Peterson 
fragmentation \cite{QGKH_peterson}. The {\small HERWIG} cluster 
fragmentation as well as
the Lund-symmetric and the modified Lund-Bowler ansatz give less
satisfactory results. In the case of {\sc Jetset ME} with Peterson
fragmentation a too narrow energy distribution indicates the lack of soft
gluon emission.

\subsection{Differences between q and g jets}
In QCD, the gluon is associated with a color charge $C_A=3$ and the quark
with a charge $C_F=4/3$. The larger color charge of the gluon means that
it is more likely to radiate an additional gluon than a quark, leading to
differences in the expected properties of quark- and gluon-induced jets.
For quark and gluon jets produced with the same energy and under the same
conditions, gluon jets are expected to have a larger mean particle
multiplicity than quark jets~\cite{QGBG-L2th1}. The larger multiplicity of
the gluon jet implies that its particle energy spectrum, known as the
fragmentation function, is softer. A related prediction is that the mean
opening angle of particles in a gluon jet is larger than in a quark
jet~\cite{QGBG-L2th2}: thus the gluon jets are broader. Much experimental
effort has been invested in an attempt to observe these predicted
differences (for a recent compilation, see~\cite{QGBG-L2jwg} and references
therein). Before LEP~1, there were experimental indications that gluon jets
were indeed broader than quark jets, based on measurements of the mean
transverse momentum of particles in a jet with respect to the jet axis, or
similar variables. However, contradictory results were published concerning
differences between the quark and gluon jet fragmentation functions, while
no evidence was found for a multiplicity difference between the two jet
types. In general, it proved difficult to obtain conclusive results on
quark-gluon jet differences at facilities before LEP~1 either because biases
were introduced by assuming the gluon jets to be the lowest energy jets in
e$^+\mbox{e}^-$ three-jet events or else because there was no event-by-event
identification of gluon jets with a resulting lack of sensitivity.

Due to large event statistics and good detector capabilities, the LEP
experiments have been able to settle the experimental question of quark and
gluon jet differences~\cite{QGBG-L2op1,QGBG-L2de}. Three aspects of
the LEP~1 studies allow this success. (1)~Symmetric events were selected in
which the quark and gluon jets being compared had the same energy and angles
relative to the other jets, allowing a direct, model independent comparison
of the jet properties. (2)~The quark jets were tagged, leading to
identification of the gluon jets with better than 90\% purity through
anti-tagging. (3)~The anti-tagged gluon jet data were combined algebraically
with the quark and gluon jet data from the untagged, symmetric events, 
leading to separated quark and gluon jet measurements with essentially 
no biases except from the jet definition. In the first LEP~1 studies, 
the quark jet samples were the natural ones for Z$^0$ decay, given by the 
Z$^0$ coupling strength to the individual flavors, corresponding to 
roughly 20\% d, u, s, c and b quarks. In
later studies, b quark jets and uds quark jets were explicitly selected to
compare to gluon jets~\cite{QGBG-L2op3}.

These studies resulted in a confirmation of the qualitative differences
between quark and gluon jets given above. Selecting 24~GeV jets in a
so-called ``Y'' symmetric event topology, it was shown that gluon jets were
60--80\% broader than quark jets as measured by the full width at half maximum
of the differential energy and multiplicity profiles~\cite{QGBG-L2op2}. The
fragmentation function of the gluon jet was observed to be much softer than
that of the quark jet. The mean charged particle multiplicity of gluon jets
was found to exceed that of quark jets by 20--25\%. Besides the Y events,
DELPHI~\cite{QGBG-L2de} studied 30~GeV jets from three-fold symmetric
``Mercedes'' events and obtained similar results. The comparison of the
fragmentation function of quark and gluon jets in Y and Mercedes events shows
the expected stronger energy dependence for gluon jets. Extensive comparisons
of Monte Carlo predictions to the quark and gluon jet data are presented
in~\cite{QGBG-L2op2} and~\cite{QGBG-L2op3}. {\sc Ariadne}, {\small HERWIG} and
{\sc Jetset} were found to be in good agreement with the measurements. The
{\sc Cojets} agreement was somewhat less good.

ALEPH~\cite{QGBG-L2al2} extended these studies by including a measurement of
sub-jet multiplicities~\cite{QGBG-L2sj}. For small values of the sub-jet
resolution scale, $y_0$ (defined using the $k_{\perp}$ jet finder), the ratio
of the gluon to quark jet mean sub-jet multiplicity was found to be similar
to the hadron level value of about 1.2 discussed above. After subtracting one
from the mean sub-jet multiplicities to account for the contributions of the
initiating quarks and gluons, the sub-jet multiplicity ratio of gluon to quark
jets was observed to reach a much larger value of about 2.0 as $y_0$
approached the resolution scale $y_1$ at which the jets were defined. The
explanation for this is that the mean sub-jet multiplicity of the quark jets
approaches unity slightly before that of the gluon jets as 
$y_0\rightarrow y_1$. {\sc Ariadne}, {\small HERWIG}, {\sc Jetset} and 
{\sc NLLjet} were all found to reproduce the measurement.

Beyond these studies based on symmetric events, ALEPH and DELPHI have examined
quark and gluon jet properties in non-symmetric three-jet event 
configurations. The DELPHI approach~\cite{QGBG-L2de} is to identify gluon jets
in
three-jet events using anti-tagging methods as mentioned above. The gluon jet
properties were compared to those of quark jets with similar energies found in
radiative QED q$\QGqbar\gamma$ events. The qualitative differences discussed
above between quark and gluon jets were observed to be present for jet 
energies
between 5 and 40~GeV and were well reproduced by {\sc Jetset}.
ALEPH~\cite{QGBG-L2al3} introduced a new method to study the multiplicity
difference between quark and gluon jets in three-jet events, by examining the
mean charged particle multiplicity of the entire event as a function of the
energies and opening angles of the jets in the event. Assuming each event to 
be
composed of a gluon jet and two quark jets, and that every particle in an 
event
could be associated with one of these jets, a fit was made to extract a value
for the ratio of the mean charged particle multiplicity values of gluon 
to quark
jets, $r_{\QGindx{ch}}$. The result for all jet energies and event topologies 
was $r_{\QGindx{ch}}=1.48$. The fit results were found to agree well with 
those from the
symmetric Y analyses when they were restricted to that geometric situation.

Thus the basic differences expected between quark and gluon jets --- a larger
mean multiplicity, a softer fragmentation function and a larger angular width
of gluon relative to quark jets --- are now all well established by the LEP~1
experiments. The QCD models are in good overall agreement with the measured
differences. Future effort in this field at LEP~1 will probably include 
studies
of differences in the identified particle rates in gluon and quark jets,
differences in particle correlation phenomena and attempts to reduce the
reliance of the analysis method on the jet definition (as the ALEPH
study~\cite{QGBG-L2al3} discussed above attempts to do). Already, L3 has
presented results which indicate an enhanced $\eta$ meson production rate in
gluon jets compared to the rates predicted by {\small HERWIG} and 
{\sc Jetset}~\cite{QGYPl3et}. This suggests that the models for 
gluon jets may need
to be modified to allow for an enhanced production of isosinglet
mesons~\cite{QGBG-L2pw}.

\subsection{Coherence}
Gluon radiation in the parton shower should be coherent. However, gluon
interference only becomes apparent when one goes beyond the Leading Log
Approximation (LLA). A number of such effects are found in the next
simplest approximation, the Modified LLA (MLLA) \cite{QGcohrev}. Due to
the non-abelian nature of QCD, the overall result of this interference
is ``angular ordering'' of the gluon radiation \cite{QGWJMao}, which
constrains the angles between the radiator and the radiated gluon to
decrease as the evolution proceeds to lower scales.

In parton-shower Monte Carlos gluon interference is either: imposed as an
{\it a posteriori}\/ constraint on gluon opening angles as in {\sc Jetset}
\cite{QGjetset}; built into the choice of evolution variable as in {\small
HERWIG} \cite{QGherwig}; or neglected in independent fragmentation models
such as {\sc Cojets} \cite{QGWJMcojets}. {\sc Ariadne} \cite{QGariadne}, on
the other hand, employs a formulation based on a cascade of q$\QGqbar$, qg
and gg dipoles which naturally incorporates interference phenomena. In 
{\sc Jetset} the angular-ordering constraint can be turned off. 
By comparing {\sc Jetset} with and without angular ordering one 
can obtain an idea of the importance of the effect. 

Some consequences of gluon interference have been calculated directly in
perturbative QCD as well as by Monte Carlo.  Such calculations apply,
strictly speaking, only to partons. Comparison with data relies on the
additional assumption of Local Parton Hadron Duality (LPHD)
\cite{QGprecon,QGWJMlphd}, which posits that many distributions of hadrons
rather closely follow the corresponding parton distribution, with
non-perturbative effects affecting mainly the normalization rather than the
shape of the distributions. However, we shall not emphasize such calculations
here, since our main purpose is to evaluate the adequacy of current Monte
Carlo programs.

The first effect to be explained \cite{QGWJMstringcoh} as a consequence of
gluon interference was the so-called string effect; first predicted using
(non-perturbative) string fragmentation phenomenology \cite{QGWJMlundstring}
and later discovered by the JADE experiment \cite{QGWJMjadestring}. In terms
of gluon interference it is explained as a purely perturbative effect at the
parton level. The string effect has been extensively studied, most recently
by DELPHI \cite{QGBG-L2de}, L3 \cite{QGWJMl3string} and OPAL
\cite{QGWJMopalstring}. These analyses have compared q$\QGqbar$g 
and q$\QGqbar\gamma$ events taking care to have samples of comparable 
kinematic
configurations. The string effect appears as a smaller particle flow in the
region between the quark jets in q$\QGqbar$g than in q$\QGqbar\gamma$ events.
ALEPH \cite{QGWJMalephstring} instead compared the particle flow between the
quarks with that between quark and gluon. The string effect is found to be
rather well reproduced by the coherent Monte Carlo models but not by the
incoherent ones. However this success is not entirely due to the coherence
at parton level; the non-perturbative modelling in the programs also
contributes.

It is also worth mentioning that evidence of gluon interference is also seen
in p$\QGbar{p}$ interactions at the Tevatron. Using events with 3
high-$p_{\perp}$ jets CDF examined the differences in rapidity and in
azimuthal angle between quark and gluon jets \cite{QGWJMcdfstring}. {\small
HERWIG}, which incorporates coherence in both space-like and time-like
showers, reproduced the data well. {\sc Pythia/Jetset}, with coherence only
in time-like showers did less well, although it improved when modified to
partially incorporate coherence in space-like showers. {\sc Isajet}, with no
coherence, performed poorly.

As is well known \cite{QGcohrev}, gluon interference leads to suppression
of soft gluons in the shower, which in turn should lead to a suppression of
soft hadrons. The distribution of $\xi_p=-\ln x_p=-\ln p/p_{\QGindx{jet}}$
is expected to have a roughly Gaussian shape and its peak position,
$\xi^\ast$, should increase  with $\sqrt{s}$. The dependence of $\xi^\ast$
on $\sqrt{s}$ is strikingly different in the MLLA from that in the LLA.
Assuming LPHD, $\xi^\ast$ is expected to show similar behaviour. Many
comparisons have been made, for many types of particle, using data from
PETRA/PEP, TRISTAN, and LEP; they support the form predicted by MLLA and
clearly reject the LLA form.

{}From MLLA+LPHD it is expected \cite{QGWJMxistar_m_dep} that $\xi^\ast$
decrease with the mass of the hadron. This is indeed found to be the case
with $\xi^\ast$ being approximately proportional to $-\ln
M_{\QGindx{hadron}}$. However, the proportionality constant is quite
different for mesons and baryons \cite{QGlepk}. This difference is due to
decays. When the $\xi_p$ distributions are corrected for decays
\cite{QGYPakp,QGlepk,QGWJMopalspect}, using {\sc Jetset}, the $\xi^\ast$
values of mesons and baryons are found to lie on a universal curve
\cite{QGlepk}. The conclusion is clearly that we must be cautious about
the interpretation of LPHD, in particular with the inclusion of decays.

The $\sqrt{s}$ dependence of $\xi^\ast$ is support for MLLA, but says little
about the quality of the Monte Carlo programs, since they are retuned at each
value of $\sqrt{s}$. However, accepting the validity of MLLA, the improvement
seen in the previous paragraph supports the description of non-perturbative
hadronization in the model.

The angular ordering resulting from gluon interference effectively moves the
radiated gluons closer to the jet axis.  The size of the effect depends on the
colour charge of the radiator and on the initial configuration of the event
(i.e. 2 or 3 jets, there being interference effects in the interjet region for
3-jet events as seen in the string effect). The total number of (sub-)jets
found in an event has been calculated \cite{QGBG-L2sj} in the Next-LLA (NLLA)
as a function of the jet resolution parameter, $y_{\QGindx{cut}}=y_0$, for 2-
and 3-jet events classified using $y_{\QGindx{cut}}=y_1>y_0$. Of course, if
$y_1$ becomes too small non-perturbative processes become important and the
calculation breaks down. The perturbative and non-perturbative regions are
rather clearly separated and the sub-jet multiplicities thus provide a test 
not
only of perturbative QCD calculations and their incorporation into Monte Carlo
programs, but also of the non-perturbative models in the programs.  

Sub-jet multiplicities have been studied \cite{QGWJMsubjetl3,QGWJMsubjetopal} 
at LEP~1. Quite good qualitative agreement is found between the data and the
NLLA calculations in the perturbative region while a simple ${\cal O}(
\QGalphas)$ calculation clearly disagrees. Of the Monte Carlo
programs, {\sc Ariadne} does quite well; {\small HERWIG 5.5} and 
{\sc Jetset 6.3} perform somewhat less well; and the incoherent model 
{\sc Cojets} gives
the worst agreement. Both versions 6.12 and 6.23 of {\sc Cojets} disagree in
the perturbative region while only 6.23 disagrees in the non-perturbative
region. {\sc Jetset} was compared \cite{QGWJMsubjetopal} using various
combinations of fragmentation and parton shower schemes. Incoherent parton
showers resulted in poor agreement in both perturbative and non-perturbative
regions independently of the fragmentation scheme. Coherent showers gave much
better agreement in the perturbative region. In the non-perturbative region
agreement was poor for independent fragmentation whilst good for string
fragmentation.

The MLLA predicts \cite{QGWJMhfmult} a suppression of gluon emission within
a cone of angle $\theta<\theta_0=M_{\QGindx{q}}/E_{\QGindx{q}}$ about the
quark direction in a parton shower. This should lead to a lower {\em primary}
multiplicity for heavy quark events. However, the total multiplicity is higher
because of the high multiplicity of heavy flavour decays. The difference in
multiplicity between heavy and light quark events is predicted to be
independent of $\sqrt{s}$, contrary to the na\"\i ve expectation that the
difference would decrease as the quark mass difference becomes smaller 
compared
to the total energy. Results from PEP/PETRA, TRISTAN, and LEP/SLC agree
reasonably well with the MLLA value, both for charm and beauty, particularly
when the recent work of Petrov and Kisselev \cite{QGWJMhfmult_mod} is taken
into account. The results \cite{QGWJMhfmult_D,QGWJMhfmult_O,QGWJMhfmult_S} at
$\sqrt{s}= M_{\QGindx{Z}}$ agree reasonably well with the predictions of 
{\sc Jetset}, with the possible exception of light (uds) quarks.

Given the appearance of angular ordering in MLLA, the effects of gluon
interference should be apparent in angular correlations. Assuming LPHD, the
correlations should persist in the hadrons. Besides the angular ordering in
the polar angle, also the azimuthal angular distribution is affected by gluon
interference. OPAL \cite{QGWJMazcor}, has studied two-particle correlations
in the azimuthal angle within restricted rapidity intervals. To avoid defining
a jet axis they, and more recently ALEPH \cite{QGWJMcoh_aleph}, have also
studied such correlations using the Energy-Multiplicity-Multiplicity
Correlation (EMMC) \cite{QGWJMemmc}. Taking in turn each track's direction as
an axis the correlation calculated, the EMMC is the average of these
correlations weighted by the axis defining track's energy. The EMMC has been
calculated analytically in leading \cite{QGWJMemmc} and next-to-leading
\cite{QGWJMemmc2} order; the corrections are large. LPHD must be assumed to
apply these calculations to those calculated from hadrons. Nevertheless,
qualitative agreement is obtained for $\phi>\pi/2$, where Monte Carlo models
show hadronization to be relatively unimportant.  Agreement with the data is
even better for Monte Carlo models which incorporate gluon interference.
Models not incorporating this interference fail to describe the data.

ALEPH \cite{QGWJMcoh_aleph,QGWJMaleph_note} and L3
\cite{QGWJMalythesis,QGWJMppca_l3} have studied two-particle angular
correlations in the full spatial angle using the Asymmetry in the
Particle-Particle Correlation (APPC). In addition, L3 has studied the Asymetry
in the Energy-Energy Correlation AEEC. The APPC is defined in analogy to the
well-known AEEC by simply removing the energy weighting. This results in a
correlation which is sensitive to all branchings of the shower, whereas the
AEEC is primarily sensitive to the earliest branchings. The APPC is less
sensitive to systematics in the correction for detector effects. On the other
hand, the energy weighting makes the AEEC less sensitive to the Bose-Einstein
effect. The use of the asymmetry serves to cancel some of the correlations
arising from other effects as well as some detector effects and Monte Carlo
uncertainties.

These correlations have been compared with Monte Carlo models. The conclusion
is that the models containing gluon interference agree much better with the
data than do the incoherent models. However neither version of {\sc NLLjet}
can be said to agree well.

All of these studies favour the Monte Carlo models {\sc Ariadne}, {\small
HERWIG} and {\sc Jetset}, which incorporate the gluon interference expected
in MLLA. In general the agreement of data with these models is quite good.
On the other hand, models that do not incorporate gluon interference, such
as {\sc Cojets} and incoherent {\sc Jetset} do not in general agree well
with the data.  Both coherent and incoherent versions of {\sc NLLjet} have
been found also not to agree well with data.

\subsection{Prompt photons}
The principal source of observable prompt photons in hadronic decays of the
Z (i.e. those with energies greater than a few GeV) is final state radiation
(FSR) emitted at an early stage in the parton evolution process initiated by
the primary quark--antiquark pair. To reduce large backgrounds from non-prompt 
sources, the first measurements reported by OPAL~\cite{QGJCTone} and followed
later by the other LEP experiments selected events with photons well isolated
from the hadrons by a geometrical cone followed by a 2-step jet reconstruction
process. In this procedure, the candidate photon is first removed from the
event and all the hadrons reconstructed into jets. Then, the photon is
replaced and its isolation from the jets tested in a second application of the
clustering algorithm. It was soon realized that the cross sections are
substantially less than those predicted from fractionally charged fermion
pairs due to the influence of gluons. Thus, the measurement of prompt photons
has become a sensitive test of the predictions of both perturbative QCD matrix
element calculations and the Monte Carlo shower models free from the direct
effects of fragmentation. 

After tuning the parameters of these models in recent versions, namely 
{\sc Ariadne 4.2}, {\small HERWIG 5.4} and {\sc Jetset 7.3} 
to the properties of
the hadrons observed in non-FSR events, there is no freedom to adjust the
photon emission parameters with the exception of the infra-red cut-off. In
the following reported analyses, these cut-off values are chosen to be similar
to those employed to terminate the parton evolution, but in any case do not
significantly influence the isolated hard photon rates.

All the published high statistics analyses from ALEPH, L3 and
OPAL~\cite{QGJCTtwo} show that both {\sc Ariadne} and {\small HERWIG} give
acceptable descriptions of the total and individual $n$-jet + $\gamma$ cross
sections as a function of the jet resolution parameter, 
$y_{\QGindx{cut}}$ (JADE E0),
as well as the distributions in $p_{\perp}$ and fractional energy 
$z_{\gamma}$ of
the photon. A more critical test to differentiate between these two models
is based on their predictions for the rate of low energy photons ($<15$ GeV) 
at large angles ($>75^{\circ}$) to the event thrust axis, where the evolution
scale ordering used in {\small HERWIG} predicts a larger cross section than
{\sc Ariadne} \cite{QGJCTthree}. Preliminary data from ALEPH indicate that
{\sc Ariadne} gives the better description but more statistics are needed. 
However, the above published results show that {\sc Jetset} is less
satisfactory predicting cross sections that are 20-30\% low (3$\sigma$'s).
ALEPH showed that this can be improved by either switching off the
$\cal{O}(\QGalphas)$ matching or by keeping $\QGalphas$ constant indicating
that virtuality as the scale controlling the parton evolution is not the best
choice. More recently, DELPHI has also shown~\cite{QGJCTfour} that their data
are in excess of {\sc Jetset} by 18$\pm$7\% in the low energy region of the
photon spectrum below 15 GeV. After clustering the hadrons with the Durham
($k_{\perp}$) algorithm in a similar 2-step procedure, their respective jet 
rates above $y_{\QGindx{cut}}\geq 0.01$ are in reasonable agreement with 
{\sc Jetset}. The
excess is largely eliminated since most low energy photons are no longer
isolated when the clustering algorithm is applied a second time. This appears
to be a different conclusion from the other experiments. However, careful
examination shows that the discrepancy between {\sc Jetset} and data for
ALEPH and L3 are largely at low $y_{cut}$ in the total cross section where
the use of different algorithms for jet-finding makes comparison difficult
with DELPHI. It should be noted also that DELPHI compare with {\sc Jetset}
at {\em hadron} level before fragmentation corrections are applied.

The 2-step analysis procedure to select isolated photon events does not
prevent a significant number of non-isolated hard photons from contaminating
the $\gamma+1$-jet event topology. Each of these photons remain within the
hadron jet formed from the remnant of the radiating quark and are better
separated from the isolated radiation by a ``democratic''
analysis~\cite{QGJCTfive}. Here, the prompt photon candidate is not removed
from the event and thus becomes a member of a hadron jet with fractional
energy $z_\gamma$ of its total energy. The true isolated component is now
concentrated at $z_\gamma=1$ broadened downwards in $z_\gamma$ by
hadronization effects to overlap with the high energy tail of the collinear
quark fragmentation component. For the  $\gamma+1$-jet (ie: 2-jet) cross
section, this is well separated from the fragmentation tail when $z_\gamma
\geq 0.95$. Fig.~\ref{rdelta_2j} shows the comparison as a function of
$y_{\QGindx{cut}}$ (Durham E0 scheme) between the data measured by ALEPH and 
the predictions of {\sc Ariadne} and {\sc Jetset} for this isolated component.
The continuous curve is a prediction of a leading order calculation dominated
by perturbative terms which are derived from a pure QED calculation. {\small
HERWIG} (not shown) is in close agreement with the data. {\sc Jetset} falls
well below the data in this case showing that its treatment of radiation as
independent emission from either quark at the first branching is quite
inadequate. This discrepancy diminishes as the jet multiplicity increases.
There is more satisfactory agreement in the fragmentation region below
$z_\gamma=0.95$.  

\begin{figure}[tb]
\vspace{-1.cm}
\begin{minipage}[t]{7.5cm}
  \vspace{-1cm}
  \mbox{\epsfig{file=qgpub_fig6.eps,width=9cm}}
  \end{minipage}
\hfill
\begin{minipage}[t]{7.5cm}
 \caption{
Integrated 2-jet rate above $z_{\gamma}$ = 0.95 as a function of 
$y_{\QGindx{cut}}$,
compared with {\sc Ariadne, Jetset} and a QCD calculation.}
 \label{rdelta_2j}
\end{minipage}
\end{figure}

Overall, the conclusion is that {\small HERWIG} gives the best description of
all prompt photon data at the Z closely followed by {\sc Ariadne}.
 
In this review  it is appropriate to mention the difficulties faced in 
determining the non-prompt photon background coming from hadrons decaying
into $\gamma$'s (mainly $\pi^0$). The isolation and energy cuts applied to the
prompt photon candidates in the 2-step analyses are insufficient to eliminate 
this background entirely even when the full granularity of the electromagnetic
calorimeters is exploited to recognize single from multiple $\gamma$ showers.
Hence, an irreducible non-prompt component must be subtracted statistically
using QCD models or inferred from other data. However, the selection cuts
applied choose a region of phase space that is not well understood in these
models as they correspond to tails in the fragmentation process which cannot
be tuned precisely.
 
The early analyses made at LEP~1 showed a clear discrepancy in the hadronic
background yield predicted by the {\small HERWIG} and {\sc Jetset}
models~\cite{QGJCTsix}. The magnitude of the differences depends strongly on
the isolation and energy cuts. A substantial effort has been made to quantify
these discrepancies in detail, most recently  by L3~\cite{QGJCTseven}. They
are able to reconstruct well resolved $\pi^0$s and $\eta$s from two
identified photons isolated by a geometric cone in which no other particles
are found with energies above 50 MeV. {\sc Jetset} reproduces the observed
rate of $\pi^0$s and $\eta$s with energies above 3 GeV for 10$^\circ$
isolation, but significantly underestimates the rate for 25$^\circ$ isolation.
This study was restricted to 8 GeV maximum energy where the direct meson
reconstruction procedure is efficient, but has been extended to 45 GeV using a
neural network. The observed background rate of non-prompt photons is about a
factor 2 larger than the predicted rate over the full energy range and the
discrepancy increases with tighter isolation cuts. {\small HERWIG} tends to
give a slightly better prediction but still underestimates the rate.

In other studies at the Z of the non-prompt photon background both
ALEPH~\cite{QGJCTtwo} and DELPHI~\cite{QGJCTfour} have reported that 
{\sc Jetset} underestimates the isolated $\pi^0$ yields but only in the lower 
part of the energy spectrum below 20 GeV. In these analyses, the limit
allowed for the maximum particle energy accompanying the photon in the cone
is set to 500 MeV. They are not inconsistent with the L3 results but instead
demonstrate that the comparison with the generators is sensitively dependent
on the isolation parameters. In the alternative ``democratic'' analysis 
without isolation cones of ALEPH~\cite{QGJCTfive} some activity is allowed in
the vicinity of the $\gamma$ which results in a better description by 
{\sc Jetset} of the region of phase-space considered for the fragmentation.

\subsection{Bose--Einstein effects}

Most of the Bose--Einstein interference studies at LEP~1 have concentrated on
two-particle correlations between identical charged pions \cite{QGAB_PICH}
using the quantity
\begin{equation}
R(M) = \frac{\rho_2(M)}{\rho_1 \otimes \rho_1(M)}
\end{equation}
Here, $\rho_2(M)$ is the two-particle correlation function, usually given as 
a function of $Q,\;Q^2=M^2-4m_\pi^2$, and $\rho_1 \otimes \rho_1(M)$ is a
reference sample. This sample should resemble $\rho_2(M)$ except for the 
Bose-Einstein correlations being studied. 

Two choices for the reference sample are made, unlike-sign pion pairs or
uncorrelated pairs from track mixing. Both alternatives have disadvantages.
Unlike-sign pion pairs suffer from correlations due to resonances not present
in like-sign pion pairs and the contribution of resonances with poorly known
rates, especially $\eta$ and $\eta'$ at low $Q$. Furthermore residual effects
of Bose--Einstein interference may also be visible in the unlike-sign pairs
(see below). The track mixing has the disadvantage that correlations, other
than from Bose-Einstein interference, are missing. In addition cuts to
suppress gluon radiation must be applied. For both methods the systematic 
uncertainties are reduced using the double ratio 
$R^{\QGindx{data}}(M)/R^{\QGindx{MC}}(M)$. 
Additional corrections for background, e.g., Coulomb interactions are
applied.

Assuming a spherical and Gaussian source the enhancement at low $Q$ is 
parameterized as $R(M) \sim 1+\lambda \exp (-r^2 Q^2)$. The chaoticity
parameter $\lambda$ is expected to vary between 0 and 1, and is extracted
from data in the range from 0.4 to 1.5; the radius $r$ of the source is
measured to be $0.4\,\mbox{fm}$ to $1\,\mbox{fm}$. In
Fig.~\ref{QGAB_fig_l_r} the background-corrected measurements are displayed 
for the mesons $\pi^\pm,\;\pi^0,\;\mbox{K}^\pm$, and K$^0$. 

Only identical mesons, that are prompt, i.e. do not originate from long-lived 
resonances, can contribute to the enhancement at low $Q$. It has been pointed
out that the measured value of $\lambda$ is about the maximum you could expect
from direct pairs or even higher \cite{QGAB_HAYWOOD}. 

In more recent analyses the fraction $f(Q)$ of direct pions as a function 
of $Q$ has been parameterized using Monte Carlo and included in the fit. For
example DELPHI uses $f(Q)=0.17 + 0.26 Q-0.12Q^2$, obtained from {\sc Jetset},
to fit $\lambda$ and $r$ for charged pions: $R(M) \sim 1+\lambda f(Q) \exp
(-r^2 Q^2)$ \cite{QGAB_PICH}. Whilst the
change in the radius is small, $\lambda$ is changed by a factor 3. A bigger
change is reported by L3 for $\pi^0-\pi^0$ correlations \cite{QGAB_PINE}. The
corrections are very sensitive to the model used. The corrections for 
non-prompt mesons are indicated by arrows in Fig.~\ref{QGAB_fig_l_r}. 
The kaons have higher
chaoticity values than pions before correction \cite{QGAB_KAON}. Only DELPHI
has estimated the corrections for non-prompt kaons. The correction for kaons
from c- and b-decay increases $\lambda$ by $\approx 25$ to $30\%$.

\begin{figure}[tb]
\vspace{-1.cm}
\begin{center}
\mbox{\epsfig{file=qgbec.eps,height=8cm}}
\end{center}
\vspace{-0.3cm}
\caption{
Chaoticity parameter $\lambda$ versus radius $r$ measured at LEP~1. Measured
values are corrected for background with statistical (solid line) and total
errors (dots) shown. The arrows indicate the changes, when corrected for
non-prompt meson-pairs estimated with {\small HERWIG} or {\sc Jetset}, when
it is calculated by the experiment 
\protect\cite{QGAB_PICH,QGAB_PINE,QGAB_KAON}.}
\label{QGAB_fig_l_r}
\end{figure}

Three-particle correlations have been studied by DELPHI. Whilst {\sc Jetset} 
without Bose-Einstein correlations fails to describe the data, {\sc Jetset} 
with Bose-Einstein correlations enabled gives a fair description of
unlike-sign triplets; the shape is reproduced, but the magnitude is too small
\cite{QG_benew}.

Bose-Einstein correlation affect the unlike-sign spectra as well. In the 
invariant mass distribution of pions the $\rho^0$ meson appears shifted 
towards lower masses \cite{QGAB_RHO}. In the framework of the  model this can 
be interpreted as coming from Bose-Einstein correlations between like-sign 
pion pairs, which induces correlations between unlike-sign combinations, for
example seen as a distortion of the $\rho^0$ line shape. OPAL finds nice 
agreement between data and {\sc Jetset} including Bose-Einstein correlations, 
when the chaoticity parameter is set to 2.5. This value of $\lambda$ was
obtained with a fit to the ratio $R(M)$. ALEPH agrees with this observation
and extracts a $\rho^0$ rate with $\lambda$ and $r$ as free parameters. The
value of $\lambda=2.1$ is compatible with OPAL in view of the different
$\eta'$ rate and choice of the coherence time parameter $\chi$. ($\chi$
gives the minimum width of resonances whose daughters contribute to the
Bose-Einstein enhancement). DELPHI, which observes a shift of the $\rho^0$,
uses its $\lambda$ value extracted from the Bose-Einstein analysis, after
correction, for the $\rho^0$ analysis. Also with this parameter choice the
agreement of data and model mass spectra is satisfactory \cite{QGKH_Kmuenich}.

Concerns have to be raised about the implementation of Bose-Einstein 
correlations in {\sc Jetset}. The implementation treats them as a classical 
force, which violates energy-momentum conservation. The rescaling applied
to restore the total energy and momentum, however, twists the event shape
variables and the model description becomes worse. Multijet rates for larger
$y_{\QGindx{cut}}$ are reduced by up to 20\% and the tails
of the thrust and minor distributions are decreased by 5-10\%. The amount of
particles with low rapidities is depleted by $\approx$5\%. A small but
significant improvement is observed for small \QGptout. The wave structure
visible for \QGptout$ < 0.8$ GeV vanishes when well tuned BE parameters are
used and the \QGptout\ distribution here can be perfectly described
\cite{QGKH_MWdiss}. Studies on a modified implementation, which also moves
unlike-sign pairs to avoid rescaling (additional $\epsilon$ parameter)
improves the situation but the description of the $\rho^0$ mass shift is in
the wrong direction (positive). 

Another new simulation, based on the area spanned by the string, is in 
preparation. A first result with a toy Monte Carlo predicts that the 
reconstructed $\lambda$ should be 2 for $\pi^0$, when $\lambda=1$ is used 
for event generation \cite{QGAB_ANDERSON}.

At first glance, the experimental results are different, $\lambda\approx1$
for corrected direct measurements (DELPHI) and $\lambda\approx2$ for an
extraction tuning the {\sc Jetset} model. However the following
differences must be kept in mind. The use of track mixing for a reference
distribution tends always to give lower $\lambda$ values than the use of
the unlike-sign meson sample. The uncorrected values for DELPHI are lower
than for the other experiments. For kaons, corrections are estimated for
c- and b-decays only, but not for strong decays. ALEPH has used daughters
of resonances wider than $\Gamma=100\,\mbox{MeV}/c^2$ as prompt pions,
excluding the K$^*$ which seems not to be affected by Bose-Einstein
correlations. Ignoring this and correcting OPAL for the $\eta'$ rate would
bring the values down to $\lambda = 1.7$ in these two analyses.

On the model side more understanding is needed of how to include the
correlation without twisting the event shape distribution. The new 
$\epsilon$ parameter is a first step but there is no real success yet. 
Taking the decay amplitudes, i.e. string area, may be another promising 
approach. 

\newpage

\section{Extrapolation to LEP~2 Energies}
\label{QGsectextra}

A question of interest for LEP~2 is that of how well the characteristics
of QCD events are understood at large energies. By QCD events, it is here
meant those that are  produced through the $s$-channel decay of a Z$^0/
\gamma^*$ into quark and gluon jets. This question is of interest
because W$^+$W$^-$ events lead to multi-jet states for
which one of the principal backgrounds will be QCD events,
because QCD events will also form a principal source
of background for higgs, chargino and other particle searches,
and because QCD events will be interesting in their own right
as a means to test perturbation theory in a regime
with particularly small hadronization uncertainties.
The principal tools to test how well QCD event
characteristics are understood are Monte Carlo generators.
The main generators,
{\sc Ariadne}, {\sc Cojets}, {\small HERWIG} and {\sc Pythia},
have  been tuned by the LEP experiments or by the Monte Carlo
authors to describe global features of hadronic Z$^0$ data.
In many cases,
the generators have  proven able to describe
detailed features of these data as well.
It is thus relevant to extrapolate the predictions of
the QCD generators to LEP~2 energies and
to compare their level of agreement for
distributions likely to be of importance at LEP~2.
In this section, such an extrapolation and
comparison is presented.

For this study, members of each of the LEP experiments generated Monte Carlo
event samples at {\QGecm}$=175$~GeV using parameter sets determined within
their Collaboration. The Monte Carlo parameter sets used at LEP~1 are
continually revised in order to yield as accurate a description of the Z$^0$
data as possible. Therefore, the parameter sets employed for this study do
not necessarily represent official versions which will be published by the
Collaborations. The parameter sets used for {\sc Ariadne}, {\small HERWIG}
and {\sc Pythia} are given in
tables~\ref{QGL2-tab-ar406}--\ref{QGL2-tab-py57par}. For {\sc Cojets}, L3
and OPAL results were made available using the parameter values given in
table~\ref{QGL2-tab-cj623par}. There are numerous parameters and strategies
involved in the optimization of the parameters. Comparison of the results
obtained using the parameter sets of the different Collaborations therefore
provides a systematic check of effects associated with the optimization
choice. Samples of 100,000 events were generated without initial-state photon
radiation or detector simulation, treating all charged and neutral particles
with mean lifetimes greater than $3\cdot 10^{-10}$~s as stable.

\begin{table}[tp]
\centering
\begin{tabular}{|c|c|c|c@{\hspace{3mm}}c@{\hspace{3mm}}c%
@{\hspace{3mm}}c|}
  \hline
   Parameter & Name  & Default & ALEPH & DELPHI & L3 & OPAL  \\
  \hline
 $\Lambda_{\QGindx{LLA}}$  & 
PARA(1)  & 0.220  & 0.218 & 0.237 & 0.220 & 0.200 \\
 $p_\perp$ cutoff         & 
PARA(3)  & 0.60   & 0.58  & 0.64  & 1.00  & 1.00 \\
 Fragmentation function &  
MSTJ(11) &   4     &  3    &  3    &  3    &  4 \\
 Baryon model option  &  
MSTJ(12) &   2     &  2    &  3    &  2    &  2 \\
 $\cal P$(qq)/$\cal P$(q) &  
PARJ(1)  &  0.100  & 0.100 & 0.096 & 0.100 & 0.100 \\ 
 $\cal P$(s)/$\cal P$(u)  &  
PARJ(2)  &  0.300  & 0.300 & 0.302 & 0.300 & 0.300 \\ 
 ($\cal P$(us)/$\cal P$(ud))/($\cal P$(s)/$\cal P$(d)) &  
PARJ(3)  & 0.400  & 0.400 & 0.650 & 0.400 & 0.400 \\ 
 (1/3)$\cal P$(ud$_1$)/$\cal P$(ud$_0$) &  
PARJ(4)  & 0.050  & 0.050 & 0.070 & 0.050 & 0.050 \\ 
 $\cal P$($S$=1)$_{\QGindx{d,u}}$ &  
PARJ(11) &  0.500  & 0.500 & ---   & 0.500 & 0.500 \\
 $\cal P$($S$=1)$_{\QGindx{s}}$ &  
PARJ(12) &  0.600  & 0.600 & ---   & 0.600 & 0.600 \\
 $\cal P$($S$=1)$_{\QGindx{c,b}}$ &  
PARJ(13) &  0.750  &  0.750 & ---   & 0.750 & 0.750 \\
 Axial,  $\cal P$($S$=0,$L$=1;$J$=1)&  
PARJ(14) &  0.000  & 0.000 & ---   & 0.100 & 0.000 \\
 Scalar,  $\cal P$($S$=1,$L$=1;$J$=0) &  
PARJ(15) & 0.000  & 0.000 & ---   & 0.100 & 0.000 \\ 
 Axial,  $\cal P$($S$=1,$L$=1;$J$=1) &  
PARJ(16) &  0.000  & 0.000 & ---   & 0.100 & 0.000 \\
 Tensor,  $\cal P$($S$=1,$L$=1;$J$=2) &  
PARJ(17) & 0.000  & 0.000 & ---   & 0.250 & 0.000 \\ 
 Extra baryon suppression &  
PARJ(19) & 1.000  & 1.000 & 0.500 & 1.000 & 1.000 \\ 
 $\sigma_{\QGindx{q}}$               &  
PARJ(21) & 0.360  & 0.354 & 0.390 & 0.500 & 0.370 \\ 
 extra $\eta$ suppression &  
PARJ(25) & 1.000 & 1.000 & 0.650 & 0.600 & 1.000 \\ 
 extra $\eta^\prime$ suppression &  
PARJ(26) & 0.400  & 0.400 & 0.230 & 0.300 & 0.400 \\ 
 $a$                    &  
PARJ(41) & 0.300  & 0.500 & 0.391 & 0.500 & 0.180 \\ 
 $b$                    &  
PARJ(42) & 0.580  & 0.810 & 0.850 & 0.650 & 0.340 \\ 
 $\epsilon_{\QGindx{c}}$      &  
PARJ(54) & $-$0.050  & $-$0.050 & $-$0.0378 & $-$0.030 & --- \\
 $\epsilon_{\QGindx{b}}$      &  
PARJ(55) & $-$0.0050 & $-$0.0060 & $-$0.00255 & $-$0.0035 & --- \\
  \hline
\end{tabular}
\caption{ Optimized parameter sets for {\sc Ariadne}, version~4.06
(for ALEPH, version~4.05), from the LEP Collaborations.
The parameters listed are those which were changed from their
default values by at least one of the groups.
The {\sc Ariadne} events were generated using {\sc Pythia} 
version 5.7 to describe the hadronization and hadron decays.
The DELPHI Collaboration implements its own procedure to
specify the relative rate at which mesons are produced in
different multiplets~\protect\cite{QGKH_D_tune},
in place of the {\sc Pythia} parameters PARJ(11)-PARJ(17).}
\label{QGL2-tab-ar406}
\end{table}

\begin{table}[tp]
\centering
\begin{tabular}{|c|c|c|cccc|}
  \hline
   Parameter & Name  & Default & ALEPH & DELPHI & L3 & OPAL  \\
  \hline
 $\Lambda_{\QGindx{MLLA}}$ &  
QCDLAM   &  0.180 & 0.149 & 0.163 & 0.170 & 0.160 \\
 Cluster mass parameter 1 &  
CLMAX    &  3.35  & 3.90 & 3.48 & 3.20 & 3.40 \\
 Cluster mass parameter 2 &  
CLPOW    &  2.00  & 2.00 & 1.49 & 1.45 & 1.30 \\
 Effective gluon mass     &  
RMASS(13)&  0.750 & 0.726 & 0.650 & 0.750 & 0.750 \\
 Photon virtuality cutoff &  
VPCUT    &  0.40  & 1.00 & 0.40 & 0.50 & 0.40 \\
 Smearing of cluster direction &  
CLSMR    &  0.00   & 0.56 & 0.36 & 0.00 & 0.35 \\
 Weight for decuplet baryons & 
DECWT & 1.00 & 1.00 & 0.77 & 1.00 & 1.00 \\
 s quark weight   & 
PWT(3) & 1.00 & 1.00 & 0.83 & 1.00 & 1.00 \\
 diquark weight  & 
PWT(7) & 1.00 & 1.00 & 0.74 & 1.00 & 1.00 \\
  \hline
\end{tabular}
\caption{ Optimized parameter sets for {\small HERWIG}, version~5.8,
from the LEP Collaborations.
The parameters listed are those which were changed from their
default values by at least one of the groups.
}
\label{QGL2-tab-hw58par}
\end{table}

\begin{table}[tp]
\centering
\begin{tabular}{|c|c|c|c@{\hspace{2.5mm}}c@{\hspace{2.5mm}}c%
@{\hspace{2.5mm}}c|}
  \hline
   Parameter & Name  & Default & ALEPH & DELPHI & L3 & OPAL  \\
  \hline
 Fragmentation function   &  
MSTJ(11) &   4     &  3    &  3    &  3    &  3 \\
 Baryon model option &  
MSTJ(12) &   2     &  2    &  3    &  2    &  2 \\
 Azimuthal correlations   &  
MSTJ(46) &   3     &  0    &  3    &  3    &  3    \\
 $\cal P$(qq)/$\cal P$(q) &  
PARJ(1)  &  0.100  & 0.095 & 0.099 & 0.100 & 0.085 \\
 $\cal P$(s)/$\cal P$(u)  &  
PARJ(2)  &  0.300  & 0.285 & 0.308 & 0.300 & 0.310 \\
 ($\cal P$(us)/$\cal P$(ud))/($\cal P$(s)/$\cal P$(d)) &  
PARJ(3)  &  0.400  & 0.580 & 0.650 & 0.400 & 0.450 \\
 (1/3)$\cal P$(ud$_1$)/$\cal P$(ud$_0$) &  
PARJ(4)  &  0.050  & 0.050 & 0.070 & 0.050 & 0.025 \\
 $\cal P$($S$=1)$_{\QGindx{d,u}}$ &  
PARJ(11) &  0.500  & 0.550 & ---   & 0.500 & 0.600 \\
 $\cal P$($S$=1)$_{\QGindx{s}}$ &  
PARJ(12) &  0.600  & 0.470 & ---   & 0.600 & 0.400 \\
 $\cal P$($S$=1)$_{\QGindx{c,b}}$ &  
PARJ(13) &  0.750  & 0.600 & ---   & 0.750 & 0.720 \\
 Axial,  $\cal P$($S$=0,$L$=1;$J$=1) &  
PARJ(14) &  0.000  & 0.096 & ---   & 0.100 & 0.430 \\
 Scalar,  $\cal P$($S$=1,$L$=1;$J$=0) &  
PARJ(15) &  0.000  & 0.032 & ---   & 0.100 & 0.080 \\
 Axial,  $\cal P$($S$=1,$L$=1;$J$=1) &  
PARJ(16) &  0.000  & 0.096 & ---   & 0.100 & 0.080 \\
 Tensor,  $\cal P$($S$=1,$L$=1;$J$=2) &  
PARJ(17) &  0.000  & 0.160 & ---   & 0.250 & 0.170 \\
 Extra baryon suppression &  
PARJ(19) &  1.000  & 1.000 & 0.500 & 1.000 & 1.000 \\
 $\sigma_{\QGindx{q}}$               &  
PARJ(21) &  0.360  & 0.360 & 0.408 & 0.399 & 0.400 \\
 extra $\eta$ suppression &  
PARJ(25) &  1.000  & 1.000 & 0.650 & 0.600 & 1.000 \\
 extra $\eta^\prime$ suppression &  
PARJ(26) &  0.400  & 0.400 & 0.230 & 0.300 & 0.400 \\
 $a$                      &  
PARJ(41) &  0.300  & 0.400 & 0.417 & 0.500 & 0.110 \\
 $b$                      &  
PARJ(42) &  0.580  & 1.030 & 0.850 & 0.848 & 0.520 \\
 $\epsilon_{\QGindx{c}}$       &  
PARJ(54) & $-$0.050  & $-$0.050 & $-$0.038 & $-$0.030 & $-$0.031 \\
 $\epsilon_{\QGindx{b}}$       &  
PARJ(55) & $-$0.0050 & $-$0.0045 & $-$0.00284 & $-$0.0035 & $-$0.0038 \\
 $\Lambda_{\QGindx{LLA}}$  &  
PARJ(81) &  0.290  & 0.320 & 0.297 & 0.306 & 0.250 \\
 $Q_0$                    &  
PARJ(82) &  1.000  & 1.220 & 1.560 & 1.000 & 1.900 \\
  \hline
\end{tabular}
\caption{ Optimized parameter sets for {\sc Pythia}, version 5.7,
from the LEP Collaborations.
The parameters listed are those which were changed from their
default values by at least one of the groups.
The DELPHI Collaboration implements their own procedure to
specify the relative rate at which mesons are produced in
different multiplets~\protect\cite{QGKH_D_tune},
in place of the {\sc Pythia} parameters PARJ(11)--PARJ(17).
}
\label{QGL2-tab-py57par}
\end{table}

\begin{table}[tp]
\centering
\begin{tabular}{|c|c|c|cc|}
  \hline
   Parameter & Name  & Default  & L3 & OPAL  \\
  \hline
 $b_{\QGindx{g}}$ &  FRALOG(2)    & 46.6 & 100.0 & 46.6 \\
 $d_{\QGindx{g}}$ &  FRALOG(4)    & 1.52 & 2.10 & 1.52 \\
 $b_{\QGindx{q}}$ &  FRALOQ(2)    & 30.5 & 43.0 & 30.5 \\
 $d_{\QGindx{q}}$ &  FRALOQ(4)    & 1.52 & 2.10 & 1.52 \\
  \hline
\end{tabular}
\caption{ Optimized parameter sets for {\sc Cojets}, version~6.23,
from the L3 and OPAL Collaborations.
The parameters listed are those which were changed from their
default values by at least one of the groups.
}
\label{QGL2-tab-cj623par}
\end{table}

The following distributions were examined using
charged particles only:
\begin{QGenumerate}
  \item charged particle multiplicity,~{\QGnchdist},
  \item scaled particle momentum,~$x_p = 2 p/${\QGecm},
  \item component of particle momentum in the
        event plane,~{\QGptin}, and
  \item component of particle momentum out of the
        event plane,~{\QGptout}.
\end{QGenumerate}
The event plane was defined by the two vectors associated with
the two largest eigenvalues of the Sphericity tensor.

The following distributions were examined using both
charged and neutral particles:
\begin{QGenumerate}
  \item Thrust, $T$~\cite{QGbibthr},
  \item Thrust major, {\QGtmajor}~\cite{QGbibtmajor},
  \item Thrust minor, {\QGtminor}~\cite{QGbibtmajor},
  \item jet rates $R_n$ defined using the {\QGdurham}
        jet finder~\cite{QGbibkt},
  \item normalized heavy jet mass for events divided into
        hemispheres by the plane perpendicular to the
        Thrust axis,~{\QGhjet}~\cite{QGbibhjet},
  \item normalized difference between the heavy and light
        jet masses,~{\QGdjet},
  \item total jet broadening, $B_T$~\cite{QGbibbwjet},
  \item wide jet broadening, $B_W$~\cite{QGbibbwjet},
  \item Sphericity, $S$~\cite{QGbibsph},
  \item Aplanarity, $A$~\cite{QGbibapl},
  \item the modified Nachtmann-Reiter four-jet angular variable,
        {\QGcosnr}~\cite{QGbibnr}, with four-jet events
        defined using the {\QGdurham} jet finder with {\QGycut}=0.01, 
        and
  \item the cosine of the angle between the two lowest
        energy jets in the four-jet events,~{\QGleang}.
\end{QGenumerate}
In addition, the mean values of {\QGnchdist}, {\QGtfamily}
were examined as a function of~{\QGecm}.

The results for {\QGnch}, {\QGtfamily}
as a function of~{\QGecm} are shown in Fig.~\ref{QGL2-fig_ext_ecm}.
\begin{figure}[p]
\begin{center}
\begin{tabular}{ll}
  \mbox{\epsfig{file=qgl2_f1a.eps,height=7.5cm}} &
  \mbox{\epsfig{file=qgl2_f1b.eps,height=7.5cm}} \\[2.0cm]
  \mbox{\epsfig{file=qgl2_f1c.eps,height=7.5cm}} &
  \mbox{\epsfig{file=qgl2_f1d.eps,height=7.5cm}} \\
\end{tabular}
\end{center}
\vspace*{0.5cm}
\caption{
The mean values of {\QGnchdist}, Thrust {\QGtfamily}
predicted by {\sc Ariadne}, {\sc Cojets},
{\small HERWIG} and {\sc Pythia} as a function of {\QGecm}
in comparison with measurements from
PEP, PETRA, TRISTAN and LEP~1.
The LEP~2 point is indicative only,
based on the {\sc Pythia} prediction.
The total uncertainty expected at LEP~2 assuming
$10\,000$ QCD events is smaller than the symbol size.
}
\label{QGL2-fig_ext_ecm}
\end{figure}
For those cases in which the results of at least three
Collaborations are similar to each other,
the Monte Carlo predictions are shown as shaded
or hatched bands.
The widths of the bands show the maximum deviations
between the results found by the different Collaborations.
The widths of the bands are generally much larger than the
statistical uncertainties.
In a few cases,
the Monte Carlo prediction obtained by one of the Collaborations
differs significantly from those obtained by the other three groups
and is shown as a separate curve.
The {\sc Cojets} predictions are likewise shown as separate curves
for purposes of clarity.
The results found by the four LEP experiments
are labelled A, D, L and O in the figure legends.

Representative measurements from PEP, PETRA, TRISTAN
and LEP~1 are included in Fig.~\ref{QGL2-fig_ext_ecm}.
For {\QGecm}$=175$~GeV,
an indicative ``data point'' is also shown,
which is taken to be equal to the mean
of the {\sc Pythia} predictions from the four groups.
The size of the symbol for the LEP~2 point is larger than
the statistical uncertainty for $10\,000$ QCD events.
Systematic terms were generally found to dominate the
statistical ones for the experimental measurements
shown in Fig.~\ref{QGL2-fig_ext_ecm}.
The total experimental uncertainties at 175~GeV
can therefore be expected to be comparable to those
found for the LEP~1 data.

\begin{figure}[p]
\begin{center}
\begin{tabular}{ll}
  \mbox{\epsfig{file=qgl2_f2a.eps,height=7.5cm}} &
  \mbox{\epsfig{file=qgl2_f2b.eps,height=7.5cm}} \\[2.0cm]
  \mbox{\epsfig{file=qgl2_f2c.eps,height=7.5cm}} &
  \mbox{\epsfig{file=qgl2_f2d.eps,height=7.5cm}} \\
\end{tabular}
\end{center}
\vspace*{0.5cm}
\caption{
Comparison of the predictions of QCD event generators
at {\QGecm}$=175$~GeV.
}
\label{QGL2-fig_ext_evshape1}
\end{figure}

{}From the distribution of {\QGnch} versus {\QGecm}
(Fig.~\ref{QGL2-fig_ext_ecm}(a)),
it is seen that,
with the exception of the L3 {\sc Ariadne} curve,
the predictions of {\sc Ariadne}, {\small HERWIG} and
{\sc Pythia} are similar.
The widths of the {\sc Ariadne}, {\small HERWIG} and {\sc Pythia} bands
are narrow for energies at and below the Z$^0$ mass,
showing that the results from the four Collaborations
are in close agreement
(with the exception of the L3 curve for {\sc Ariadne}).
For energies above about 150~GeV,
the {\small HERWIG} band becomes broader,
indicating that there is some divergence in the predictions
obtained by the different groups.
{}From Fig.~\ref{QGL2-fig_ext_ecm}(a) it is also seen that {\sc Cojets}
predicts a substantially larger value of {\QGnch}
than the other models for energies above the Z$^0$ mass.
This difference is suggestive of coherence effects
in the parton shower,
which are absent in {\sc Cojets} but present
in the other three models.
Coherence reduces the mean soft gluon multiplicity
in the parton shower.
It is generally expected that coherence will lead to
a reduction in the mean hadron multiplicity as well.
Thus, a measurement of {\QGnch} at LEP~2 could help to establish
the existence of coherence phenomena in the data.

Figs.~\ref{QGL2-fig_ext_ecm}(b)--(d)
show the corresponding distributions for the
{\QGtfamily} variables.
Again, {\sc Ariadne}, {\small HERWIG} and {\sc Pythia} are seen
to exhibit similar behavior.
{\sc Cojets} agrees well with the other models for $T$,
but lies below them for {\QGtmajor} and above them
for {\QGtminor} in the LEP~2 energy range.
Thus the jets from {\sc Cojets} are less oblate
than those from {\sc Ariadne}, {\small HERWIG} or {\sc Pythia}.
(The Oblateness $O$ of an event is given by
$O=\,${\QGtmajor}$\,-\,${\QGtminor}.)
The differences between {\sc Cojets} and the other three
models become larger as {\QGecm} increases.

\begin{figure}[p]
\begin{center}
\begin{tabular}{ll}
  \mbox{\epsfig{file=qgl2_f3a.eps,height=7.5cm}} &
  \mbox{\epsfig{file=qgl2_f3b.eps,height=7.5cm}} \\[2.0cm]
  \mbox{\epsfig{file=qgl2_f3c.eps,height=7.5cm}} &
  \mbox{\epsfig{file=qgl2_f3d.eps,height=7.5cm}} \\
\end{tabular}
\end{center}
\vspace*{0.5cm}
\caption{
Comparison of the predictions of QCD event generators
at {\QGecm}$=175$~GeV.
}
\label{QGL2-fig_ext_evshape2}
\end{figure}

In Fig.~\ref{QGL2-fig_ext_evshape1},
the Monte Carlo predictions for {\QGnchdist}, $x_p$, {\QGptin}
and {\QGptout} at 175~GeV are shown.
The corresponding results for $T$, {\QGtmajor},
{\QGtminor} and~$R_n$,
for {\QGhjet}, {\QGdjet}, $B_T$ and~$B_W$,
and for $S,\;A$, {\QGcosnr} and~{\QGleang}
are shown in
Figs.~\ref{QGL2-fig_ext_evshape2},
\ref{QGL2-fig_ext_evshape3},
and~\ref{QGL2-fig_ext_evshape4}, respectively.
Overall, the models are seen to be in general agreement
with each other.
Some of the more notable exceptions to this agreement
are discussed below.
\begin{QGenumerate}
   \item
A striking difference is observed between {\sc Cojets} and the
other models for the {\QGnchdist} and {\QGptin} distributions
(Figs.~\ref{QGL2-fig_ext_evshape1}(a) and~(c)).
Smaller but visible differences are observed between
{\sc Cojets} and the other models for a number of the
other distributions as well.
At the Z$^0$ mass,
these differences between {\sc Cojets} and the other models
are either not present or are much smaller.
This implies that the energy scaling behavior of {\sc Cojets}
differs from that of {\sc Ariadne}, {\small HERWIG} and {\sc Pythia}.
   \item
For {\small HERWIG},
the $x_p$ distribution is much harder using the L3 parameter set
than it is using the parameter sets of the other Collaborations
(Fig.~\ref{QGL2-fig_ext_evshape1}(b)).
This feature is also observed at the Z$^0$ energy.
The primary reason for this difference between
L3 and the other groups
is the different treatment of the parameter~CLSMR
(see table~\ref{QGL2-tab-hw58par}).
   \item
{}From Fig.~\ref{QGL2-fig_ext_evshape2}(d),
it is seen that the three-jet rate from {\sc Pythia} is
significantly larger than that of the other models
for {\QGycut} values below about~0.02.
Correspondingly, the two jet rate from {\sc Pythia} is smaller.
This difference is also observed at {\QGecm}$=91$~GeV.
{}From this same figure,
{\sc Cojets} is seen to predict a three-jet rate which
is smaller than that of the other models:
this last difference is not observed at LEP~1 energies.
   \item
{\sc Cojets} exhibits a clear deviation with respect to
the predictions of the other models for the jet mass distributions,
{\QGhjet} and {\QGdjet} (Figs.~\ref{QGL2-fig_ext_evshape3}(a) and~(b)).
Less of a deviation is present for the jet broadening variables,
$B_T$ and~$B_W$ (Figs.~\ref{QGL2-fig_ext_evshape3}(c) and~(d)).
This suggests that these last two variables may be less subject
to uncertainties related to the modelling of QCD and hadronization
than the first two variables.
   \item
{\sc Cojets} and {\small HERWIG} are seen to exhibit a somewhat
flatter distribution in {\QGcosnr} than {\sc Ariadne} and {\sc Pythia}
(Fig.~\ref{QGL2-fig_ext_evshape4}(c)).
\end{QGenumerate}

The general conclusion that can be drawn from this study
is that there is relatively little uncertainty in the predictions
of QCD generators for event characteristics at LEP~2.
Such basic features of events as charged multiplicity,
Thrust and Oblateness
are described in an almost identical manner by
{\sc Ariadne}, {\small HERWIG} and {\sc Pythia}.
Only {\sc Cojets} deviates significantly from the
predictions of the other models.
On the other hand,
there is modest disagreement between the models for
variables which require use of a jet finding algorithm:
$R_n$ (Fig.~\ref{QGL2-fig_ext_evshape2}(d))
and {\QGcosnr} (Fig.~\ref{QGL2-fig_ext_evshape4}(c)).
This could have some implication for the W mass
determination based on the reconstruction of jets.

\begin{figure}[p]
\begin{center}
\begin{tabular}{ll}
  \mbox{\epsfig{file=qgl2_f4a.eps,height=7.5cm}} &
  \mbox{\epsfig{file=qgl2_f4b.eps,height=7.5cm}} \\[2.0cm]
  \mbox{\epsfig{file=qgl2_f4c.eps,height=7.5cm}} &
  \mbox{\epsfig{file=qgl2_f4d.eps,height=7.5cm}} \\
\end{tabular}
\end{center}
\vspace*{0.5cm}
\caption{
Comparison of the predictions of QCD event generators
at {\QGecm}$=175$~GeV.
}
\label{QGL2-fig_ext_evshape3}
\end{figure}

\begin{figure}[p]
\begin{center}
\begin{tabular}{ll}
  \mbox{\epsfig{file=qgl2_f5a.eps,height=7.5cm}} &
  \mbox{\epsfig{file=qgl2_f5b.eps,height=7.5cm}} \\[2.0cm]
  \mbox{\epsfig{file=qgl2_f5c.eps,height=7.5cm}} &
  \mbox{\epsfig{file=qgl2_f5d.eps,height=7.5cm}} \\
\end{tabular}
\end{center}
\vspace*{0.5cm}
\caption{
Comparison of the predictions of QCD event generators
at {\QGecm}$=175$~GeV.
}
\label{QGL2-fig_ext_evshape4}
\end{figure}

\newpage

\section{Monte Carlo descriptions}
\label{QGsectMC}

In this section we have collected brief descriptions for the 
main QCD generators and other pieces of QCD code. These writeups 
are intended to introduce the main physics ideas and give further 
references to manuals and codes --- a full coverage of all 
physics and programming aspects is excluded for space reasons. 
The compilation
below should be rather complete for programs intended for the
main QCD-related processes, such as $\QGgaZ$ and W pair production. 
Special emphasis is put on {\small HERWIG} and {\sc Pythia/Jetset},
which have been used extensively at LEP~1 and are equipped with
a simulation both of electroweak and QCD aspects. 
A few programs include QCD aspects but have still been 
judged to better belong elsewhere, e.g. {\sc Phojet} is a
$\gamma\gamma$ physics generator and {\sc Isajet} is mainly 
of interest (in the $\QGee$ sector) as a supersymmetry generator.

\subsection{ARIADNE}

\noindent{\bf Basic Facts}
\vspace{-\baselineskip} 
\begin{tabbing}
{\bf Program location:} \= \kill
{\bf Program name:}     \> {\sc Ariadne} \cite{QGariadne}          \\
{\bf Version:}          \> 4.08 of 30 November 1995                  \\
{\bf Author:}           \> Leif L{\"o}nnblad                       \\
                        \> NORDITA, Blegdamsvej 17,                \\
                        \> DK 2100 Copenhagen {\O}, Denmark        \\
                        \> Phone: + 45 -- 35325285                 \\
                        \> E-mail: leif@nordita.dk                 \\
{\bf Program size:}     \> 12853 lines                             \\
{\bf Program location:} \> http://surya11.cern.ch/users/lonnblad/ariadne/
\end{tabbing}

The {\sc Ariadne} program implements the Dipole Cascade Model (DCM)
for QCD cascades \cite{QGLLdcm}. In this model the
emission of a gluon $\QGg_1$ from a $\QGq\QGqbar$ pair created in an
$\QGee$ annihilation event can be described as radiation from the
colour dipole between the $\QGq$ and $\QGqbar$. A subsequent emission of
a softer gluon $\QGg_2$ can be described as radiation from two
independent colour dipoles, one between the $\QGq$ and $\QGg_1$ and one
between $\QGg_1$ and $\QGqbar$, neglecting the contribution from the
$\QGq\QGqbar$ dipole, which is suppressed by $1/N_C^2$.  Further gluon
emissions are given by three independent dipoles, {\it etc.} In this way, 
the end result is a chain of dipoles, where one dipole connects two
partons, and a gluon connects two dipoles.  This is in close
correspondence with the Lund string picture, where gluons act as kinks
on a string-like field stretched between the $\QGq\QGqbar$ pair.

This formulation of the partonic cascade in terms of colour dipoles
means that the coherence effects, handled by introducing angular
ordering in conventional parton showers, is correctly taken into
account. Also, the DCM has the advantage that the first gluon emission
is done according to the correct first-order matrix element, so that
an explicit matching procedure like the ones introduced in 
{\small HERWIG} and {\sc Jetset} is not needed.

Although the model has been developed a lot since the last LEP
workshop, much of this development has been related to the description
of Deep Inelastic Scattering and hadron-hadron collisions and will be
described in some detail in the report from the $\gamma\gamma$
generator working group. Here only aspects relevant to $\QGee$
annihilation will be discussed.

The basic DCM only describes gluon emission, and the process of
splitting a gluon into a $\QGq\QGqbar$ pair has therefore been added
according to \cite{QGLLdcmq}. Although this procedure reproduces
fairly well the amount of secondary c$\overline{\mbox{c}}$ production 
observed at LEP~\cite{QGLLopalc}, there has been some criticism 
\cite{QGLLmike} that the model may be overestimating the phase space 
available for this process. Therefore an extra restriction of this 
phase space suggested in \cite{QGLLmike} has been implemented 
as an option in the last versions.

The radiation of photons from quarks is handled by allowing the
process of emitting a photon from the {\em electro-magnetic} dipole
between the original $\QGq\QGqbar$ to compete with the gluon emission
from the colour dipoles \cite{QGLLannecy}. This competition is
governed by the ordering in transverse momenta of the emitted
gluons/photons, which is different from {\sc Jetset} and {\small HERWIG},
where virtuality and angle, respectively, is used for ordering.

In the latest version, a scheme for colour reconnections has been
added to the program. The model is described fully in 
\cite{QGTSrecLL} and briefly in section \ref{QGsubsectreconn}. 
Unfortunately, the manual included in the code
distribution has not yet been updated to describe this new feature,
and users who want to try it are advised to contact the author by
e-mail before doing so.

Since {\sc Ariadne} only handles the perturbative QCD cascade in an
event, it has to be interfaced to the {\sc Pythia/Jetset} programs for
generation of the hard sub-process, the hadronization and the particle
decays. Such an interface is included in the code, and only very minor
changes to the steering program is needed to replace the parton
showers in {\sc Pythia/Jetset} with the dipole shower in {\sc Ariadne}
for any type of process. In a typical steering program for running 
{\sc Pythia}, the changes needed are as follows.
\begin{QGitemize}
\item Immediately before the call to {\tt PYINIT} there should be
inserted a call to {\tt ARTUNE('4.07')} to set up default parameters
in {\sc Ariadne} and {\sc Pythia/Jetset}, followed by a call to {\tt
ARINIT('PYTHIA')} to initialize the {\sc Pythia} interface. To change
the default behavior of {\sc Ariadne}, changes may be made to the 
{\tt ARDAT1} common block {\em between} the calls to {\tt ARTUNE} and 
{\tt ARINIT}.
\item Immediately after a call to {\tt PYEVNT}, a call to 
{\tt AREXEC} should be made to perform the actual dipole cascade. 
If {\sc Pythia} is set up to handle fragmentation and decays in the 
{\tt PYEVNT} call, this is now handled in {\tt AREXEC} instead.
\end{QGitemize}
Sample programs for how to do this is included in the
distribution. The distribution also includes a subroutine {\tt AR4FRM}
which is an interface to four-fermion generators according to the
standard presented in section \ref{QGsubsectEWint}. Except for what is
needed to run {\sc Pythia/Jetset}, no additional software is required
to run {\sc Ariadne}.

\subsection{COJETS}

\noindent{\bf Basic Facts}
\vspace{-\baselineskip} 
\begin{tabbing}
{\bf Program location:} \= \kill
{\bf Program name:}     \> {\sc Cojets} \cite{QGROcj}              \\
{\bf Version:}          \> {\sc Cojets} 6.23 of 10 February 1992   \\
{\bf Author:}           \> Roberto Odorico                         \\
                        \> Department of Physics                   \\
                        \> University of Bologna                   \\   
                        \> Via Irnerio 46, I-40126 Bologna, Italy  \\
                        \> Phone: + 39 -- 51 - 24 20 18            \\
                        \> E-mail: odorico@bo.infn.it              \\
{\bf Program size:}     \> 19742 lines                             \\
{\bf Program location:} \> http://www.bo.infn.it/preprint/odorico.html
\end{tabbing}

{\sc Cojets} simulates electron-positron annihilation into jets of hadrons. 
(It also simulates proton-proton and antiproton-proton interactions.) 
The simulation is based, at the parton level, on the standard model with 
perturbative QCD treated in the leading-log approximation. QED radiation 
off beam particles is treated according to the BKJ program. Partons from 
parton showers are independently fragmented into jets of hadrons 
according to a Field-Feynman model extended to include heavy quarks and 
baryons and modified in the generation of soft particles. Gluons are 
fragmented as a pair of light quark and antiquark jets of opposite 
random flavors, each one having half the energy of the gluon and its 
same direction and with fragmentation parameters distinct from those of 
quark jets. Jet non-perturbative masses are limited by bounds originated 
by an approximate treatment of global phase-space effects at the multi-
jet level. A previous version of the program, {\sc Cojets} 6.12, in which 
quarks 
and gluons share the same fragmentation model is also available. The jet 
fragmentation model adopted goes hand in hand with the setting of the 
minimum parton-mass cutoff to a value of 3 GeV, which is substantially 
larger than those used in string- and cluster-based fragmentation 
models.

The output common block, containing the generated particle stream, has 
the standard {\tt /HEPEVT/} format, with PDG codes used for particles.

{\sc Cojets} is maintained with the PATCHY code management system. The 
appropriate FORTRAN77 codes are obtained by means of suitable pilot 
patches. The program file also includes the documentation.

Recently, the program has been mainly used to check the relevance of 
evidence for string fragmentation, parton coherence and quark/gluon 
differences in jet fragmentation. Its usage to study the 
signal/background enrichment for bottom non-leptonic decay events by 
means of neural networks has shown that differences in internal 
correlations between signal and background are fuzzier in {\sc Cojets} 
than in {\sc Jetset} \cite{QGROmaz93}. Thus for LEP~2 {\sc Cojets} 
could be useful when studying ways of disengaging events with W pairs 
decaying non-leptonically from background.

{\sc Cojets} had its fragmentation parameters sensibly tuned to reproduce 
basic experimental distributions. So far the tuning has been done by the 
author, mimicking experimental apparatus effects but without a proper 
GEANT simulation. That can be done by the user by means of the program 
TUNEMC, based on Minuit's Simplex algorithm (a more advanced version of 
the program is in preparation).

Programs {\sc Cojets} 6.12 and TUNEMC can be found at the same WWW URL 
as {\sc Cojets} 6.23.

\subsection{HERWIG}

\noindent{\bf Basic Facts}
\vspace{-\baselineskip} 
\begin{tabbing}
{\bf Program location:} \= \kill
{\bf Program name:}     \> {\small HERWIG} \cite{QGherwig}              \\
{\bf Version:}          \> {\small HERWIG} 5.9 from 1 January 1996      \\
{\bf Authors:}          \> G. Marchesini$^1$, B.R. Webber$^2$,
                           G. Abbiendi$^3$, I.G. Knowles$^4$,      \\
                        \> M.H. Seymour$^5$, L. Stanco$^3$         \\
                        \> 1, Dipartimento di Fiscia, Universita di Milano. \\
                        \> 2, Cavendish Laboratory, University of Cambridge. 
\\
                        \> 3, Dipartmento di Fisica, Universita di Padova. \\
                        \> 4, Department of Physics and Astronomy, University
                           of Glasgow.                             \\
                        \> 5, Theory Division, CERN.               \\
                        \> E-mail: webber@hep.phy.cam.ac.uk,
                                   knowles@v6.ph.gla.ac.uk,        \\
                        \>         seymour@surya11.cern.ch.        \\
{\bf Program size:}     \> 15500 lines                             \\
{\bf Program location:} \> http://surya11.cern.ch/users/seymour/herwig/
\end{tabbing}

\subsubsection{Introduction}
{\small HERWIG} (Hadron Emission Reactions With Interfering Gluons) is a 
large,
multipurpose Monte Carlo event generator which has been extensively used at
LEP~1. Version 3.2, as applied to $\QGee$ annihilation, was described in
detail for the LEP~1 workshop \cite{QGIKl1her}, here we concentrate 
principally
on program developments and new aspects of relevance to LEP~2 physics.

QCD Monte Carlo event generators utilize the fact that any hard scattering
processes can be factorized into separate components at leading twist. These
are: the hard sub-process itself; perturbative initial and final state
showers; non-perturbative hadronization; resonance decays; and beam remnant
fragmentation. In {\small HERWIG} great emphasis is placed on making available
a very sophisticated, partonic treatment of the calculable QCD showers. In
contrast the description of the at present uncalculable hadronization and
beam remnant components is in terms of very simple models. Since {\small
HERWIG} contains many hard sub-processes and supports all combinations of
hadron, lepton and photon beams this allows the physics of many types of
particle collisions to be simulated in the same package. In view of the
universality of the factorized components that build up {\small HERWIG}
events this allows experience gained at HERA and the TEVATRON, for example,
to be made directly available to LEP physicists.

Since version 3.2 was released the {\small HERWIG} code was reorganized to
isolate the shower, cluster hadronization and unstable particle decay
routines. This modularity facilitates the creation of hybrid programs in
which sections of code are replaced with interfaces to other Monte Carlo
programs. To identify {\small HERWIG} code the names of all options statements
now begin {\tt HW****}. The {\tt /HEPEVT/} standard proposed in \cite{QGIKhep}
is also now used throughout the program in {\tt DOUBLE PRECISION}. The random
number generator has been upgraded to a l'Ecuyer's algorithm as recommended
in \cite{QGIKlecuy}. Discussion of physics changes are contained in the
following sections.

\subsubsection{Hard Sub-processes}
An extensive range of hard sub-processes are available in the {\small HERWIG}
program allowing the full spectrum of standard model LEP~2 physics to be
simulated. These are illustrated in table~\ref{QGIKHWt1}; for a complete
listing see the program release notes in the text file {\tt HERWIGnm.DOC}.

The matrix elements used for the continuum processes {\tt IPROC}=100--153
now allow for arbitrary polarization of the lepton beams, an additional
$\mbox{Z}'$ including complete $\gamma^\star/\mbox{Z}^0$ interference and 
full mass effects (IQ $\neq0$). When the {\tt ZPRIME=.TRUE.} option is set 
the $\mbox{Z}'$
weak couplings used are taken from the arrays {\tt AFCH(*,2)} and {\tt
VFCH(*,2)}, see {\tt HWIGIN} for details. The arrays {\tt Q/V/AFCH(*,1)}
are used consistently throughout the program for the standard model electric,
weak vector and axial-vector fermion couplings. A running electromagnetic
coupling $\QGalphaem(Q^2)$ is used for internal photons \cite{QGIKaemrun}
with the hadronic part taken from \cite{QGIKaemhad}. It is normalized to the
Thomson limit ($Q^2=0$) value {\tt ALPHEM}. Process 107 is included to
facilitate $\QGq/\QGg$ studies; in analogy with quarks the gluons are given a
$1+\cos^2\theta$ distribution. The difference between processes 100--106
and the original 120--126 lies in the treatment of hard gluon emission. The
massless matrix element matching scheme used by {\small HERWIG} is discussed
under parton showers.

Specialist Matrix Element programs for 
$\mbox{W}^+\mbox{W}^-/\mbox{Z}^0\mbox{Z}^0$ production, more
properly four fermi\-on generators, employ the full set of gauge invariant
diagrams. In comparison {\small HERWIG} only includes the subset of diagrams
containing $\mbox{W}^+\mbox{W}^-$ (``CC03'') or 
$\mbox{Z}^0\mbox{Z}^0$ (``NC02'') pairs but does provide
realistic hadronic final states. These matrix elements are taken from the
program of Kunszt \cite{QGIKzoltan} and correctly include spin correlations
in the gauge boson decays. Additionally a model for colour re-arrangement
within the context of {\small HERWIG} is available, see \ref{QGsubsectreconn}
for details. The decays of the vector bosons are controlled via the array
{\tt MODBOS}, as detailed in {\tt HWIGIN}, and include spin correlations.
Please also see the detailed prescription, discussed in section
\ref{QGsubsectEWint}, for interfacing {\small HERWIG} to specialist four
fermion Monte Carlos.

At LEP~2 the principal Higgs production mechanism is the Bjorken process,
{\tt IPROC}=300+ID, where one or both $\mbox{Z}^0$'s may be off-shell, 
$\mbox{Z}^{0(\star)} \rightarrow \mbox{Z}^{0(\star)}\mbox{h}^0$; 
also available is vector boson fusion, {\tt
IPROC}=400+ID. In both cases the exact leading order matrix element is used
in the improved $s$-channel approximation \cite{QGIKhiggs}. At LEP~2 a
discoverable Higgs would be narrow; in the program the actual mass used is
taken from within the range 
$[M_{\QGindx{h}}-\mbox{\tt GAMMAX}\star\Gamma_{\QGindx{h}},M_{\QGindx{h}}+
\mbox{\tt GAMMAX}\star\Gamma_{\QGindx{h}}]$ (default {\tt GAMMAX=10}) using a
`Breit-Wigner' distribution corrected for an energy dependent width. The
event weight is the product of the cross-section (in nb) multiplied by the
branching ratio to the channel specified by {\tt ID}. The Higgs' partial
widths are calculated in {\tt HWDHIG}: the quark decay channels include
next--to--leading logarithmic corrections and the vector boson decay modes
allow for off-shell $\mbox{WW}/\mbox{ZZ}$ pairs.

\begin{table}[p]
\begin{center}
\begin{tabular}{|l|ll|}
\hline
IPROC & \multicolumn{2}{c|}{Process} \\
\hline
\multicolumn{1}{c|}{} & 
\multicolumn{2}{c|}{$\gamma^\star/\mbox{Z}^0/\mbox{Z}'$
Continuum Processes} \\
\hline
100+IQ & $\QGee\rightarrow \QGq\QGqbar(\QGg)$ & IQ=1--6: 
$\QGq=\mbox{d},\ldots,\mbox{t}$; IQ=0: all
flavours \\
107 & $\QGee\rightarrow \mbox{gg(g)}$ & \\
110+IQ & $\QGee\rightarrow \QGq\QGqbar\QGg$ & IQ as above, 
includes masses exactly \\
120+IQ & $\QGee\rightarrow \QGq\QGqbar$ &
IQ as above\hspace{5mm} no correction to hard \\
127 & $\QGee\rightarrow \mbox{gg}$ &
\hspace*{23mm}\raisebox{1ex}[0ex][0ex]{\LARGE\}} gluon branching \\
150+IL & $\QGee\rightarrow \ell\bar \ell$ & IL=2,3: $\ell=\mu,\tau$ \\
\hline
\multicolumn{1}{c|}{} & \multicolumn{2}{c|}{Di-Boson Production} \\
\hline
200 & $\QGee\rightarrow \mbox{W}^+\mbox{W}^-$ & 
\hspace*{5mm}$\mbox{W}^\pm/\mbox{Z}^0$ decays controlled \\
250 & $\QGee\rightarrow \mbox{Z}^0\mbox{Z}^0$ &
\raisebox{1ex}[0ex][0ex]{\LARGE\}} by MODBOS \\
\hline
\multicolumn{1}{c|}{} & \multicolumn{2}{c|}{Bjorken process:
$\QGee\rightarrow \mbox{Z}^0\mbox{h}^0$} \\
\hline
300+IQ & $+\mbox{h}^0\rightarrow \QGq\QGqbar$ & IQ as above \\
300+IL & $+\mbox{h}^0\rightarrow\ell\bar\ell$ & IL=1,2,3: $\ell=e,\mu,\tau$ \\
310,311 & $+\mbox{h}^0\rightarrow \mbox{W}^+\mbox{W}^-, 
\mbox{Z}^0\mbox{Z}^0$ & \\
312 & $+\mbox{h}^0\rightarrow \gamma\gamma$ & \\
399 & $+\mbox{h}^0\rightarrow$ anything & \\
\hline
\multicolumn{1}{c|}{} & \multicolumn{2}{c|}{Vector Boson Fusion} \\
\hline
400+ID & $\QGee\rightarrow\nu\bar\nu \mbox{h}^0+\QGee\mbox{h}^0$ & 
ID as IPROC=300+ID \\
\hline
\multicolumn{1}{c|}{} & \multicolumn{2}{c|}{Zero Resolved Gamma-Gamma:
$\QGee\rightarrow(\QGee)\gamma\gamma$} \\
\hline
500+ID & $\gamma\gamma\rightarrow \QGq\QGqbar/\ell\bar\ell/%
\mbox{W}^+\mbox{W}^-$ & ID=0-10 as for IPROC=300+ID \\
\hline
\multicolumn{1}{c|}{} & \multicolumn{2}{c|}{Gamma-$W$ Fusion:
$\QGee\rightarrow(\mbox{e}^+\nu_{\QGindx{e}})\gamma \mbox{W}^-$} \\
\hline
550+ID & $\gamma \mbox{W}^-\rightarrow \QGq\QGqbar'/\ell\bar\nu_\ell$ &
ID=0-9 as for IPROC=300+ID \\
\hline
\multicolumn{1}{c|}{} & \multicolumn{2}{c|}{Doubly Resolved Gamma-Gamma} \\
\hline
1500 & $\mbox{gg}\rightarrow \mbox{gg},\mbox{qg}\rightarrow \mbox{qg}$, 
{\it etc.} & 31 ${\cal O}(\QGalphas^2)$ two-to-two QCD scatterings \\
1700+IQ & $\mbox{gg}\rightarrow \mbox{Q}\QGbar{Q},\; 
\mbox{gQ}\rightarrow \mbox{gQ}$, {\it etc.} &
16 ${\cal O}(\QGalphas^2)$ heavy quark production processes \\
1800 & $\QGg\QGq\rightarrow\gamma \QGq,\;
\mbox{gg}\rightarrow\gamma \QGg$, {\it etc.} &
17 ${\cal O}(\QGalphas\QGalphaem,\;\QGalphas^3\QGalphaem)$ direct
photon processes\\
2200 & $\QGq\QGqbar\rightarrow \gamma\gamma,\;
\mbox{gg}\rightarrow\gamma\gamma$ &
3 ${\cal O}(\QGalphaem^2,\;\QGalphas^2\QGalphaem^2)$ di-photon processes \\
\hline
\multicolumn{1}{c|}{} & \multicolumn{2}{c|}{Singly Resolved Gamma-Gamma} \\
\hline
5000 & $\gamma \QGq\rightarrow \mbox{gq},\;
\gamma \QGg\rightarrow \QGq\QGqbar$, {\it etc.} &
3  ${\cal O}(\QGalphas\QGalphaem)$ dijet processes \\
5100+IQ & $\gamma \QGg\rightarrow \mbox{Q}\QGbar{Q}$ & 
heavy flavour pair production, IQ as above \\
5200+IQ & $\gamma \mbox{Q}\rightarrow \mbox{gQ},\;
\QGq\QGbar{Q}\rightarrow \QGq\QGbar{Q}$ & 
heavy flavour excitation, IQ as above \\
5500 & $\mbox{gg}\rightarrow V\QGg,\;\QGg\QGq\rightarrow V\QGq'$, {\it etc.} &
6 ${\cal O}(\QGalphas^2\QGalphaem)$ light (u,d,s)$\;L=0$ meson production \\
5510,5520 & & $J(=S)=0,1$ mesons only \\
\hline
8000 & & Minimum bias soft collision \\
\hline
\multicolumn{1}{c|}{} & \multicolumn{2}{c|}{Charged lepton Deep Inelastic
Scattering} \\
\hline
9000+IQ & $\mbox{eq}\rightarrow \mbox{eq}$,
e$\QGqbar\rightarrow\mbox{e}\QGqbar$ & NC DIS on flavour IQ as above \\
9010+IQ & $\mbox{eq}\rightarrow\nu_{\QGindx{e}} \QGq'$,
e$\QGqbar\rightarrow\nu_{\QGindx{e}}\QGqbar'$ & CC DIS on flavour IQ
as above \\
\hline
10000+IP & & As IPROC=IP but with suppressed SUE \\
\hline
\end{tabular}
\end{center}
\vspace{-4.5mm}
\caption{The principal {\small HERWIG} hard sub-process of importance for 
LEP~2
physics. In QCD scatterings {\tt IHPRO} labels the actual sub-process, 
allowing
for colour decomposition, generated.\label{QGIKHWt1}}
\end{table}

The cross-sections for $\gamma\gamma$ interactions rise with c.m. energy to
become the commonest physics processes at LEP~2. When considering hadronic
final states each photon may be viewed as interacting either as a point-like
particle or as being resolved into constituent (anti-)quarks and gluons.
This leads to three basic sets of hard sub-processes (zero, singly or doubly
resolved), a division adopted in the wide selection of sub-processes made
available in {\small HERWIG}. Note that this separation is in fact artificial
and all three components must be combined to obtain the full cross-section.
Discussion of these processes can be found in the {\small HERWIG} description
provided in the gamma-gamma section of this report.

\subsubsection{Initial State Radiation}
In $\QGee$ scattering real photons are radiated from the incoming lepton
lines. At LEP~1 any effects were mitigated against by the penalty involved
in going off the $\mbox{Z}^0$ resonance. However photon bremsstrahlung is 
expected
to be an important feature at LEP~2 energies where the basic cross-section
typically rises as $\hat s$ decreases. {\small HERWIG} uses an electron
structure function approach to write the total cross-section for a process
as:
\begin{equation}
\sigma(s)=\int^1_0\mbox{d}x_1\int^1_0\mbox{d}x_2
f^{\QGindx{e}}_{\QGindx{e}}(x_1)
f^{\QGindx{e}}_{\QGindx{e}}(x_2)\hat\sigma(x_1x_2s)
\end{equation}
Employing a natural choice of variables, $\tau=x_1x_2(=\hat s/s)$ and $x=x_1$,
this can be written:
\begin{equation}
\sigma(s)=\int^1_T\mbox{d}\tau\hat\sigma(\tau s)\int^1_\tau\frac{\mbox{d}x}{x}
f^{\QGindx{e}}_{\QGindx{e}}(x)
f^{\QGindx{e}}_{\QGindx{e}}\left(\frac{\tau}{x}\right)
\end{equation}
where $\tau>T$ ({\tt TMNISR}) is a physical cut-off used to avoid the $1/s$
pole in the cross-section's photon exchange term.

The actual structure function used is the second order solution to the
full Altarelli-Parisi equation with exponentiated coefficients:
\begin{eqnarray}
f^{\QGindx{e}}_{\QGindx{e}}(x) & = & 
\beta(1-x)^{\beta-1}\exp\left\{\beta\frac{x}{2}
\left(1+\frac{x}{2}\right)\right\}\left(1-\beta^2\frac{\pi^2}{12}\right)
\hspace{2cm} 
\beta=\frac{\QGalphaem}{\pi}\left[\log\left(
\frac{Q^2}{m^2_{\QGindx{e}}}\right)-1\right]
\nonumber \\
& + & \frac{\beta^2}{8}\left[(1+x)[(1+x)^2+3\log x]-\frac{4\log x}{1-x}\right]
+{\cal O}(\QGalphaem^3)
\end{eqnarray}
This is equivalent\footnote{A discrepancy in the coefficient of the $\pi^2$
term, a factor 2 too large, is believed to be their typographic error.} to
the expression, eqs.~(58,60), given on p.~34 of \cite{QGlep1}. This
means that the single photon emission allowed for in {\small HERWIG} gives
equivalent energy and $p^\perp$ spectra as multiple photon emission. In the
soft photon limit $f^{\QGindx{e}}_{\QGindx{e}}(x)$ simplifies significantly 
to the following form,
used to efficiently generate the $\{x_i\}$ via importance sampling:
\begin{equation}
f^{\QGindx{e}}_{\QGindx{e}}(x,Q^2)\approx\beta(1-x)^{\beta-1}
\end{equation}
In practical situations one has: $\QGalphaem/\pi\ll\beta\ll1$, so 
that $f^{\QGindx{e}}_{\QGindx{e}}(x)$
has an integrable singularity in the soft photon limit, $x\rightarrow 1$;
$1-x$ is the energy fraction carried by the photon. To regularize this
divergence {\small HERWIG} employs a resolution parameter, $x<X$, called {\tt
ZMXISR} (default $1-10^{-6}$), and includes a fraction of events with no
emission, so that:
\begin{equation}
f^{\QGindx{e}}_{\QGindx{e}}(x)\mapsto\bar f(x)=
\Theta(X-x)f(x)+\delta(1-x)(1-X)^\beta
\end{equation}
Here $X$ is only an internal parameter and unphysical in the sense that
cross-sections should not depend on it. Observe that even for $1-X={\cal
O}(10^{-6})$ the non-emission probability is $\approx 45$\% at LEP~2. Note
also setting {\tt ZMXISR}=0 has the effect of switching off the initial
state photon radiation.

After the emission of a photon the electron entering the hard sub-process
is off-shell. In {\small HERWIG} its negative virtuality is selected from a
logarithmic distribution, $\mbox{d}q^2/q^2$, bounded in magnitude by
eq.(\ref{QGIKlim}).

Allowing for the virtualities of the electron lines and treating $x$ as a
lightcone momentum fraction $\hat s$ is reconstructed as:
\begin{equation}
\hat s=\tau s-q^2_1-q^2_2+\frac{q^2_1q^2_2}{s}-2\underline{p}^\perp_1
\underline{p}^\perp_2
\end{equation}
Since $\hat\sigma$ is a rapidly varying function of $\hat s$ near the 
$\mbox{Z}^0$
{\small HERWIG} slightly shifts the $\{x_i\}$ fractions to preserve $\hat s=
\tau s$. Specifically the highest $p^\perp$ photon is taken to be emitted
first and its $x\mapsto x'=x+q^2/s$ ($x'$ is the energy fraction) so that
$\hat s$ would be preserved in the absence of emission from the other lepton.
The $x$ of the lower $p^\perp$ photon is then shifted so as to give exactly
$\hat s$. For simplicity the program requires photon emission to be in the
forward hemisphere which imposes the condition:
\begin{equation}
\label{QGIKlim}
q^2<\frac{x'}{1+x'}(1-x)s
\end{equation}
This inequality is applied to both leptons. Note that this still allows the
possibility for on resonance $\gamma \mbox{Z}^0$ states to be produced but 
only to the accuracy of the leading logarithm approximation.

The use of the Equivalent Photon Approximation for the case of virtual photon
emission in which it is the photon which enters the hard sub-process is again
discussed in the report of the gamma-gamma working group.

\subsubsection{Parton Showers}
{\small HERWIG} employs highly developed parton shower algorithms to provide
an accurate description of the perturbative QCD jet evolution. Coherence,
due to leading infrared singularities \cite{QGcohrev}, is automatically
included through the choice of evolution variables, ordering in which
naturally restricts the branching phase space to an angular ordered region.
Further angular screening due to heavy quark masses, the {\it dead cone},
is also fully included \cite{QGIKmass}. At each branching the azimuthal angles
are distributed according to the eikonal dipole distribution for soft gluons
\cite{QGIKsoft}, including mass effects, and to the full collinear leading
logarithm accuracy for hard emission \cite{QGIKhard}. At large momentum
fractions the coherent algorithm used also correctly describes next-to-leading
contributions \cite{QGIKcat}. By using a two-loop expression for 
$\QGalphas$ this allows the Monte Carlo $\Lambda$ to be related to 
$\Lambda_{\overline{\QGindx{MS}}}$ as
$x\rightarrow 1$
\begin{equation}
\Lambda_{\QGindx{MC}}=
\exp\left\{\frac{C_A(67-3\pi^2)-10n_f}{6(11C_A-2n_f)}\right\}
\Lambda_{\overline{\QGindx{MS}}}\approx|_{n_f=5}1.569
\Lambda^{(5)}_{\overline{\QGindx{MS}}}
\end{equation}

Since the time of the LEP~1 workshop significant progress has been made in
the study of final states involving photons \cite{QGanacy}, leading to
the implementation of final state photon radiation in {\small HERWIG}
\cite{QGIKmike}. The momentum sharing in $\QGq\rightarrow \QGq\gamma$ 
branchings
and relative rate compared to $\QGq\rightarrow \QGq\QGg$ branchings are 
controlled by the following splitting function and Sudakov form factor:
\begin{eqnarray}
P_{\QGindx{q}\rightarrow \QGindx{q}\gamma}(z) & = & e^2_{\QGindx{q}}
\frac{\QGalphaem}{2\pi} \frac{1+z^2}{1-z}
\nonumber \\
\log\Delta_s(Q^2,Q_0^2) & = & -e^2_{\QGindx{e}}\frac{\QGalphaem}{\pi}[
(\log(Q/Q_\gamma)-3/4)^2 - (\log(Q_0/Q_\gamma)-3/4)^2]
\end{eqnarray}
where, since the photon is in the final state, a fixed $\QGalphaem$ is used
(allowing analytic integration of the Sudakov form factor) and 
$Q_0=Q_{\QGindx{q}}+Q_\gamma$ with $Q_{\QGindx{q}}$ and $Q_\gamma$ 
the cut-offs on the quark and photon scales respectively. The
branching $\gamma\rightarrow \QGq\QGqbar$ is expected to be small and is not
included. Competition between the two types of quark branching is handled in
the standard way. That is the $Q^2$ scales at which the two types of branching
attain a preselected probability of occurring are found, the larger is taken
to occur first and if its $Q^2$ is above $(Q_{\QGindx{q}}+Q_\gamma)^2$ 
it is accepted.
The scale $Q^2$ of any branching is bounded above by that of the last 
emission,
irrespective of type. However the opening angle is bounded from above by the
opening angle of the last emission of the same type; this is exact in the
case when azimuthal photon-gluon correlations are integrated out.

Due to the choice of evolution variables in {\small HERWIG}, configurations in
which a very hard gluon or photon recoils against the $\QGq\QGqbar$ pair are
not generated by the showering algorithm, that is a `dead zone' exists
\cite{QGIKmike}. This is particularly important in the photon case due to the
relative ease with which they can be identified in the final state. The
{\small HERWIG} solution is to find what fraction of events are missing by
integrating the three parton matrix element over the dead zone and then add
back this fraction starting the evolution from a correctly distributed
$\QGq\QGqbar \QGg/\gamma$ configuration. The algorithm of \cite{QGIKron} 
is used to
exactly include initial/final state correlations starting from a massless
$\QGq\QGqbar$ configuration.

The matching of a hard gluon or photon to the exact matrix element is
controlled by the logical {\tt HARDME} (default {\tt .TRUE.}). Additionally
there is a `soft' matrix element correction, where soft here means 
inside the phase space region accessible to the branching algorithm;
{\tt SOFTME} (default {\tt .TRUE.}) controls the matching of the hardest 
emission, not necessarily the first, to the exact matrix element
\cite{QGIKmikesonlyref}.

\subsubsection{Hadronization}
The basic preconfinement inspired \cite{QGprecon} cluster hadronization model
used in {\small HERWIG} remains little changed from its original formulation
\cite{QGIKclus}. The principle criterion for selecting the flavours and spins
of the primary hadrons in the cluster two body decays is the phase space
available; though weights {\tt PWT(1-6), VECWT, TENWT} and {\tt DECWT} can
be used to alter the flavour/spin compositions. The cluster decays are
isotropic, in their own rest frame, except when a perturbative quark is
involved, that is one from the hard sub-process or a 
$\QGg\rightarrow \QGq\QGqbar$
splitting. If ({\tt CLDIR}=1), the default, then the hadron containing this
quark is aligned with the quark direction in the cluster rest frame. The
main effect is to stiffen the spectrum of heavy charm and bottom hadrons.
It is possible to partially decorrelate this direction retention using the
parameter {\tt CLSMR} (default 0), the width of an exponential distribution
in $1-\cos\theta_{\QGindx{qh}}$; thus increasing {\tt CLSMR} increases 
the smearing.

New parameters have been introduced to control the treatment of clusters
with anomalous masses. {\tt CLPOW} (default 2) influences the decision on
whether a heavy cluster $\QGq_1\QGqbar_2$ should first be split in two prior
to hadronization according to if its mass satisfies the inequality:
\begin{equation}
M^{CLPOW}_{cl}>CLMAX^{CLPOW}+(m_{\QGindx{q}_1}+
m_{\overline{\QGindx{q}}_2})^{CLPOW}
\end{equation}
Using smaller values of {\tt CLPOW} leads to an increased yield of heavy
clusters containing heavy quarks and thence to more heavy baryons; light
quark clusters are affected less. The parameter {\tt B1LIM} (default 0)
can be used to increase the number of relatively light bottom clusters that
undergo a one-body decay. If $M_{thr}$ is the threshold for two-body decay
then the probability of a one-body decay becomes:
\begin{equation}
{\cal P}=\left\{\begin{array}{lrcccl}
1 & & & M_{cl} & < & M_{thr} \\
1-\frac{M_{thr}-M_{cl}}{B1LIM\star M_{thr}} & M_{thr} & < & M_{cl} & < &
(1+B1LIM)M_{thr} \\
0 & (1+B1LIM)M_{thr} & < & M_{cl} & & \end{array}\right.
\end{equation}
For light quark clusters the one-body decay criterion remains equivalent
to the above with {\tt B1LIM}=0. In practice {\tt CLPOW} proves more
effective in controlling the spectrum of both bottom and charm hadrons.
When one-body decays do occur a Lorentz covariant treatment is now used
to effect the necessary momentum rearrangement.

In the default version of {\small HERWIG} the quark--antiquark pairs which
form the colour singlet clusters are taken to be nearest neighbour pairs,
in a sense defined by the shower. However a colour reconnection model is
now available. It is  based upon minimizing the spatial sizes of pairs of
clusters as determined from the semi-classical positions of the partons
at the end of the showers. This model is discussed more fully
in \ref{QGsubsectreconn}.

More recently the number of hadrons supported has been enlarged to incorporate
all $L=0,1$ mesons (including the $0^{+(+)}$ and $1^{+(+)}$ states) composed
of d,u,s,c,b quarks and all $J=1/2$ (`octet') and $J=3/2$ (`decuplet')
baryons composed of the form $\QGq_1\QGq_2\QGq_3$ or 
$\mbox{Q}\QGq_1\QGq_2,\; (\mbox{Q}=\mbox{c,b})$. Should the
user wish to add any new particles it is sufficient to simply specify their
properties: name, PDG code number, mass, spin and flavour compositions in the
arrays {\tt RNAME, IDPDG, RMASS, RSPIN} and {\tt IFLAV} and they will be
included automatically in cluster decays. Using the array {\tt VTOCDK} it is
also possible to veto a particular hadron's production in cluster decays.

\subsubsection{Decay Tables}
The {\small HERWIG} decay routines have been largely re-written to make them
more user friendly and to adopt the proposals made in section
\ref{QGsubsectdecay}. Up to five body decays are supported with a number
of standard matrix elements made available. Specific hadronic decay channels
for B hadrons can now be included. This is in addition to the original
partonic model based on spectator decays \cite{QGIKmass}; note this may
involve some double counting. The production of a selected particle via
unstable particle decays can be vetoed by specifying it in the array {\tt
VTORDK}; any branching ratio sums affected because of excluded channels are
automatically reset to unity. The subroutine {\tt HWIODK} has been added to
allow the {\small HERWIG} decay tables to be inputted and outputted in the
proposed standard format. When read in the program checks that the decay is
kinematically allowed and does not violate electric charge conservation; if
necessary the sum of branching ratios is reset to one. The use of this
subroutine makes it simple for the users to adapt the provided tables for
their own use. The subroutine {\tt HWMODK} allows individual channels in the
decay tables to be added or modified between events. The actual default decay
tables themselves have also been updated to include modes at the one {\it per
mille} level.

Interfaces to the {\sc eurodec} \cite{QGIKeuro} and {\sc cleo} \cite{QGIKcleo}
B hadron decay packages are also built into {\small HERWIG}. The selection is
made by setting {\tt BDECAY='EURO','CLEO'} or {\tt 'HERW'} (the default is
of course {\tt 'HERW'}).

The production vertices of hadrons are now calculated by {\small HERWIG} and
stored using the {\tt VHEP} array of {\tt /HEPEVT/}. This is based on the
particle lifetimes in the {\tt RLTIM} array. A particle is set unstable if
its lifetime is less than {\tt PLTCUT} however when {\tt MAXDKL=.TRUE.} all
decays are tested in the routine {\tt HWDXLM} and required to occur within
a volume specified by {\tt IOPDKL} else left undecayed. If {\tt B0MIX=.TRUE.}
then neutral $B^0_{\QGindx{d},\QGindx{s}}$ mesons are allowed to mix before 
decaying.

\subsubsection{Source Code}

In addition to the WWW site quoted above copies of the {\small HERWIG} source
code and supporting files are maintained in the following VAX directories:
\begin{tabbing}
xxxxxxxxxxxxxxxxxxxxx\= \kill
                     \>{\tt CBHEP::DISK\$THEORY:[THEORY.HERWIG]HERWIGnm.*} \\
                     \>{\tt FNALV::USR\$ROOT2:[BWEBBER.HERWIG]HERWIGnm.*} \\
                     \>{\tt VXCERN::DISK\$CR:[WEBBER.HERWIG]HERWIGnm.* }
\end{tabbing}
The files supplied are {\tt HERWIGnm.COM, *.DOC, *.FOR, *.INC, *.MSG, *.SUD}
and {\tt *.TST}. The command file {\tt HERWIGnm.COM} runs a test job {\tt
*.TST} containing the main program. This uses the source code subroutines
found in {\tt *.FOR} with the declarations and common blocks in {\tt *.INC}
and default Sudakov form factors in {\tt *.SUD}. Release notes are found in
{\tt *.MSG} and more complete documentation in {\tt *.DOC}.

\subsection{NLLjet}
\noindent{\bf Basic Facts}
\vspace{-\baselineskip} 
\begin{tabbing}
{\bf Program location:} \= \kill
{\bf Program names:}    \> {\sc NLLjet}  \cite{QGKKrefa} \\
{\bf Versions:}         \> {\sc NLLjet} 3.0 of  September 1992 \\
{\bf Author:}           \> Kiyoshi Kato                  \\
                        \> Kogakuin University \\   
                        \> Nishi-Shinjuku 1-24, Shinjuku, %
Tokyo 160, Japan\\
                        \> Phone: + 81 -- 3 - 3342 - 1211       \\
                        \> E-mail: kato@sin.cc.kogakuin.ac.jp    \\
                        \> Tomo Munehisa\\   
                        \> Yamanashi University \\   
                        \> Takeda 4-3, Kofu 400, Japan\\   
                        \> Phone: + 81 -- 552 - 20 - 8584       \\
                        \> E-mail: munehisa@top.esb.yamanashi.ac.jp \\
{\bf Program size:}     \> 7742 lines                     \\
{\bf Program location:} \> ftp.kek.jp : kek/minami/nlljet \\
\end{tabbing}

{\sc NLLjet} is a Monte Carlo code for the generation of jet
events in $\QGee$ annihilation based on the parton shower method.
The events are parton final states in the
form of a list with particle codes and four-momenta.
Connection to the hadronization is open for the user, and
a standard interface to Lund hadronization is provided.

Generation of QCD jets by the parton-shower method was born of
Konishi, Ukawa and Veneziano in 1979 as the ``jet calculus'' in 
which the method to make systematic summation of the collinear 
singularity in QCD was given.  Here, the factorization of the mass 
singularity works well and the choice of physical gauge leads to a
suppression of interference terms, so that a stochastic treatment 
for jets becomes possible.

Soon after that, models of the QCD parton shower in the 
leading-logarithmic (LL) approximation were developed.  These models
are good for the description of jets in high energy.  However,
they have no chance to determine the fundamental parameter of
QCD, $\QGalphas(\mu^2)$ (or QCD $\Lambda$), because
starting from {\it any} renormalization scheme in QCD, you obtain
the {\it same} formula for physical quantities in LL approximation. 
This limits the analysis for the determination of the strong
coupling constant in jet phenomena only to the calculation 
based on the QCD matrix elements.  
However, the Monte-Carlo simulation of jets
by matrix elements is not appropriate for the global description of jets
since it has an avoidable defect, the discontinuity between
$n$- and $(n+1)$-parton states.

The idea of {\sc NLLjet} was spawned from observation above.
In this parton-shower model, 
the collinear singularity of QCD is summed up to
the next-to-leading logarithmic(NLL) order.
All components in NLL order are computed in the $\overline{\rm MS}$
scheme, and they are implemented in the model.
Thus {\sc NLLjet} has the potential to determine the QCD 
$\Lambda_{\overline{\rm MS}}$
through a comparison of generated events with experiments
\cite{QGKKrefb}.
The basic ingredients of {\sc NLLjet} are as follows:
\begin{QGitemize}
 \item Sudakov factor which is defined by the integral
 of the $P$ function up to $O(\QGalphas^2)$.
 \item Two-body branching by the two-body vertex function
 up to $O(\QGalphas^2)$.
 \item Three-body branching by the three-body vertex function
 in $O(\QGalphas^2)$.
 \item Hard cross section of the primary $\QGq\QGqbar\QGg$ process 
 up to $O(\QGalphas^2)$.
 \item Kinematical conditions and correction terms.
\end{QGitemize}

The effect of soft-gluon contribution is an important issue
in perturbative QCD.
In {\sc NLLjet}, the strong coupling constant in the Sudakov
factor is defined
to be $\QGalphas(x(1-x)Q^2)$, and it corresponds to the inclusion
of soft gluon resummation.  The angular ordering is not
introduced to all branchings but only to those 
in which the angular ordering is really required.

The important point of the formulation beyond LL order is that
each kinematical modification is always controlled properly 
through the introduction of a correction term in the NLL order 
functions.  
The three-body vertex functions become positive with the
correction for the angular ordering in $\QGq \to \QGq+\QGg+\QGg$ 
and $\QGg\to \QGg+\QGg+\QGg$ and that for the momentum conservation.
The double cascade scheme, which is necessary to recover the
symmetry between $\QGq$ and $\QGqbar$, also gives another
correction term.

The parton shower method still has a few ambiguous points which are 
hard to determine from the theoretical view point in perturbative QCD.
For example, the virtuality of partons in final states should be
less than a cutoff value, $Q_0^2$.  Normally, one sets it 
equal to $Q_0^2$. However, sometimes
better agreement with experiments is found by taking it to be 0.
In this version, this modification is included by setting
{\tt KINEM -1} parameter.

The effect of a quark mass is only counted kinematically 
by replacing $Q^2$ by $Q^2+m_{\QGindx{q}}^2$. 
Neither azimuthal correlations nor the parton polarization
are considered.

Essential input parameters of {\sc NLLjet} are
$W$, $\Lambda$, $Q_0^2$, $\delta$, and $C$.
Here $W$ stands for the center-of-mass energy.
Physics should not depend strongly on the cutoff $Q_0^2$
and its dependence is to be counted as a systematic error
of the theory. Parameter $\delta$ is specific to
NLL parton shower and it is absent in LL order. 
The distribution is expected to be independent of $\delta$.
However, detailed study shows that there is small bend
at the region connected by $\delta$. If one sets $\delta$
large ($\sim 0.5$), the events are free from the bend
at the expense of the exclusion of $\QGq\QGqbar\QGg$ primary
vertex. The scheme parameter $C$ is available in order 
to replace $\mu^2$ in
$\QGalphas(\mu^2)$ from $\mu^2=Q^2$ to $\mu^2=CQ^2$.
However, it is not possible to change $C$ in large. In the
matrix element, QCD is studied at $Q^2=W^2$ while in parton
shower, it is done for $Q^2=W^2 \sim Q_0^2$.

\subsection{PYTHIA/JETSET}
\label{QGsubsectPythia}

\noindent{\bf Basic Facts}
\vspace{-\baselineskip} 
\begin{tabbing}
{\bf Program location:} \= \kill
{\bf Program names:}    \> {\sc Pythia} and {\sc Jetset} \cite{QGjetset} \\
{\bf Versions:}         \> {\sc Pythia} 5.720 of 29 November 1995 \\
                        \> {\sc Jetset} 7.408 of 23 August 1995    \\
{\bf Author:}           \> Torbj\"orn Sj\"ostrand                  \\
                        \> Department of Theoretical Physics       \\
                        \> University of Lund                      \\   
                        \> S\"olvegatan 14A, S-223 62 Lund, Sweden \\
                        \> Phone: + 46 -- 46 - 222 48 16           \\
                        \> E-mail: torbjorn@thep.lu.se             \\
{\bf Program size:}     \> 19936 + 11541 lines                     \\
{\bf Program location:} \> http://thep.lu.se/tf2/staff/torbjorn/
\end{tabbing}

\subsubsection{Introduction}
The {\sc Jetset} program has been used frequently for QCD physics 
studies at LEP~1. For applications at LEP~2, {\sc Jetset} should be
complemented with the {\sc Pythia} program. While {\sc Jetset} only
gives access to one hard process, $\QGee \to \gamma^*/\mbox{Z}^0 \to 
\QGq\QGqbar$, {\sc Pythia} contains a wealth of 
different processes. The two programs are fully integrated, in that a 
call to {\sc Pythia} will not only generate a hard process 
but also automatically call {\sc Jetset} routines to
perform (timelike) parton showers and fragmentation. Output is in
the normal {\tt LUJETS} commonblock (with easy translation to the 
{\tt HEPEVT} standard) and can be studied with the
{\sc Jetset} analysis routines. The emphasis of the {\sc Pythia/Jetset}
package is to provide a realistic description of varying hadronic 
final states, but also non-hadronic processes may be generated.

In addition to the briefer published description of the programs,
there is a complete manual and physics description of over 300 pages
\cite{QGjetset}. The programs, the manual, update notes and sample 
main programs can be picked up from the
web address given above; additionally the CERN program library
provides the programs and hardcopies of the manual. The description
given here therefore only contains some highlights, with special
emphasis on the aspects of relevance for LEP~2 applications. 

For the description of a typical high-energy event, a
generator should contain a simulation of several physics aspects.
If we try to follow the evolution of an event in some semblance of
a time order, one may arrange these aspects as follows:
\begin{QGenumerate}
\item Initially the e$^+$ and e$^-$ are coming in towards each other.
An electron contains virtual fluctuations into photons, quarks, gluons, 
and so on. It is useful to employ the same parton-distribution 
and parton-shower language as for hadrons. Thus also electrons and
photons are included in the parton concept. An initial-state parton
shower develops by branchings such as $\mbox{e} \to \mbox{e} \gamma$,
$\gamma \to \QGq \QGqbar$ and $\QGq \to \QGq \QGg$.
\item One parton from each of the e$^+$- and e$^-$-initiated showers
enters the hard process, where then a number of outgoing 
partons/particles are produced. It is the nature of this process that 
determines the main characteristics of the event. (Also some soft 
processes are included in the program; since much of the same framework
can be used we do not here belabour the differences.)
\item If the hard process produces massive electroweak particles, such 
as the Z$^0$, the W$^{\pm}$ or a Higgs, the decay into lighter objects 
must be considered. 
\item The outgoing partons may branch, to build up
final-state showers.
\item Further semihard interactions may occur between the other partons 
in the case of two incoming resolved photons.
\item When a shower initiator is taken out of a beam particle,
a beam remnant is left behind. 
\item The QCD confinement mechanism ensures that the outgoing quarks 
and gluons are not observable, but instead fragment to colour-neutral 
hadrons.
\item Many of the produced hadrons are unstable and decay further.
\end{QGenumerate}
The time-order above does not have to coincide with the generation
sequence. Typically the hard process is selected first.
 
\begin{table}[t]
\caption{Main LEP~2 physics processes available in {\sc Pythia}.
\label{QGTSpyhard}}
\vspace{2mm}
\hfill
\begin{tabular}[t]{|r|l|}
\hline
\rule{0mm}{5mm}ISUB & Process  \\[1mm]
\hline
\multicolumn{2}{|c|}{\rule{0mm}{5mm}Gauge boson production} \\[1mm]
\hline
\rule{0mm}{5mm}%
~1 & $\QGee \to \QGgaZ$ \\
18 & $\QGee \to \gamma\gamma$ \\
19 & $\QGee \to \gamma(\QGgaZ)$ \\
22 & $\QGee \to (\QGgaZ)(\QGgaZ)$ \\
25 & $\QGee \to \mbox{W}^+ \mbox{W}^-$ \\
35 & $\mbox{e} \gamma \to \mbox{e} (\QGgaZ)$ \\
36 & $\mbox{e} \gamma \to \nu \mbox{W}$ \\
69 & $\gamma \gamma \to \mbox{W}^+ \mbox{W}^-$ \\
70 & $\gamma  \mbox{W} \to \mbox{Z}^0 \mbox{W}$ \\ 
\hline
\multicolumn{2}{|c|}{\rule{0mm}{5mm}Higgs production} \\[1mm]
\hline
\rule{0mm}{5mm}%
~24 & $\QGee \to \mbox{Z}^0 \mbox{h}^0$ \\
103 & $\gamma\gamma \to \mbox{h}^0$ \\
110 & $\QGee \to \gamma \mbox{h}^0$ \\ 
123 & $\QGee \to \QGee \mbox{h}^0$ \\
124 & $\QGee \to \nu_{\mbox{e}} \overline{\nu}_{\mbox{e}}
      \mbox{h}^0$ \\
141 & $\QGee \to \gamma^*/\mbox{Z}^0/\mbox{Z}'^0 \to 
      \mbox{H}^+ \mbox{H}^-, \mbox{h}^0\mbox{A}^0, 
      \mbox{H}^0 \mbox{A}^0$ \\
171 & $\QGee \to \mbox{Z}^0 \mbox{H}^0$ \\
173 & $\QGee \to \QGee \mbox{H}^0$ \\
174 & $\QGee \to \nu_{\mbox{e}} \overline{\nu}_{\mbox{e}}
      \mbox{H}^0$ \\
\hline
\multicolumn{2}{|c|}{\rule{0mm}{5mm}Other processes} \\[1mm]
\hline
\rule{0mm}{5mm}%
~10 & $\QGee \to \QGee, \nu_{\mbox{e}} \overline{\nu}_{\mbox{e}}$ \\
141 & $\QGee \to \gamma^*/\mbox{Z}^0/\mbox{Z}'^0$ \\
\hline
\end{tabular}
\hfill
\begin{tabular}[t]{|r|l|}
\hline
\rule{0mm}{5mm}ISUB & Process  \\[1mm]
\hline
\multicolumn{2}{|c|}{\rule{0mm}{5mm}$\gamma\gamma$ physics} \\[1mm]
\hline
\rule{0mm}{5mm}%
58 & $\gamma \gamma \to \QGq \QGqbar, \ell^+ \ell^-$ \\
33 & $\gamma \QGq \to \QGq \QGg$ \\
54 & $\gamma \QGg \to \QGq \QGqbar$\\
11 & $\QGq \QGq' \to \QGq \QGq'$ \\
12 & $\QGq \QGqbar \to \QGq' \QGqbar'$ \\
13 & $\QGq \QGqbar \to \QGg \QGg$ \\
14 & $\QGq \QGqbar \to \QGg \gamma$ \\
18 & $\QGq \QGqbar \to \gamma \gamma$ \\
28 & $\QGq \QGg \to \QGq \QGg$ \\
29 & $\QGq \QGg \to \QGq \gamma$ \\
53 & $\QGg \QGg \to \QGq \QGqbar$ \\
68 & $\QGg \QGg \to \QGg \QGg$ \\
91 & $\gamma \gamma \to V V'$ \\
92 & $\gamma \gamma \to X V$ \\
93 & $\gamma \gamma \to V X$ \\
94 & $\gamma \gamma \to X_1 X_2$ \\
95 & $\gamma \gamma \to \mbox{low-}p_{\perp}$ \\
85 & $\gamma \gamma \to \mbox{Q}\overline{\mbox{Q}}, \ell^+ \ell^-$ \\
84 & $\gamma \QGg \to \mbox{Q}\overline{\mbox{Q}}$ \\
81 & $\QGq \QGqbar \to \mbox{Q}\overline{\mbox{Q}}$ \\
82 & $\QGg \QGg \to \mbox{Q}\overline{\mbox{Q}}$ \\
\hline
\multicolumn{2}{|c|}{\rule{0mm}{5mm}DIS} \\[1mm]
\hline
10 & $\mbox{e} \QGq \to \mbox{e} \QGq$ \\ 
\hline
\end{tabular}
\hfill
\end{table}

\subsubsection{Hard processes}
Close to a hundred subprocess cross sections have been encoded 
in {\sc Pythia}. Lepton, hadron and photon beams are allowed;
thus the program can be used for p$\overline{\mbox{p}}$/pp 
physics at the Tevatron or LHC or for ep physics 
at HERA. Here we concentrate on processes of relevance for LEP~2.
Some of the more interesting ones are listed in 
table~\ref{QGTSpyhard} and discussed below. Further comments 
may be found in other sections of this report.

It it important to note that {\sc Pythia} is not intended to be a
precision program for electroweak physics. The philosophy is to
provide sensible first approximations to a wide selection of hard
processes, as a starting point for a detailed 
simulation of the subsequent QCD steps, i.e. parton showers, 
fragmentation and decay. It is therefore orthogonal in philosophy
to many dedicated electroweak generators, that attempt to provide
the hard-scattering cross section with very high precision but
do not go beyond a parton-level description.  

Subprocess 1 is the familiar $\gamma^*/\mbox{Z}^0$ process that
dominates LEP~1 physics. The full interference structure between
the $\gamma$ and Z$^0$ propagators is included. It supersedes
the {\tt LUEEVT} generator of {\sc Jetset}. The main
differences are: 
\begin{QGitemize}
\item {\tt LUEEVT} uses a matrix-element approach to generate at 
most one initial-state photon, while {\sc Pythia} allows for 
multiple photon emission in a parton-shower approach; 
\item {\tt LUEEVT} allows only hadronic final states, while {\sc Pythia}
also includes leptonic ones;
\item {\tt LUEEVT} contains a simple Breit-Wigner with the width
$\Gamma_{\QGindx{Z}}$ as input, while {\sc Pythia} contains
an $s$-dependent Breit-Wigner that is dynamically calculated from
electroweak parameters; and
\item the option to simulate first- or second-order MEs currently only
exists with {\tt LUEEVT}.
\end{QGitemize}
Subprocess 19 contains a photon in addition to
the $\gamma^*/\mbox{Z}^0$. This means double counting, since already 
process~1 can contain initial-state-radiation photons, so results
from the two processes should not be added. The usage of process 19 
should be restricted to events that contain a high-$p_{\perp}$ photon,
where generation then is more efficient (and accurate) than what is 
offered by process 1.

Subprocess 25 describes W pair production, including subsequent decay
into four fermions with full angular correlations. The formalism 
includes $s$-dependent widths in the Breit-Wigners and options to
pick the set of independent electroweak parameters. However, it is 
restricted to the basic graphs of W pair production (``CC03''). 

Subprocess 22 describes $\QGgaZ$ pair production in a similar 
approximation (``NC02''). Note that interference terms between process 
22 and 25 are not
found anywhere. This is in accordance with the basic philosophy of a
reasonable but not exhaustive description of electroweak processes. 

Subprocesses 35 and 36 describe the production of a single $\QGgaZ$ 
or W in the approximation of an effective photon flux.
A process such as $\mbox{e}^+ \gamma \to \overline{\nu}_{\QGindx{e}}
\mbox{W}^+$ thus is convoluted with the parton-shower approximation
of the $\mbox{e}^- \to \mbox{e}^- \gamma$ branching to give an effective
process $\QGee \to \overline{\nu}_{\QGindx{e}} \mbox{e}^- \mbox{W}^+$.
Process 35 has a singularity when the scattered electron has vanishing 
$p_{\perp}$ (in principle this is regularized by the electron 
mass, but in practice the $m_{\QGindx{e}}$ has been neglected).
Therefore it is necessary to run with some minimum $p_{\perp}$ cut-off; 
for numerical reasons at least 0.01~GeV. An alternative description
can be obtained by using an electron-inside-photon-inside-electron
parton distribution ({\tt MSTP(12)=1}) in process 1. For process 36 the 
decay of the W is assumed isotropic since the appropriate matrix elements
have not been coded. Also subprocesses 69 and 70 assume isotropic
W/Z decay. Furthermore, process 70 does not include contributions from
$\gamma^*$ but only from Z$^0$. The process implementations in this 
paragraph thus are less sophisticated than the single $\QGgaZ$, W pair
and $\QGgaZ$ pair processes above.  

{\sc Pythia} is equipped with an extensive selection of
production processes for the standard model Higgs, here denoted ``h$^0$''.
(It is called ``H$^0$'' in the program, which confuses matters when two
Higgs doublets are introduced, but for this report we stay with the
conventional terminology.) Not all available processes have been listed, 
but only those of some interest. The most important by far (at LEP~2) 
is process 24, $\QGee \to \mbox{Z}^0 \to \mbox{Z}^0 \mbox{h}^0$.
Both Z$^0$'s in the graph have been included with a Breit-Wigner
shape, so there is no formal restriction that either of them need be 
on or close to the mass shell. For a Higgs with 
$m_{\QGindx{h}} + m_{\QGindx{Z}} > E_{\QGindx{cm}}$
process 124 takes over, $\QGee \to 
\nu_{\mbox{e}} \overline{\nu}_{\mbox{e}} \mbox{W}^+ \mbox{W}^-
\to \nu_{\mbox{e}} \overline{\nu}_{\mbox{e}} \mbox{h}^0$, but
at a much smaller rate. Processes 110, 123 (Z$^0$Z$^0$ fusion) and 
103 are even further suppressed. 

All major h$^0$ decay modes are included: 
$\mbox{h}^0 \to \QGq\QGqbar$, $\mbox{h}^0 \to \ell^+\ell^-$,
$\mbox{h}^0 \to \mbox{W}^+\mbox{W}^-$,  
$\mbox{h}^0 \to \mbox{Z}^0\mbox{Z}^0$,
$\mbox{h}^0 \to \mbox{gg}$, $\mbox{h}^0 \to \gamma\gamma$ 
and $\mbox{h}^0 \to \gamma\mbox{Z}^0$. The branching ratios are
automatically recalculated based on the Higgs mass. One point 
that should be noted is that the parton-shower algorithm matches
to the same three-jet matrix element that is used for $\QGgaZ$
decays. This gives a somewhat incorrect rate for three-jet 
production in Higgs decay.

In the minimal supersymmetric extension to the standard model
the number of production processes is further increased.
The full set of Higgs particles is included in the program: 
h$^0$, H$^0$, A$^0$ and H$^{\pm}$. Masses and couplings can be 
set by the user; this is a somewhat lengthy process, however,
since currently the one-loop mass relations are not built into
the program. The H$^0$ have the same production processes as the
h$^0$; the list in table~\ref{QGTSpyhard} only shows the more 
interesting. Higgs pair production, $\mbox{H}^+ \mbox{H}^-$, 
$\mbox{h}^0\mbox{A}^0$ and $\mbox{H}^0 \mbox{A}^0$, proceeds only
through $s$-channel graphs and has been included as part of 
$\gamma^*/\mbox{Z}^0/\mbox{Z}'^0$ decays, process 141.
The Z$'^0$ part can easily be switched off ({\tt MSTP(44)=4}),
so that processes 1 and 141 become identical except for the larger
selection of decay modes in the latter. (Technically, this way the 
program can distinguish the Z$'$ decaying to Higgses from a Z
produced in Higgs decay, and accommodate different decay modes for 
the two.)
 
Subprocess 141 is also useful for the study of virtual corrections
caused by the existence of a Z$'^0$ somewhere above the LEP~2
energy range. Vector and axial couplings may be set freely to 
simulate various scenarios, and interference with $\QGgaZ$
is automatically included.

$\gamma\gamma$ physics is a large area, in that a wealth of different
subprocesses is involved. A photon may act as a pointlike particle
or as a resolved, hadronlike state. A simple subdivision of processes
is therefore into direct (58, 85), once-resolved (33, 54, 84)
and twice-resolved (the rest, that is all processes allowed e.g.
in pp collisions). The resolved part of the photon may be further
subdivided into a VMD (vector meson dominance) and anomalous part.
In total therefore six classes of events can be 
separated \cite{QGTSgamgam}. An automatic
mix to provide a ``minimum bias'' sample of events is obtainable
as an option ({\tt MSTP(14)=10}). Processes 81--85 include masses
in the matrix elements, and thus are convenient to study e.g.
heavy-flavour production. It should be noted that several
aspects remain to be solved, for instance that of (slightly) off-shell
incoming photons. Furthermore, on a technical note, {\sc Pythia} is 
originally designed for fixed energies of the incoming particles, 
and so the process of having $\gamma\gamma$ ``hadronic'' collisions 
at varying energies is not yet fully automated.
 
Finally, note that process 10 can be used both as a Bhabha and a
deep-inelastic-scattering generator. In neither respect is it 
competitive with dedicated programs, but it may be useful for 
first estimates.

\subsubsection{Hard process generation}
The cross section for a process $ij \to k$ is given by
\begin{equation}
\sigma_{ij \to k} = \int \mbox{d}x_1 \int \mbox{d}x_2 \, 
f^{\QGindx{e}^+}_i(x_1,Q^2) \,
f^{\QGindx{e}^-}_j(x_2,Q^2) \, 
\hat{\sigma}_{ij \to k}(\hat{s}) ~.
\end{equation}
Here $\hat{\sigma}$ is the cross section for the hard partonic process,
as codified in the matrix elements for each specific process. For 
processes with several particles in the final state it would be replaced 
by an integral over the allowed final-state phase space. 
The $f_i(x,Q^2)$ are the parton distribution functions, which 
describe the probability to find a parton $i$ inside an e$^{\pm}$ beam 
particle, with parton $i$ carrying a fraction $x$ of the total 
e$^{\pm}$ momentum, when the e$^{\pm}$ is probed at some squared 
momentum scale $Q^2$ that characterizes the hard process. The hard 
scattering therefore only involves a squared invariant mass 
$\hat{s} = x_1 x_2 s = x_1 x_2 E_{\QGindx{cm}}^2$, where 
$E_{\QGindx{cm}}$ is the c.m. energy of the event.

The electron-inside-electron parton distributions are  based on a
next-to-leading order exponentiated description, see \cite{QGLEPoneQED}. 
The approximate behaviour is
\begin{equation}
f_{\QGindx{e}}^{\QGindx{e}}(x,Q^2) \approx 
\frac{\beta}{2} (1-x)^{\frac{\beta}{2}-1};
\hspace{1cm} \beta = \frac{2 \QGalphaem}{\pi}
\left( \ln \frac{Q^2}{m_{\QGindx{e}}^2} -1 \right).
\end{equation}
The form is divergent but integrable for $x \to 1$, i.e. the electron
likes to keep most of the energy. To handle the numerical precision
problems for $x$ very close to unity, the parton distribution is set, by
hand, to zero for $x > 0.999999$, and is rescaled upwards in the range
$0.9999 < x < 0.999999$, in such a way that the total area under the
distribution is preserved.
 
In the $\gamma$e or $\gamma\gamma$ processes, an equivalent flow of 
photons is assumed, based on first-order formulae. There is some 
ambiguity in the choice of $Q^2$ range over which emissions should 
be included. In the probably most appropriate alternative 
({\tt MSTP(13)=2}) the form is
\begin{equation}
f_{\gamma}^{\QGindx{e}}(x,Q^2) = 
\frac{\QGalphaem}{2 \pi} \, \frac{1+(1-x)^2}{x}
\, \ln \left( \frac{Q^2_{\QGindx{max}} (1-x)}%
{m_{\QGindx{e}}^2 \, x^2} \right).
\end{equation}
Here $Q^2_{\QGindx{max}}$ ({\tt PARP(13)}) is a user-defined cut for 
the range of scattered electron kinematics that is counted as
photoproduction. Note that we now deal with two different $Q^2$
scales, one related to the hard subprocess itself, which appears
as the argument of the parton distribution, and the other related to
the scattering of the electron, which is reflected in 
$Q^2_{\QGindx{max}}$. In the default alternative ({\tt MSTP(13)=1}) 
only one scale is assumed, i.e. $Q^2_{\QGindx{max}} (1-x) / x^2$ is 
replaced by $Q^2$ above.
 
Resolved photoproduction also involves the distributions of
quarks and gluons inside the photon inside the electron. By
default the SaS 1D set \cite{QGTSSaS} is used for the parton 
distributions of the photon, but several alternatives are available.

\subsubsection{Parton showers}
In every process that contains coloured and/or charged objects
in the initial or final state, gluon and/or photon radiation
may give large corrections to the overall topology of events.
The philosophy of {\sc Pythia} is to stay with the lowest-order 
cross sections (modulo trivial loop corrections such as the 
running of coupling constants) and then generate higher-order
corrections in the parton-shower approach. This is less exact than
the explicit calculation of higher-order matrix elements, but
has the advantage that it can be applied also to processes where
higher orders have not yet been calculated; additionally it includes 
multiple emissions.

Showers may be subdivided into initial- and final-state ones,
depending on whether they precede or follow the hard scattering.
Of course, the subdivision often contains an element of arbitrariness, 
since interference terms may exist. 
In both initial- and final-state showers, the structure is given in
terms of branchings $a \to bc$, specifically e $\to$ e$\gamma$, 
q $\to$ qg, q $\to$ q$\gamma$, g $\to$ gg, and g $\to$ q$\QGqbar$. 
The kernel $P_{a \to bc}(z)$ of a branching gives the probability
distribution of the energy sharing, with daughter $b$ taking a 
fraction $z$ and daughter $c$ the remaining $1-z$ of the $a$ energy.
Once formed, the daughters $b$ and $c$ may branch in their turn, and 
so on.
 
Each parton is characterized by some virtuality scale $Q^2$, which 
gives an approximate sense of time ordering to the cascade. 
In the initial-state shower, spacelike $Q^2$ values are
gradually increasing as the hard scattering is approached, while
timelike $Q^2$ values are decreasing in the final-state showers.
Shower evolution is cut off at some lower scale $Q_0$, typically 
around 1~GeV for QCD branchings and around $m_{\QGindx{e}}$ for
initial-state QED ones. From above, a maximum scale $Q_{\QGindx{max}}$ 
is introduced, where the showers are matched to the hard interaction 
itself. Unfortunately the selection of $Q_{\QGindx{max}}$ for a given
hard scattering is not unique, but gives rise to some slop.
 
Despite a number of common traits, the initial- and final-state
radiation machineries are in fact quite different. 
The {\sc Jetset} final-state algorithm has been used extensively for 
Z$^0$ hadronic decays at LEP~1, and is not significantly altered since 
the LEP~1 writeup \cite{QGLEPone}. 
 
Initial-state radiation is handled within the backwards evolution
scheme \cite{QGTSbackwards}. In this approach, the
choice of the hard scattering is based on the use of evolved
parton distributions, which means that
the inclusive effects of initial-state radiation are already
included. What remains is therefore to construct the exclusive
showers. This is done starting from the two incoming partons
at the hard interaction, tracing the showers ``backwards in time'',
back to the two shower initiators. In other words,
given a parton $b$, one tries to find the parton $a$ that branched
into $b$. The evolution in the Monte Carlo is therefore in
terms of a sequence of decreasing space-like virtualities $Q^2$
and increasing momentum fractions $x$. Branchings on
the two sides are interleaved in a common sequence of
decreasing $Q^2$ values.
The definition of the $x$ and $z$ variables for off-mass-shell partons 
is not unique; in {\sc Pythia} the $z = x_b/x_a$ of a branching
tells how much the scattering subsystem invariant mass-squared is 
reduced by the branching. If originally parton $b$ was assumed to 
have vanishing $p_{\perp}$, the reconstruction of the branching 
$a \to bc$ introduces a $p_{\perp}$ for $b$, which is compensated 
by $c$. 

\subsubsection{Beam remnants and multiple interactions}
The initial-state radiation algorithm reconstructs one shower initiator
in each beam. Together the two initiators delineate an interaction
subsystem, which contains all the partons that participate in
the initial-state showers, the hard interaction, and the final-state 
showers. Left behind are two beam remnants. In some cases a 
remnant is a single object, as when a $\gamma$ is taken out of an e 
beam, leaving behind an e. When taking an e out of an e, a soft 
$\gamma$ is left behind, which is then more related to the cutoff of 
$f_{\QGindx{e}}^{\QGindx{e}}(x,Q^2)$ at $x=0.999999$ than
to the ordinary beam-remnant concept, but is handled with the same
machinery. In other cases a remnant consists of two objects, as when a 
q is taken out of an e, leaving behind $\mbox{e} + \QGqbar$. The latter
example has a coloured remnant, meaning that the fragmentation of the
hard-process partons is connected with that of the beam remnants.

A resolved photon contains many partons. In a twice-resolved 
$\gamma\gamma$ event there is thus the possibility of multiple
interactions, i.e. of multiple semi-hard 
parton--parton processes in the same event. A model for this
phenomenon is included in {\sc Pythia} \cite{QGTSgamgam}, and will be 
further developed to better represent differences between the VMD and 
anomalous states.

\subsubsection{Fragmentation and decay}
The Lund string fragmentation description \cite{QGlund} and the 
decay routines in {\sc Jetset} have not changed 
significantly since the LEP~1 writeup \cite{QGLEPone}, and so
are not described here. The string fragmentation approach has been 
generally successful in comparisons with LEP data, although some 
shortcomings have shown up. See section~\ref{QGsectexper} for 
further details. 

The issue of Bose-Einstein effects has received increased attention
in recent years, e.g. in connection with possible consequences
for the W mass determinations \cite{QGTSboei}. The existing 
algorithm \cite{QGLEPone} works well in many respects, but is by no 
means to be considered as a definite solution to the problem.
A somewhat different approach has been implemented to allow some
cross-checks, and further alternatives may appear in the future.    
In the current standard algorithm, identical particles are
pulled closer together in such a way as to enhance the
two-particle correlation at small relative momentum separation.
This makes jets slightly narrower, so that fragmentation parameters
have to be retuned for reasonable agreement with data.
In the alternative, the shift of identical particles is somewhat 
reduced, while non-identical particles are pushed apart a bit,
so that the average properties of jets remain unchanged.
This alternative does not yet come with {\sc Jetset}, but is 
available as a plug-in replacement for the {\tt LUBOEI} routine,
at http://thep.lu.se/tf2/staff/torbjorn/test/main10.f.

Also colour rearrangement has been extensively discussed in recent
years. Code that allows this has not yet been integrated in
the standard {\sc Pythia/Jetset} libraries, but is 
obtainable separately, see section~\ref{QGsubsectreconn}.  

\subsubsection{Final comments}
{\sc Pythia/Jetset} are likely to be among the major event generators
at LEP~2: access to a  broad selection of hard scattering subprocesses 
is combined with a well-tested description of parton showers and
fragmentation. Limitations exist, however. {\sc Pythia} is not a program 
for precision extraction of electroweak parameters; for instance,
no (non-trivial) loop corrections are included in the matrix elements.
One may well imagine hybrid arrangements, where dedicated generators
are used to provide an improved description of some especially 
interesting hard scattering processes, such as four-fermion final 
states, while the rest of the {\sc Pythia/Jetset} machinery is used to  
turn a simple parton configuration into a complex hadronic final state.
An example of such an interface is discussed in section 
\ref{QGsubsectEWint}. Furthermore, the {\sc Ariadne} program for colour 
dipole radiation offers an alternative to the parton-shower description 
of {\sc Jetset}, and can be used for all the hard processes in 
{\sc Pythia}.

While there are no major additions planned for {\sc Pythia/Jetset},
the intention is to continue a steady development and support 
activity. The $\gamma\gamma$ sector maybe is the area where most
further studies are required to complete the picture, but also other
aspects deserve attention.   

\subsection{UCLA ansatz}

\noindent{\bf Basic Facts}
\vspace{-\baselineskip} 
\begin{tabbing}
{\bf Program location:} \= \kill
{\bf Program names:}    \> {\sc UCLA} \cite{QGucla,QGCBuclathree}  \\
{\bf Versions:}         \> {\sc UCLA} 7.41 of 1 October 1995       \\
{\bf Author:}           \> Sebong Chun and C.D.~Buchanan           \\
                        \> Department of Physics                   \\
                        \> UCLA                                    \\   
                        \> 405 Hilgard Ave, Los Angeles, CA 90024  \\
                        \> USA                                     \\
                        \> Phone: (310) 815-1992, 7466             \\
                        \> E-mail: \= chun@physics.ucla.edu,       \\
                        \>         \> buchanan@physics.ucla.edu    \\
{\bf Program size:}     \> 1922 lines                              \\
{\bf Program location:} \> http://www.physics.ucla.edu/$\sim$chuns \\
\end{tabbing}
 
The goal of the UCLA hadronization modeling is to study and develop the 
underlying principles of $\QGee$ annihilation into hadrons, constructing 
a simple phenomenology which can be used both as a ``target'' for 
non-perturbative QCD calculations and also to accurately predict 
data.

The UCLA7.41 program, a spin-off of the Lund relativistic string 
Monte Carlo program {\sc Jetset}, is the manifestation of this 
modelling to be used in comparing predictions with $\QGee$ data.  
As {\sc Jetset} has upgraded to new versions, the UCLA program has 
likewise been adapted with a parallel nomenclature.

The modern UCLA modeling \cite{QGucla} presumes that, 
by making a few assumptions which can be rationalized within a QCD 
context (for example, a strong coupling expansion in lattice QCD), 
one can construct a Weight Function for any specified 
$\QGee \to$~hadrons event.  That is, given the 
center-of-mass energy of the $\QGee$ system and the flavor and 
momenta of the primary hadrons produced, the UCLA modeling attaches 
a weight to the entire event, to be used in comparison with other 
possible events at that {\QGecm}.

The general structure of the Weight Function (in addition to 
kinematics of energy/momen\-tum conservation and phase space with 
limited transverse momentum) depends on (a) an area law in 
space--time, (b) possible suppression factors at 
the vertices where a virtual q$\QGqbar$ pair is created from the 
colorfield, (c) ``knitting factors'' to knit a quark and antiquark 
together into the spatial wave function of a meson (or quark and 
diquark into a baryon), and (d) Clebsch--Gordon coefficients to 
knit the quark and antiquark (diquark) together into the flavor 
and spin state of the meson (baryon).  

a)  The area law is $\exp(-b'A)$ where $b'$ is a constant and $A$ is 
the area enclosed by the quark and antiquark trajectories in a 
space-time plot of the event. Almost any strong-coupling 
interaction will, in fact, give this sort of dependence.

b)  The UCLA modeling assumes that there is no significant vertex 
suppression for q$\QGqbar$ pairs if the quark mass is less than the 
hadronic scale of $\simeq 1$ GeV; that is, u$\overline{\mbox{u}}$, 
d$\overline{\mbox{d}}$ {\em and} s$\overline{\mbox{s}}$ all have 
probability $\simeq 1.0$ of virtual creation from the colorfield.

c)  The UCLA modeling assumes that all knitting factors are 
comparable, whether the hadron to be constructed is a spin 0 
or 1 meson or a spin $1/2$ or $3/2$ baryon. (Probability 
normalization on the fragmentation function derived
below yields a value of the knitting factor of 
$\simeq (40~\mbox{Mev})^{-2}$.)   

d)  The Clebsch-Gordon coefficients are simply the relevant 
flavor/spin coupling of a quark and antiquark (diquark) into a 
meson (baryon).  Note that (c) and (d) taken together describe the 
coupling of a quark and antiquark 
or diquark into the complete state function of a hadron.

Although, in principle, knowing the Weight Function for the final 
state is enough to select an event, it is practically impossible 
to implement in this form.  In order to implement this simple event 
Weight Function approach into a working Monte Carlo program, it is 
necessary to derive a fragmentation function for an ``outside-in 
iterative one-particle-at-a-time'' implementation such as
{\sc Jetset} uses.  

By somewhat lengthy but straightforward algebra, this can be 
accomplished.  The result so derived turns out to be the Lund 
Symmetric Fragmentation Function (LSFF) \cite{QGCBuclathree}, with 
normalizing parameters of the vertex suppression, the 
spatial knitting factor, and the Clebsch-Gordon coefficient 
(see \cite{QGucla}). That is, the UCLA modeling simply 
amounts to using the LSFF as a hadronic production density weighted 
by Clebsch-Gordon coefficients, where the suppression of heavy mass 
particle production arises entirely from the $\exp(-bm^2/z)$ factor 
in the LSFF. (Note: the general structure of the Weight Function 
and the subsequent derivation of the fragmentation function
can also be used to describe the Lund {\sc Jetset} treatment.  
The difference is that {\sc Jetset} presumes an 
s$\overline{\mbox{s}}$ vertex suppression of about 0.3 and a 
knitting factor for vector mesons of about 30\% of that for 
pseudoscalar mesons, does not in general use Clebsch-Gordon 
coefficients, and adopts a normalization scheme that does not 
incorporate the $\exp(-bm^2/z)$ factor.)  

The UCLA7.41 program uses the parton shower and decay table parts 
of {\sc Jetset}, but replaces the flavor and momentum selection part 
with the UCLA modeling ansatz described above.  Default values for 
the parton shower are $\Lambda = 0.2$~GeV and $Q_0 = 1.0$~GeV.  
Meson production is controlled by the two natural parameters of 
the LSFF with default values of  $a = 2.1$  and  
$b = 1.1$~GeV$^{-2}$. Local transverse momentum compensation is 
approximated by a factor of $\exp(-{n\over {n-1}}b p_{\perp}^2/z)$, 
where $n$ is a parameter of default value 2.0.  
For baryon production, with ``popcorn'' mesons produced between 
baryon and antibaryon, an additional popcorn suppression factor 
of $\exp(-\eta m_{\QGindx{pop}})$ is introduced with the default 
value of $\eta = 10$~GeV$^{-1}$.  For more details, 
please refer to refs. \cite{QGucla}.

This structure and values gives a rather good description of 
multiplicities, inclusive distributions, and correlations for 
hadron production from {\QGecm} of 10 to 91 GeV, with the possible 
exception of the spin $3/2$ baryons at 91 GeV. The description of 
heavy flavor (c, b) production distributions also seems
reasonably good, with no additional parameterization or 
parameters.

For a detailed instruction on how to set up parameters and use 
the program, please refer to the manual at WWW location
http://www.physics.ucla.edu/$\sim$chuns.
A short set of instructions is available in the header 
to the actual program.

\subsection{Colour reconnection codes}
\label{QGsubsectreconn}

One of the QCD questions that has attracted attention in recent years 
is that of colour reconnection (or colour rearrangement)
\cite{QGTSrecLL,QGTSrecGPZ,QGTSrecSK,QGTSrecGH}. This issue
has implications for W mass studies, but is also of interest for
our general understanding of QCD.

The concept may be illustrated by the process
$\mbox{e}^+ \mbox{e}^- \to \mbox{W}^+ \mbox{W}^- \to
\mbox{q}_1 \overline{\mbox{q}}_2 \, \mbox{q}_3 \overline{\mbox{q}}_4$.
To first approximation, the hadronic final state can be viewed
as coming from the incoherent superposition of two sources of
particle production: the $\mbox{q}_1 \overline{\mbox{q}}_2$ and 
the $\mbox{q}_3 \overline{\mbox{q}}_4$ ones. If colours are
reconnected, the sources would instead be 
$\mbox{q}_1 \overline{\mbox{q}}_4$ and 
$\mbox{q}_3 \overline{\mbox{q}}_2$. 
The picture is complicated by the possibility of gluon emission. 
Gluons with an energy above the W width can be viewed as independently 
emitted from the respective W source, to a good first approximation:
propagator effects ensure that interference terms are suppressed. 
No similar suppression exist for soft 
gluons or in the nonperturbative r\'egime. Therefore standard
calculational techniques are of limited interest, and the phenomenon 
mainly has to be studied within the context of specific models.
By now, several independent codes exist, some part of
existing QCD generators, others available as add-ons. Below we 
list the known ones and give some specific details. 

\subsubsection{A PYTHIA-based implementation} 
\vspace{-5mm}
(code by T.~Sj\"ostrand)

The code used for the studies in \cite{QGTSrecSK} has not (yet)
been incorporated in the {\sc Pythia/Jetset} programs.
A sample main program and the colour rearrangement 
subroutines can be obtained at web address
http://thep.lu.se/tf2/staff/torbjorn/test/main01.f.
Several different options are available, among others:
\begin{QGitemize}
\item scenario I, where strings are considered as extended colour 
flux tubes and the reconnection probability is proportional (up to
saturation corrections) to the space--time overlap of the W$^+$ 
and W$^-$ strings;
\item scenario II, where strings are considered as thin vortex lines
and reconnection may occur when strings cross;
\item scenario II$'$, a variant of scenario II where only those 
reconnections are allowed that reduce the total string length; and
\item the instantaneous scenario, where reconnections are allowed 
before the parton-shower evolution \cite{QGTSrecGPZ}; unphysical but
handy for comparisons. 
\end{QGitemize} 
At most one reconnection is performed per event, in scenario II the
one that occurs first in time, in scenario I selected according to 
relative probabilities given by the overlaps. Reconnections within a 
W system are not considered.

\subsubsection{Another PYTHIA-based implementation}
\vspace{-5mm}
(code by \v{S}.~Todorova)

The code follows the physical approach of \cite{QGTSrecSK}. 
The reconnection phenomenon is simulated with the help of the 
string model, where strings are considered to be either flux tubes 
or vortex lines with arbitrary diameter of the core.
In order to get a more realistic estimation of the effect of colour 
reconnection, the following features (not found in the preceding 
code) were incorporated in the simulation:
\begin{QGitemize}
\item space-time evolution of parton shower; 
\item multiple reconnections; and
\item self-interaction of a string (production of glueballs).
\end{QGitemize}
It should be noticed that the space-time evolution of a shower together 
with the self-interaction of a string allow the study of string 
reconnection effects in a single parton shower (decay of a 
single Z$^0$).
  
The search for candidates for reconnection is processed in parallel with
the shower development (reconnection can take place before the emission 
of the last partons).
Overlaps of the colour fields of flux tubes are calculated numerically
(using multichannel MC integration with importance sampling). The method
is slow but this is the price to be paid for (relative) accuracy and 
individual treatment of each event.
The minimal distance between ``vortex lines'' is found by a 
minimization procedure based on parabolic fit.
                                               
The code is available in the directory 
crnvax:[nova.colour\_ reconnection].

\subsubsection{An ARIADNE-based implementation} 
\vspace{-5mm}
(code by J.~H\"akkinen)

The aim of the simulation program presented in~\cite{QGTSrecGH} is not 
so much to study the effect of recoupling on the average events, but to 
study if rare recoupled events can be identified. Perturbative QCD 
favours states which correspond to ``short strings'', i.e. parton states 
which produce few hadrons. This string ``length'' can here be specified 
by the $\lambda$ measure, defined in~\cite{QGJHba88}, which correspond 
to an effective rapidity range. If recoupling occurs it is conceivable 
that it is favoured when the recoupling produces a state with lower 
$\lambda$ measure, and such states may also be more easy to identify.
For this reason the program produces recoupling such that the $\lambda$ 
measure for the reconnected final state is minimized. Gluons with 
$E~\raisebox{0.5ex}{$>$}~\hspace{-1.5em}~\raisebox{-0.5ex}{$\sim$}~\!
\Gamma_{\QGindx{W}}$ are emitted independently within the original 
q$\QGqbar$ systems~\cite{QGJHyd93,QGTSrecSK}. This emission is simulated 
using the Dipole Cascade Model~\cite{QGLLdcm} implemented in 
{\sc Ariadne}~\cite{QGariadne}. For gluons with c.m. energy below 
$\Gamma_{\QGindx{W}} \approx 2$~GeV there may be unknown interference 
effects due to emission from the two W systems. These low-energy gluons 
give very  little effect on the hadronic final state, however, if the 
hadronization phase is described by the Lund string model~\cite{QGlund} 
implemented in {\sc Jetset}~\cite{QGjetset}. They are therefore 
disregarded in the parton states, which implies that a small fraction of 
the energy ($\approx \! 4$\%) will be lost in the simulations. During 
minimization of $\lambda$ all possible final state configurations, 
obtained by cutting the original gluon chains in one place only before 
reconnecting to two new systems, are compared with each other. In this 
program reconnection between the two strings occur once in every event 
while reconnections within the strings are not considered. Thus, the 
program is not expected to reproduce average events, but possibly a small 
admixture of recoupled events. 

The C code used in~\cite{QGTSrecGH} is available at 
http://thep.lu.se/tf2/hep/hep.html or through anonymous ftp at 
thep.lu.se:/pub/LundPrograms/Misc/wwpair.tar.Z. The code contains two 
more models; the instantaneous scenario of~\cite{QGTSrecGPZ}, and 
random reconnection of the strings. These models are only used for 
comparison with the ``main'' model. 

\subsubsection{Another ARIADNE-based implementation} 
\vspace{-5mm}
(code by L.~L\"onnblad)

The model in \cite{QGTSrecLL} for colour reconnections,
implemented in the {\sc Ariadne} program \cite{QGariadne}, is similar to
the one in \cite{QGTSrecGH} in that it reconnects colour dipoles
within the framework of the Dipole Cascade Model (DCM)
\cite{QGLLdcm} with a probability $1/N_C^2$ only if the
total string length becomes reduced. The main differences are that
reconnections within each W system (and also among the partons from
a Z decay) is allowed, that several such reconnections are allowed
in each event, and that reconnections are allowed during the
perturbative cascade.

To achieve this, colour indices are assigned to each dipole, and after
each emission, dipoles with identical indices are allowed to
reconnect. The indices are chosen randomly, but restrictions are made
to ensure physical colour flows, e.g.\ two gluons created by a gluon
splitting should not be allowed to form a colour singlet. In the DCM,
however, a gluon is radiated coherently by the dipole between two
partons, and a procedure has to be introduced, where the emitted gluon
is said to have been radiated off one of the two emitting partons with
some probability depending on which is closer in phase space.

In the case of $\mbox{e}^+ \mbox{e}^- \to \mbox{W}^+ \mbox{W}^-$
reconnections are initially only allowed within each W system
separately. After all gluons with $E_{\QGindx{g}} > \Gamma_{\QGindx{W}}$ 
have been emitted, reconnections between the W systems is switched on 
and gluon emission with $E_{\QGindx{g}} < \Gamma_{\QGindx{W}}$ is 
performed in the possibly reconnected systems before hadronization.

\subsubsection{A HERWIG-based implementation} 
\vspace{-5mm}
(code by B.R.~Webber)

A model for colour reconnection has been implemented in a package of
subroutines that can be used with {\small HERWIG} (version 5.8).  The new 
integer parameter {\tt IRECO=0,1,2} determines the reconnection option 
used. {\tt IRECO=0} means no reconnection and {\tt IRECO=2} gives
``immediate'' reconnection of the quark--antiquark pairs in
hadronic WW events, before parton shower generation, with
probability {\tt PRECO} (default value = 1/9).  In {\small HERWIG}
this changes the evolution of the showers, as well as the colour
connections, because the initial opening angles are different.

The most serious option is {\tt IRECO=1}, which invokes a model
based on the assumption that reconnection occurs locally in space--time.
First, a space--time structure is computed for each parton shower in the
event.  This is done using a package written by Mike Seymour to store
the internal lines of showers, which is turned on by setting
{\tt INTLIN=.TRUE.}.  The algorithm is semi-classical, but qualitative
features and orders of magnitude should be correct.  In the case of W
hadronic decays, each W decay point is generated with the appropriate
exponential decay distribution.  Then the locations of all the vertices
in the showers are computed by assigning a space-time separation
$\Delta x_i=q_i/(q_i^2-m_i^2)$ to vertices joined by an internal
line $i$ of 4-momentum $q_i$ and on-shell mass $m_i$.

The {\small HERWIG} cluster hadronization model, normally called immediately 
after
showering has terminated, involves splitting each final-state gluon into
a quark--antiquark pair.  For each quark $i$ there is a colour partner
antiquark $j$, with which the quark would normally be paired to form a
colour singlet cluster $(ij)$.  The {\tt IRECO=1} option introduces a
reconnection phase before cluster formation.  In this phase the program
looks for another colour-connected quark-antiquark pair $(kl)$ such that
$(il)$ and $(kj)$ could be colour singlets and
\begin{equation}
|\Delta x_{il}|^2 + |\Delta x_{kj}|^2 <
|\Delta x_{ij}|^2 + |\Delta x_{kl}|^2 ~,
\end{equation}
where $\Delta x_{ij}$ is the $(ij)$ cluster size, defined as the separation
of the {\em production} vertices of $i$ and $j$ (note that this can
be zero, e.g.\ if $i$ and $j$ come from a W decay that did not radiate
any gluons). If such a pair exists, the reconnection $(ij)(kl)\to (il)(kj)$
would reduce the cluster sizes, and so it is performed with probability
{\tt PRECO}.  Note that reconnection can happen inside a single shower
and not just between different showers.  Thus some retuning of parameters
to fit data on $\QGee \to \mbox{Z}^0\to$ hadrons will be necessary when
using {\tt IRECO=1}.

The code can be obtained by anonymous ftp from \\[1mm]
\hspace*{1cm}{\tt hep.phy.cam.ac.uk} $\equiv$ 131.111.66.27 \\[1mm]
The following files should be copied from directory 
{\tt disk\$alpha1:[public.herwig]}:
\begin{QGitemize}
\item {\tt hwwmas58.for} -- sample main program and analysis routines;
\item {\tt hwreco58.for} -- modified {\small HERWIG} routines {\tt HWBFIN},
{\tt HWBJCO}, {\tt HWCFOR} which {\em replace} those in {\small HERWIG}
version 5.8, plus {\em new} routines {\tt HWGCLU}, {\tt HWGCMO},
{\tt HWUPIP}, {\tt HWVHEP}.
\end{QGitemize}
There is a new common block containing relevant parameters and 
counters:\\[1mm]
\hspace*{1cm}{\tt COMMON/HWRECO/PRECO,EXAG,IRECO,MEVTS,MCLUS,MRECO,%
MSWCH,INTLIN}\\[1mm]
{\tt PRECO} is the reconnection probability (default 1/9);
{\tt EXAG} is an `exaggeration factor' for the W lifetime,
to study effects of the WW separation (default 1.0);
{\tt IRECO} is the reconnection option (see above, default 1);
{\tt MEVTS} {\it etc.} are integer counters for number of events, clusters,
reconnections, and WW reconnections; {\tt INTLIN} is set to {\tt .TRUE.}
when {\tt IRECO=1} (see above).

The code is still under development; please notify
{\tt webber@hep.phy.cam.ac.uk} of any problems, bugs
and/or peculiarities.

\newpage
\subsection{Monte Carlo Implementations of Exact Next-to-Leading
Order Calculations}

\noindent{\bf Basic Facts}
\vspace{-\baselineskip} 
\begin{tabbing}
{\bf Program name:} \= nason@surya11.cern.ch \= E.W.N.Glover@dur.ac.uk
\= \kill
{\bf Program name:} \> {\small EVENT} \> {\small EERAD} \> {\small EVENT2} \\
{\bf Authors:}      \> Zoltan Kunszt  \> Walter Giele \> Stefano Catani \\
                    \> Paolo Nason    \> Nigel Glover \> Mike Seymour \\
{\bf email:}        \> nason@surya11.cern.ch \> E.W.N.Glover@dur.ac.uk \>
                                                  seymour@surya11.cern.ch \\
\end{tabbing}

There are now three publicly-available programs for calculating
next-to-leading order corrections to arbitrary infrared-safe two- and
three-jet quantities in $\QGee$ annihilation.  Although these
use Monte Carlo integration techniques, they should be contrasted with
Monte Carlo Event Generators in several ways.  Firstly, they calculate
the {\em exact} result in perturbation theory for the 
${\cal O}(\QGalphas)$ corrections to a given quantity~--- no more nor 
less. Secondly, the phase-space configurations generated do not have
positive-definite weights, so a probabilistic interpretation is not
possible.  Finally, for both these reasons, the programs only ever
consider the partonic final state, and no treatment of hadronization is
attempted.

There are many advantages of implementing higher-order QCD calculations
as matrix-element Monte Carlo programs.  For all but the simplest
observables, the required phase-space integrals are not analytically
tractable, and some form of numerical integration becomes mandatory.
Since each phase-space point sampled by the program has a direct
correspondence to a set of final-state momenta, any infrared-safe jet or
event-shape definition may be used, and can be implemented exactly as in
an experimental analysis.  Many event properties can be analyzed
simultaneously, simply by adding code to the analysis routine of the
program to histogram the quantity of interest.

However as is well-known, the real and virtual corrections are
separately divergent but with finite sum, so na\"\i ve numerical
integration of each matrix element would fail.  Thus a regularization
scheme must be used to render the integrals finite.  It is principally
in the definitions of regularization scheme that the three programs
differ, although there are other important differences.

The difference between the regularization schemes can be illustrated
using a simple one-dimensional example.  In dimensional regularization
using $(4-2\epsilon)$ dimensions, the integrals we encounter are
typically of the form
\begin{equation}
  \langle O \rangle = \int_0^1 \frac{\mbox{d}x}{x^{1+\epsilon}} O(x)
  +\frac1\epsilon O(0),
\end{equation}
where the first part represents the real cross-section, the second the
virtual, and $x$ would typically be a gluon energy or parton-parton
invariant mass.  The function $O(x)$ represents a final-state observable
that is infrared safe, i.e.~with the requirement that $O(x)$ tends
smoothly to $O(0)$ as $x$ tends to 0.  If the integral were analytically
tractable, it would yield an $\epsilon$-pole that canceled the virtual
term, leaving a finite result.  However, in general it is not, and we
must manipulate it into a form in which the physical limit
$\epsilon \to 0$ can be taken {\em before} numerical integration, without
making any assumption about $O(x)$.

The phase-space {\em slicing method} does this by introducing an
unphysical parameter $x_0$,
\begin{equation}
  \langle O \rangle = \int_0^{x_0} \frac{\mbox{d}x}{x^{1+\epsilon}} O(x)
  +\int_{x_0}^1 \frac{\mbox{d}x}{x^{1+\epsilon}} O(x)
  +\frac1\epsilon O(0)
  \approx
  \int_{x_0}^1 \frac{\mbox{d}x}{x} O(x)
  +\log(x_0) O(0)
  +{\cal O}\left(x_0O'(0)\right).
\end{equation}
The result becomes exact in the limit $x_0\to0$, which practically means
$x_0\ll x_{\QGindx{physical}}$, where $x_{\QGindx{physical}}$ is the
smallest physical scale in the problem.

The {\em subtraction method} works by subtracting and adding a term
derived by projecting each point in four-parton phase-space onto some
point in three-parton phase-space, and calculating the observable at
this phase-space point together with an approximate matrix element.
This must be such that it matches all the divergent terms of the full 
matrix element.  In our simple example this corresponds to
\begin{equation}
  \langle O \rangle = \int_0^1 \frac{\mbox{d}x}{x^{1+\epsilon}} O(x)
  -\int_0^1 \frac{\mbox{d}x}{x^{1+\epsilon}} O(0)
  +\int_0^1 \frac{\mbox{d}x}{x^{1+\epsilon}} O(0)
  +\frac1\epsilon O(0)
  =
  \int_0^1 \frac{\mbox{d}x}{x} \left(O(x)-O(0)\right).
\end{equation}
Note that this is exact and does not depend on any unphysical
parameters.

The matrix elements for $\gamma^* \to \QGq\QGqbar\QGg$ have been known
to next-to-leading order for many years\cite{QGMHSert}.  These were
later checked by other groups, and used for specific calculations of a
variety of event shapes.  For the `QCD at LEP' report \cite{QGMHSlep1},
Kunszt and Nason wrote a general-purpose Monte Carlo program using the
subtraction method that could calculate the next-to-leading correction
to any event shape or jet definition, {\small EVENT}\cite{QGMHSkn}.
This has been considered the standard calculation for many years, but
has two significant shortcomings owing to the matrix elements used: they
have been summed over permutations of the outgoing partons, which means
that quarks and gluons cannot be distinguished in the final state; and
they consider the decay of a virtual photon, so can only predict
quantities averaged over orientations of hadronic events, losing all
information on their lab-frame directions and lepton-hadron
correlations.  Furthermore they neglect specific axial-axial
contributions that as a point of principle are essential for describing
Z$^0$ decays, although in practice these are never numerically
significant.

More recently two groups have proposed general algorithms for
calculating next-to-leading order corrections in arbitrary processes,
and both have used three-jet production in $\QGee$
annihilation as a simple first proving ground for their methods.  These
have resulted in the {\small EERAD} program by Giele and
Glover\cite{QGMHSgg}, which uses the slicing method, and the
imaginatively-titled {\small EVENT2} program by Catani and
Seymour\cite{QGMHScs}, which uses the subtraction method.  Both of these
use the full next-to-leading order matrix elements for
$\QGee \to \QGq\QGqbar\QGg$, avoiding the shortcomings of 
{\small EVENT}.  Although {\small EVENT2} uses the matrix elements of the
Leiden group\cite{QGMHSlei} by default, it has options to use the same
matrix elements as {\small EVENT} or {\small EERAD} as a cross-check.
Numerical results of the three algorithms are discussed in
\cite{QGMHSlep2} and shown to be in good agreement.  Since they are
supposed to be exact calculations of the same quantity, rather than
models, any differences between them should be treated as bugs.

\newpage

\section{Standardization}
\label{QGsectstandard}

\subsection{Particle codes and {\tt /HEPEVT/} update}
The {\tt /HEPEVT/} standard \cite{QGIKhepevt} has been widely adopted by Monte
Carlo authors for storing information on generated events. In practice the
real variables are commonly declared to be {\tt DOUBLE PRECISION} and 
often the size is expanded to {\tt NMXHEP=4000}. We propose that these are 
now added to the standard. 

In {\tt /HEPEVT/} it was intended for particles to be identified using the PDG
numbering scheme \cite{QGIKcodes}. However the conventional numbers assigned
have deficiencies, particularly concerning the neglect of particles expected
according to the quark model but not yet identified in experiment, for example
the $\mbox{h}_{\QGindx{b}}$. This proves troublesome for those program authors 
who include such
states and has lead to {\it ad hoc} solutions. Further the higher, orbitally
excited $L=2,3,\ldots$ and radially excited $n=2,3,\ldots$ mesons are labelled
in a somewhat unsystematic way. In order to preserve the concept of 
uniqueness,
allow for the missing quark model states, systematize the numbering and remain
true to the spirit of the PDG scheme we suggest the following revised 
numbering.

\begin{table}[tb]
\begin{center}
\begin{tabular}{|c|c|c|c|c|c|c|c|r|}
\cline{3-8}
\multicolumn{2}{c|}{} & \multicolumn{2}{c|}{$L=0$} &
\multicolumn{4}{c|}{$L=1$} & \multicolumn{1}{c}{} \\
\cline{3-8}
\multicolumn{2}{c|}{} & $S=0$ & $S=1$ & $S=0$ & \multicolumn{3}{c|}{$S=1$} &
\multicolumn{1}{c}{} \\
\cline{3-8}
\multicolumn{2}{c|}{} & $J=0$ & $J=1$ & $J=1$ & $J=0$ & $J=1$ & $J=2$ &
\multicolumn{1}{c}{} \\
\cline{1-8}
$\QGq_1$ & $\QGqbar_2$  & $\star\star$1 & $\star\star$3 & 10$\star\star$3 &
10$\star\star$1 & 20$\star\star$3 & $\star\star$5 & \multicolumn{1}{c}{} \\
\hline
d & $\QGbar{d}$ & $\pi^0$ & $\rho^0$ & $\mbox{b}^0_1$ & $\mbox{a}^0_0$ & 
$\mbox{a}^0_1$ & $\mbox{a}^0_2$ &
11 \\
& $\QGbar{u}$ & $\pi^-$ & $\rho^-$ & $\mbox{b}^-_1$ & $\mbox{a}^-_0$ & 
$\mbox{a}^-_1$ & $\mbox{a}^-_2$ &
$-21$ \\
& $\QGbar{s}$ & $\mbox{K}^0$ & $\mbox{K}^{\star 0}$ & $\mbox{K}^0_1(1270)$ & 
$\mbox{K}^{\star 0}_0$ &
$\mbox{K}^0_1(1400)$ & $\mbox{K}^{\star 0}_2$ & 31 \\
& $\QGbar{c}$ & $\mbox{D}^-$ & $\mbox{D}^{\star -}$ & $\mbox{D}^-_1(2420)$ & 
$\mbox{D}^{\star -}_0$ &
$\mbox{D}^-_1(H)$ & $\mbox{D}^{\star -}_2$ & $-41$ \\
& $\QGbar{b}$ & $\mbox{B}^0$ & $\mbox{B}^{\star 0}$ & $\mbox{B}^0_1(L)$ & 
$\mbox{B}^{\star 0}_0$ &
$\mbox{B}^0_1(H)$ & $\mbox{B}^{\star 0}_2$ & 51 \\
\hline
u & $\QGbar{u}$ & $\eta$ & $\omega$ & $\mbox{h}_1(1170)$ & 
$\mbox{f}_0(980)$ & $\mbox{f}_1(1285)$
& $\mbox{f}_2(1270)$ & 22 \\
& $\QGbar{s}$ & $\mbox{K}^+$ & $\mbox{K}^{\star +}$ & $\mbox{K}^+_1(1270)$ & 
$\mbox{K}^{\star +}_0$ &
$\mbox{K}^+_1(1400)$ & $\mbox{K}^{\star +}_2$ & 32 \\
& $\QGbar{c}$ & $\QGbar{D}^0$ & $\QGbar{D}^{\star 0}$ & 
$\QGbar{D}^0_1(2420)$ &
$\QGbar{D}^{\star 0}_0$ & $\QGbar{D}^0_1(H)$ & 
$\QGbar{D}^{\star 0}_2$ & $-42$ \\
& $\QGbar{b}$ & $\mbox{B}^+$ & $\mbox{B}^{\star +}$ & $\mbox{B}^+_1(L)$ & 
$\mbox{B}^{\star +}_0$ &
$\mbox{B}^+_1(H)$ & $\mbox{B}^{\star +}_2$ & 52 \\
\hline
s & $\QGbar{s}$ & $\eta^\prime$ & $\phi$ & $\mbox{h}_1(1380)$ & 
$\mbox{f}_0(1300)$ &
$\mbox{f}_1(1510)$ & $\mbox{f}^\prime_2(1525)$ & 33 \\
& $\QGbar{c}$ & $\mbox{D}^-_{\QGindx{s}}$ & $\mbox{D}^{\star -}_{\QGindx{s}}$ 
& 
$\mbox{D}^-_{\QGindx{s}1}(2536)$ & $\mbox{D}^{\star -}_{\QGindx{s}0}$ &
$\mbox{D}^-_{\QGindx{s}1}(H)$ & $\mbox{D}^{\star -}_{\QGindx{s}2}$ & $-43$ \\
& $\QGbar{b}$ & $\mbox{B}^0_{\QGindx{s}}$ & $\mbox{B}^{\star 0}_{\QGindx{s}}$ 
& 
$\mbox{B}^0_{\QGindx{s}1}(L)$ & $\mbox{B}^{\star 0}_{\QGindx{s}0}$ &
$\mbox{B}^0_{\QGindx{s}1}(H)$ & $\mbox{B}^{\star 0}_{\QGindx{s}2}$ & 53 \\
\hline
c & $\QGbar{c}$ & $\eta_{\QGindx{c}}$ & J$/\psi$ & $\mbox{h}_{\QGindx{c}}$ & 
$\chi_{\QGindx{c}0}$ & $\chi_{\QGindx{c}1}$ &
$\chi_{\QGindx{c}2}$ & 44 \\
& $\QGbar{b}$ & $\mbox{B}^+_{\QGindx{c}}$ & $\mbox{B}^{\star +}_{\QGindx{c}}$ 
& 
$\mbox{B}^+_{\QGindx{c}1}(L)$ & $\mbox{B}^{\star +}_{\QGindx{c}0}$ &
$\mbox{B}^+_{\QGindx{c}1}(H)$ & $\mbox{B}^{\star +}_{\QGindx{c}2}$ & 54 \\
\hline
b & $\QGbar{b}$ & $\eta_{\QGindx{b}}$ & $\Upsilon(1S)$ & 
$\mbox{h}_{\QGindx{b}}$ & 
$\chi_{\QGindx{b}0}$ & $\chi_{\QGindx{b}1}$ & $\chi_{\QGindx{b}2}$ & 55 \\
\hline
\end{tabular}
\end{center}
\caption{Proposed numbering scheme for the lowest lying mesons: for example
the a$^-_1$ has the number $-20213$. The names of the pseudovector particles
are distinguished by their masses, if these are not currently established
$L$ and $H$ are used to indicate light and heavy. The pseudo-vector states
K$_1(1270)$ and K$_1(1400)$ are believed to be admixtures of the K$_{1B}$
1$^1P_1$ and K$_{1A}$ 1$^3P_1$; in such situations the lighter state is given
the lower number.}
\label{QGIKtmes}
\end{table}

Table~\ref{QGIKtmes} lists the $n=1,\;L=0,1$ mesons and indicates their
numbering, for these states this is largely in accord with the PDG scheme
and with the {\sc stdhep} ({\sc Jetset}) implementation \cite{QGIKstd}. In the
pairs of $I=0$, (u, d, s) mesons: 
$(\eta,\eta'),\;(\omega,\phi),\;(\mbox{h}_1(1170),
\mbox{h}_1(1380)),$ {\it etc.} the lighter state is labelled 22 
and the heavier 33,
reflecting the naive, dominant quark contents. Bound states involving
top quarks are not expected, due to the quark's high mass, and therefore are
not considered. The mixed $\mbox{K}^0_S$ and $\mbox{K}^0_L$ states are still 
labelled 310
and 130 respectively. The table should be extended to include $n=1,\;L=2,3,
\ldots$ states; this leads to up to four mesons of the same total spin. It
is proposed to reserve the fifth digit to differentiate these states by
continuing the sequence established for the $L=1$ mesons. That is, for a
given $J>0$ the numbers would be: $(L,S)=(J-1,1): \;\star\star m,\;(J,0):10
\star\star m,\;(J,1):\;20\star \star m$ and $(J+1,1):\;30\star\star m$,
where as usual $m=2J+1$. The $J=0$ states represent an exceptional case,
here we propose $(L,S)=(0,0):\;\star\star1$ and $(1,1):\;10\star\star1$, as
done in table~\ref{QGIKtmes}; this may be thought of as $L0\star\star1$.
Radially excited mesons, $n=2,3\ldots$, are effectively copies of the the
above states, it is proposed to introduce a sixth digit to differentiate them
as follows: $n=1:\;\star\star\star\star\star,\;n=2:\;1\star\star\star\star
\star,\;n=3:\;2\star\star\star\star\star$, {\it etc}. Thus for example the
K$^{*+}(1680)$, a 1$^3D_1$ state would be numbered 30323 and the 
$\rho^0(1450)$, a 2$^3S_1$ state 100113. The numbering of excited mesons
suggested here differs significantly from the original PDG scheme.

\begin{table}[htb]
\hfill
\begin{tabular}{|c|c|c|c|c|}
\cline{2-4}
\multicolumn{1}{c|}{}& \multicolumn{2}{c|}{$J=1/2$} & $J=3/2$ \\
\cline{1-4}
$\QGq_1\QGq_2\QGq_3$ & $\star n_3n_2$2 & $\star n_2n_3$2 & 
$\star\star\star$4 \\
\hline
ddd & & & $\Delta^-$ & 111 \\
udd & & n & $\Delta^0$ & 211 \\
uud & & p & $\Delta^+$ & 221 \\
uuu & & & $\Delta^{++}$ & 222 \\
\hline
sdd & & $\Sigma^-$ & $\Sigma^{\star -}$ & 311 \\
sud & $\Lambda$ & $\Sigma^0$ & $\Sigma^{\star 0}$ & 321 \\
suu & & $\Sigma^+$ & $\Sigma^{\star +}$ & 322 \\
\hline
ssd & & $\Xi^-$ & $\Xi^{\star -}$ & 331 \\
ssu & & $\Xi^0$ & $\Xi^{\star 0}$ & 332 \\
sss & & & $\Omega^-$ & 333 \\
\hline
cdd & & $\Sigma^0_{\QGindx{c}}$ & $\Sigma^{\star 0}_{\QGindx{c}}$ & 411 \\
cud & $\Lambda^+_{\QGindx{c}}$ & $\Sigma^+_{\QGindx{c}}$ & 
$\Sigma^{\star +}_{\QGindx{c}}$ & 421 \\
cuu & & $\Sigma^{++}_{\QGindx{c}}$ & $\Sigma^{\star ++}_{\QGindx{c}}$ & 422 \\
csd & $\Xi^0_{\QGindx{c}}$ & $\Xi^{\prime 0}_{\QGindx{c}}$ & 
$\Xi^{\star 0}_{\QGindx{c}}$ & 431 \\
csu & $\Xi^+_{\QGindx{c}}$ & $\Xi^{\prime +}_{\QGindx{c}}$ & $
\Xi^{\star +}_{\QGindx{c}}$ & 432 \\
css & & $\Omega^0_{\QGindx{c}}$ & $\Omega^{\star 0}_{\QGindx{c}}$ & 433 \\
\hline
\end{tabular}
\hfill
\begin{tabular}{|c|c|c|c|c|}
\cline{2-4}
\multicolumn{1}{c|}{}& \multicolumn{2}{c|}{$J=1/2$} & $J=3/2$ \\
\cline{1-4}
$\QGq_1\QGq_2\QGq_3$ & $\star n_3n_2$2 & $\star n_2n_3$2 & 
$\star\star\star$4 \\
\hline
ccd & & $\Xi^+_{\QGindx{cc}}$ & $\Xi^{\star +}_{\QGindx{cc}}$ & 441 \\
ccu & & $\Xi^{++}_{\QGindx{cc}}$ & $\Xi^{\star ++}_{\QGindx{cc}}$ & 442 \\
ccs & & $\Omega^+_{\QGindx{cc}}$ & $\Omega^{\star +}_{\QGindx{cc}}$ & 443 \\
ccc & & & $\Omega^{\star ++}_{\QGindx{ccc}}$ & 444 \\
\hline
bdd & & $\Sigma^-_{\QGindx{b}}$ & $\Sigma^{\star -}_{\QGindx{b}}$ & 511 \\
bud & $\Lambda^0_{\QGindx{b}}$ & $\Sigma^0_{\QGindx{b}}$ & 
$\Sigma^{\star 0}_{\QGindx{b}}$ & 521 \\
buu & & $\Sigma^+_{\QGindx{b}}$ & $\Sigma^{\star +}_{\QGindx{b}}$ & 522 \\
bsd & $\Xi^-_{\QGindx{b}}$ & $\Xi^{\prime -}_{\QGindx{b}}$ & 
$\Xi^{\star -}_{\QGindx{b}}$ & 531 \\
bsu & $\Xi^0_{\QGindx{b}}$ & $\Xi^{\prime 0}_{\QGindx{b}}$ & 
$\Xi^{\star 0}_{\QGindx{b}}$ & 532 \\
bss & & $\Omega^-_{\QGindx{b}}$ & $\Omega^{\star -}_{\QGindx{b}}$ & 533 \\
\hline
bcd & $\Xi^0_{\QGindx{bc}}$ & $\Xi^{\prime 0}_{\QGindx{bc}}$ & 
$\Xi^{\star 0}_{\QGindx{bc}}$ & 541 \\
bcu & $\Xi^+_{\QGindx{bc}}$ & $\Xi^{\prime +}_{\QGindx{bc}}$ & 
$\Xi^{\star +}_{\QGindx{bc}}$ & 542 \\
bcs & $\Omega^0_{\QGindx{bc}}$ & $\Omega^{\prime 0}_{\QGindx{bc}}$ & 
$\Omega^{\star 0}_{\QGindx{bc}}$
& 543 \\
bcc & & $\Omega^+_{\QGindx{bcc}}$ & $\Omega^{\star +}_{\QGindx{bcc}}$ & 544 \\
\hline
bbd & & $\Xi^-_{\QGindx{bb}}$ & $\Xi^{\star -}_{\QGindx{bb}}$ & 551 \\
bbu & & $\Xi^0_{\QGindx{bb}}$ & $\Xi^{\star 0}_{\QGindx{bb}}$ & 552 \\
bbs & & $\Omega^-_{\QGindx{bb}}$ & $\Omega^{\star -}_{\QGindx{bb}}$ & 553 \\
bbc & & $\Omega^0_{\QGindx{bbc}}$ & $\Omega^{\star 0}_{\QGindx{bbc}}$ & 554 \\
bbb & & & $\Omega^{\star -}_{\QGindx{bbb}}$ & 555 \\
\hline
\end{tabular}
\hspace*{\fill}
\caption{The proposed numbering scheme for the baryons. In the first $J=1/2$
column the order of the light quark numbers is reversed; for example $\Lambda$
has number 3122  whilst $\Sigma^0$ is 3212.}
\label{QGIKtbar}
\end{table}

Table~\ref{QGIKtbar} lists the lowest lying $J=1/2,~3/2$ baryons, including 
the anticipated charm and bottom states. Two $J=1/2$ states exist for baryons
containing three different flavours of quarks. When the two lighter flavours
are in a symmetrical ($J=1$) state the baryon is called a $\Sigma,\;\Xi'$ or
$\Omega'$ and a $\Lambda,\;\Xi,$ or $\Omega$ if they are in an antisymmetric
($J=0$) state. To distinguish the lighter, antisymmetric states the light
quark numbers are reversed; for example\footnote{Actually the PDG naming rules
do not make it clear which state to put the prime on, we have provisionally
chosen to place it on the heavier state.} $\Xi^{\prime -}_{\QGindx{b}}$ has 
number 5312 and the $\Xi^-_{\QGindx{b}}$ number 5132. This extends the
convention that heavier states are given larger numbers. Excited states are
not yet incorporated into event generators and thus are not covered here.

Increasingly supersymmetric particles are found in event generators, we
therefore take his opportunity to put forward the following numbering scheme
for them. A seventh digit is added being: either 1 ($1\star\star\star\star
\star\star$) for the partner of a boson or left-handed fermion; or 2 ($1\star
\star\star\star\star\star$) for the partner of a right-handed fermion.
When left-right mixing occurs the ordering should be by mass. Examples
include:
\begin{displaymath}
\eqalign{
1000011 &\hspace{10mm}\tilde{\mbox{e}}^-_L \cr
2000011 &\hspace{10mm}\tilde{\mbox{e}}^-_R \cr
1000012 &\hspace{10mm}\tilde{\nu}_{\QGindx{e}} \cr
2000006 &\hspace{10mm}\tilde{\mbox{t}}_R \cr}
\hspace{20mm}
\eqalign{
-2000006 &\hspace{10mm}\bar{\tilde{\mbox{t}}}_R \cr
1000021 &\hspace{10mm}\tilde{\mbox{g}} \cr
1000024 &\hspace{10mm}\tilde{\mbox{W}}^+/\tilde\chi^+_1 \cr
1000037 &\hspace{10mm}\tilde{\mbox{H}}^+/\tilde\chi^+_2 \cr}
\hspace{20mm}
\eqalign{
1000022 &\hspace{10mm}\tilde\gamma/\tilde\chi^0_1 \cr
1000023 &\hspace{10mm}\tilde{\mbox{Z}}^0/\tilde\chi^0_2 \cr
1000025 &\hspace{10mm}\tilde{\mbox{H}}^0_1/\tilde\chi^0_3 \cr
1000035 &\hspace{10mm}\tilde{\mbox{H}}^0_2/\tilde\chi^0_4 \cr}
\end{displaymath}
The possibility of numbering potential SUSY mesons and baryons in the same
spirit is left open at present.

\subsection{Decay Tables}
\label{QGsubsectdecay}

The study of identified particle production and the physics underlying the
hadronization mechanism continues to be an active area of research at LEP~1.
In hadronic Monte Carlo event generators final state particles are produced
in two stages. Primary hadrons come directly from the clusters/strings/{\it
etc}. that model the non-perturbative parton to hadron transition. 
Subsequently
chains of secondary particles arise from the decays of the unstable primary
hadrons. Note this separation is well defined in the context of MC programs
but in reality for the short lived, strongly decaying resonances it may be
only semantics.

Thus a major common component of hadronic MCs are routines to do the decay
of unstable particles. These are based on the use of tabulated branching
ratios and basic matrix elements, though in the case of $\tau$'s and 
b-hadrons
specialized packages are also available. The construction of these decay
tables is not a simple task and requires much per-in-spiration to fill in
gaps in present measurements \cite{QGIKPDG} and deal with problematic cases.
It would save much duplication of effort if one basic table could be used
by all programs. This implies the ability to swap decay tables and thereby
would allow some control over a (spurious) source of apparent variation in
the rates of primary hadron production in the different hadronization models.
A common, user friendly, interface would also enable easy maintenance and
modification of the tables by users.

To achieve such a goal requires a unique way of identifying the particles,
and any associated matrix elements, together with a standard format for
outputting and inputting the tables. The revised PDG codes above provide a
unique and logical means of identifying the particles. To identify the
matrix elements we propose developing a set of standard three-digit integer
codes, following the convention of table~\ref{QGIKtcod}.
\begin{table}[htb]
\begin{center}
\begin{tabular}{|c|l|}
\hline
Code & Matrix Element \\
\hline
0 & Isotropic decay \\
\hline
1-99 & Standard codes to be agreed \\
\hline
$\geq100$ & Program specific options \\
\hline
\end{tabular}
\end{center}
\caption{Proposed convention for matrix element codes}
\label{QGIKtcod}
\end{table}

It is reasonable to restrict both the number of decay products to five,
using zeros to complete an entry, and also numerical branching ratios to
five decimal places. A more than five body decay can be stored, realistically,
as a sequence of decays involving intermediate resonances. In studies
involving very rare decays it is sensible to use a higher branching ratio
and then apply a compensating normalization factor. It is then proposed
to write out the following information,
\begin{tabbing}
xxxxxxxx\= \kill
\>Number of decays listed \\
\>Decaying particle, branching ratio, matrix element code, 1--5 decay products
\end{tabbing}
using the following FORMAT statements,
\begin{tabbing}
xx\= \kill
\>{\tt 100 FORMAT(1X,I4)} \\
\>{\tt 200 FORMAT(1X,I8,1X,F7.5,1X,I3,5(1X,I8))}
\end{tabbing}
An example is provided by the $\pi^0$ decays:
\begin{tabbing}
000\= \kill
\>2 \\
-0000\=000x\=0.00001x12\=3x-00000\=01x-00000\=02x-000000\=3x-000000
\=4x-000000\=5xxxx\=\kill
    \>111 \>0.98800   \>0     \>22     \>22      \>0      \>0      \>0   \>
($\pi^0\rightarrow\gamma\gamma$) \\
-0000\=000x\=0.00001x\=123x-00000\=01x-00000\=02x-0000\=003x-000000
\=4x-000000\=5xxxx\=\kill
    \>111 \>0.01200   \>101   \>22     \>11    \>$-$11      \>0      \>0   \>
($\pi^0\rightarrow\gamma\mbox{e}^-\mbox{e}^+$)
\end{tabbing}

It must be recognised that b-hadrons represent a special case. In the absence
of detailed knowledge about a significant fraction of their decays MC programs
resort to models based on partonic decays and fragmentation. Partonic decay
modes may also be stored in the above format. Suppose in a b$\QGqbar$ hadron
(here $\QGqbar$ may represent a diquark) the decay is 
$\mbox{b} \rightarrow \mbox{c} \mbox{W}^- \rightarrow \mbox{c} 
\QGq_1 \QGqbar_2$, this can be coded in one of two sequences either
`$\QGq_1,\QGqbar_2,\mbox{c}$' or `$\mbox{c},\QGqbar_2,\QGq_1$'. 
These two options can be exploited
to refer to the two possible colour connections separately: 
$(\mbox{c}\QGqbar)(\QGq_1\QGqbar_2)$ and 
$(\mbox{c}\QGqbar_2)(\QGq_1 \QGqbar)$ respectively, at the discretion 
of program
authors\footnote{Observe that if the $V-A$ matrix element was constructed as
$(p_0.p_2)(p_1.p_3)$ these would both give 
$(\mbox{b}.\QGqbar_2)(\QGq_1.\mbox{c})$.}.

It is now simply a matter of providing a {\tt .DAT} file containing the decay
table listed in the above format. To standardize the interface to the
individual MC programs the following two subroutines are proposed:
\begin{tabbing}
xxxxxxxxxxx\= $\star\star${\tt IODK(IDK,BR,ME,IPRD1,IPRD2,IPRD3,IPRD4,IPRD5)}
\=\kill
\>{\tt $\star\star$IODK(IUNIT,IFORMAT,IOPT)} \> and\\
\>$\star\star${\tt MODK(IDK,BR,ME,IPRD1,IPRD2,IPRD3,IPRD4,IPRD5)}
\end{tabbing}
were $\star\star$ identifies the MC program. The first is used to read {\tt
IUNIT<0} or write {\tt IUNIT>0} the decay table to the given unit number with
{\tt IFORMAT} specifying how the particles are identified. The standard is
{\tt IFORMAT=1}, that is use the revised PDG codes; nonportable program
specific options may include: =2 use the internal numbering or =3 use the
internal character string names. Authors and users may prefer the later
options as more transparent than the PDG numbers. If {\tt IOPT=1} then
matrix element codes $\geq100$ (program specific) are accepted, if {\tt 
IOPT=0}
then such codes are treated as not recognised and set to zero, isotropic 
decay.
The subroutine {\tt $\star\star$MODK} is intended to allow individual lines
of the table to be modified or added, before or during event generation;
the arguments follow the standard format. Note that when a new mode is added
or an existing branching ratio modified the sum of the remaining branching
ratios should be rescaled to preserve unit sum. This means that when two
modes of the same particle are altered the order of the calls is important
for their resultant branching ratios.

The provision of such an interface rests with the actual program authors who
need to convert between the standard format and their own internal structures.
These interfaces may be expected to be robust against unrecognized or blank
particle names and provide basic checks of the allowed kinematics, electric
charge conservation and unit sum of branching ratios. Such an interface has
been established for {\small HERWIG} and successfully used to import the
{\sc Jetset} decay table. However if program users do modify the provided
decay tables then they must accept responsibility for them making sense.

\subsection{Interfaces to electroweak generators}
\label{QGsubsectEWint}

A number of dedicated four-fermion generators are being written for
LEP~2 applications. The good ones will do the electroweak 
theory much better than standard general-purpose QCD generators. 
On the other hand, they do not contain any 
QCD physics aspects, i.e. neither perturbative parton showers nor 
nonperturbative hadronization. This makes the electroweak (EW)
generators well suited for some applications, such as total cross 
sections and leptonic final states, but generally unsuited for the study
of hadronic or mixed hadronic--leptonic final states. It is therefore 
logical to interface them with parton-shower and hadronization programs. 
To some extent, this is already happening. However, in writing these
interfaces there are certain dangers involved. There may also be a lot of
work involved. 

It would therefore be advantageous if the event generator authors involved 
could agree on a common approach: EW authors provide the four-fermion 
configuration in a standard format and QCD authors provide a standard
interface that converts this to a set of final hadrons. Then only one 
interface needs to be written for each program, instead of one for each
combination of EW and QCD programs. In this section we propose such
a standard and report on progress in implementing it.

\subsubsection{The basic problem}
In the electroweak sector, fermions can be viewed as asymptotically free
final state particles. This means that the production of a specific
final state is fully calculable perturbatively. Many different
intermediate states can contribute to the same final state, without
any ambiguities being introduced by this. The total probability for a 
final state is given by squaring the sum of amplitudes
\begin{equation}
   |A|^2 = | A_1 + A_2 + \ldots + A_n |^2
\end{equation}
(suppressing the issue of helicity sums, {\it etc.} --- these aspects are not
important for the general discussion). Interference effects therefore
are included automatically.

QCD is different. Quarks are {\em not} asymptotic states. The final state
consists of colour singlet hadrons, not coloured partons. The transition
from perturbative to non-perturbative physics is not understood from
first principles, but is at present modelled. The model used describe well 
what happens to a simple quark-antiquark pair, e.g. 
$\mbox{Z}^0 \to \QGq\QGqbar$ at LEP~1. At LEP~2, four-quark states 
$\QGq_1\QGqbar_2\QGq_3\QGqbar_4$ have to be mastered. If we want to make 
use of our hard-won phenomenological experience, it is therefore 
essential that the $\QGq_1\QGqbar_2\QGq_3\QGqbar_4$ system can be 
subdivided into two colour singlet subsystems, either
$\QGq_1\QGqbar_2 + \QGq_3\QGqbar_4$ or 
$\QGq_1\QGqbar_4 + \QGq_3\QGqbar_2$. Each 
subsystem can then be described in the same way as a LEP~1 event.
On the contrary, if we are not allowed to use such a subdivision
into singlets, a completely new hadronization formalism would have to
be invented (with brand new parameters to be tuned to the LEP~2 data).

Unfortunately, there are complications. As a simple illustration, 
consider a system 
$\mbox{u}\overline{\mbox{d}}\mbox{d}\overline{\mbox{u}}$. 
The production obtains contributions from several possible intermediate 
states. One is a $\mbox{W}^+\mbox{W}^-$ pair, with 
$\mbox{W}^+ \to \mbox{u}\overline{\mbox{d}}$ and 
$\mbox{W}^- \to \mbox{d}\overline{\mbox{u}}$. Another is a 
$\mbox{Z}^0\mbox{Z}^0$ pair, with the first 
$\mbox{Z}^0 \to \mbox{u}\overline{\mbox{u}}$ and the second 
$\mbox{Z}^0 \to \mbox{d}\overline{\mbox{d}}$.
These two alternative intermediate states correspond to different colour
singlets, and therefore would differ with respect to the treatment of 
subsequent parton showers and hadronization. That is, the final state
contains a ``memory'' of the intermediate state. Furthermore, $|A|^2$ 
contains an interference term between the two alternatives, where the 
colour flow is not well-defined in perturbation theory. This is 
reflected in a relative colour factor $1/(N_C^2-1)$ for the interference 
term, meaning e.g. that both $\mbox{u}\overline{\mbox{u}}$ and 
$\mbox{u}\overline{\mbox{d}}$ are in relative
colour singlet states. Kinematical factors are not likely to compensate 
for the colour suppression, so numerically the interference terms may not 
be large. However, when the aim is to make a precision measurement of 
the W mass (to better than one {\it per mille}), one cannot rashly neglect 
their possible contribution. Since we know of no ``correct'' procedure
to calculate it the reasonable approach is to adopt a ``good bet'' default
with a method to define a ``band of uncertainty''.  

\subsubsection{Flavour and kinematics specification}
It is natural to use the {\tt HEPEVT} common block specification
\cite{QGIKhepevt} to transfer flavour and kinematics information from 
the electroweak generator to the QCD one. After all, the {\tt HEPEVT}
standard was devised specifically with this kind of tasks in mind.
The original standard has been changed so that real variables are
given in {\tt DOUBLE PRECISION}. 

For the current interface, only {\tt NHEP}, {\tt IDHEP} and
{\tt PHEP} are actually mandatory. EW generator authors are invited 
to fill also the other information, such as mother--daughter pointers,
but that is optional. Furthermore, any number of entries may be used 
in the event record to indicate the incoming $\QGee$ pair and 
intermediate states, but the only objects allowed to have status code 
{\tt ISTHEP}$ = 1$ are the two final fermion--antifermion pairs and 
an arbitrary number of photons. The fermions may be interspersed
with photons in the listing, but the relative order of fermions is 
strict: 
\begin{QGitemize}
\item[1] one outgoing fermion, i.e. $\QGq / \ell^- / \nu_{\ell}$; 
\item[2] one outgoing antifermion, i.e. 
         $\QGqbar / \ell^+ / \overline{\nu}_{\ell}$;
\item[3] another outgoing fermion; and
\item[4] another outgoing antifermion.
\end{QGitemize}

The pairing of the outgoing fermions and antifermions should be done
so that, when $\mbox{W}^+\mbox{W}^-$ intermediate states can contribute, 
the pair 1 and 2 corresponds to a possible decay of the $\mbox{W}^+$,
and the pair 3 and 4 to a possible decay of the $\mbox{W}^-$. An 
example of an allowed order is 
$\mbox{u}\overline{\mbox{d}}\mbox{d}\overline{\mbox{u}}$, while 
$\mbox{u}\overline{\mbox{u}}\mbox{d}\overline{\mbox{d}}$ is not correct. 
When a $\mbox{W}^+\mbox{W}^-$ pair cannot contribute, the
ordering should be instead made consistent with the decay of of one 
$\mbox{Z}^0$ to the pair 1 and 2, and another 
$\mbox{Z}^0$ to the pair 3 and 4, 
e.g. $\mbox{u}\overline{\mbox{u}}\mbox{c}\overline{\mbox{c}}$. 
This way, coloured and uncoloured fermions can not be mixed in a pair, 
i.e. $\mbox{u}\mbox{e}^-\overline{\nu}_{\QGindx{e}}\overline{\mbox{d}}$
is not an allowed ordering.

Of course, adopting the fixed order above is not crucial, but it
avoids the need for QCD generators to do a lot of rearrangements,
and establishes a standard for the colour flow weights in the
next section.

\subsubsection{Colour flow specification}
As was already mentioned above, the colour flow is not uniquely
specified when both outgoing fermion pairs are of quark-antiquark type.
A QCD event generator is therefore required to make a choice. We 
propose the following procedure. 

The acceptance of a kinematical configuration by the electroweak
generator (including specific helicities for some generators) is 
based on the total squared amplitude, $|A|^2$, so
this number is available ``for free''. Internally, a generator
has access to the subamplitudes, $A = A_1 + A_2 + \ldots + A_n$.
Each subamplitude does correspond to a well-defined colour flow,
so split the amplitudes into two classes, I and II, with 
I corresponding to colour singlets $1+2$ and $3+4$, and II to
singlets $1+4$ and $3+2$. The total squared amplitude can then be 
written as 
\begin{equation}
|A|^2 = |A_{\QGindx{I}} + A_{\QGindx{II}}|^2 =
|A_{\QGindx{I}}|^2 + |A_{\QGindx{II}}|^2 + 
2 \mbox{Re}(A_{\QGindx{I}}A^*_{\QGindx{II}})
= |A_{\QGindx{I}}|^2 + |A_{\QGindx{II}}|^2 + \Delta ~.
\end{equation}    
This subdivision should be gauge invariant.
 
Each electroweak event generator should return the three (positive)
numbers $|A|^2$, $|A_{\QGindx{I}}|^2$ and $|A_{\QGindx{II}}|^2$.
Then the colour-suppressed interference term $\Delta$ is easily found
as $\Delta = |A|^2 - |A_{\QGindx{I}}|^2 - |A_{\QGindx{II}}|^2$.

A ``good bet'' approach to the colour assignment problem is for
the QCD generator to neglect the interference term, and use the
relative magnitude of $|A_{\QGindx{I}}|^2$ and $|A_{\QGindx{II}}|^2$
to make a choice at random between the two possible colour flows.

More sophisticated recipes are used for QCD processes like 
$\QGq\QGg \to \QGq\QGg$ in {\small HERWIG} and {\sc Pythia}, where the 
interference 
terms are split between the non-interference ones in accordance with 
the pole structure. However, such an approach presupposes a detailed
study for each specific combination of allowed graphs, and so cannot
be part of a generic interface. Should a generator provide such a 
subdivision, maybe as an option, it would be easy to represent
by modified numbers $|A_i|^2 \to |A_i|^2 + \Delta_i$ so that
$\Delta \to \Delta - \Delta_{\QGindx{I}} - \Delta_{\QGindx{II}} = 0$.

When $\Delta$ is nonvanishing,
the uncertainty can be estimated by assigning the interference terms
so that either class I or class II is maximized. Specifically, class
I is maximized when the choice of colour flow is based on the relative
magnitude of $R_{\QGindx{I}}$ and $R_{\QGindx{II}}$, where
\begin{equation}
\begin{tabular}{rlrll}
$R_{\QGindx{I}}$  & $= |A_{\QGindx{I}}|^2 + \Delta$ & $R_{\QGindx{II}}$ &
$= |A_{\QGindx{II}}|^2$ & if $\Delta > 0$ \\[2mm]
& $= |A_{\QGindx{I}}|^2$ & & $= \max(0, |A_{\QGindx{II}}|^2 + \Delta)$ &
else  \\
\end{tabular}
\end{equation}
and correspondingly with I$\leftrightarrow$II for class II maximized.
If the difference between these two extremes is small, then presumably
the default procedure can be trusted.

\subsubsection{Further problems}
A number of potential problems exist, where the current approach may not 
be enough. These are discussed in the following.

It has implicitly been assumed that the scale of perturbative QCD
parton-shower evolution is set by the mass of the respective colour
singlet. An example of a process where this need not be the case
is $\QGee \to \QGgaZ \to \mbox{u}\overline{\mbox{u}} \to %
\mbox{u}\overline{\mbox{u}}\mbox{Z}^0 \to %
\mbox{u}\overline{\mbox{u}} \mbox{d}\overline{\mbox{d}}$. 
The $\mbox{u}\overline{\mbox{u}}$ pair here has a large original mass, 
which is reduced by the emission of a $\mbox{Z}^0$. It is not clear 
whether the QCD radiation can be well approximated by that of the final
$\mbox{u}\overline{\mbox{u}}$ mass, or whether the original mass is 
somehow felt e.g. by a larger rate of hard-gluon emission. A study of 
the matrix element for 
$\QGee \to \QGq\QGqbar \to \QGq\QGqbar\mbox{Z}^0\QGg$ 
would here be necessary. However, these graphs are not expected to give 
a major contribution, so presumably the uncertainty from this source is 
not significant.

When QCD processes are introduced, interference terms need not be 
colour-sup\-pressed. Specifically, the graph 
$\QGee \to \QGgaZ \to \mbox{u}\overline{\mbox{u}} \to %
\mbox{u}\overline{\mbox{u}}\QGg \to\mbox{u}\overline{\mbox{u}} %
\mbox{d}\overline{\mbox{d}}$ gives two colour singlets 
$\mbox{u}\overline{\mbox{d}}$ and $\mbox{d}\overline{\mbox{u}}$, 
just like a $\mbox{W}^+\mbox{W}^-$ intermediate state would. 
Therefore a suppression of interference contributions
has to be based entirely on kinematical considerations.

Since separation of quark and gluon jets is very difficult on an 
individual basis, also $\QGq\QGqbar\QGg\QGg$ gives a background to 
four-fermion final states. Here, of course, there can be no 
interference with the other processes.

The addition of parton showers to a QCD four-jet event, either
$\QGq\QGqbar\QGg\QGg$ or 
$(\QGq\QGqbar\QGg \to) \QGq\QGqbar\QGq'\QGqbar'$, has to follow 
quite different rules from that of other four-fermion events, e.g.
with respect to angular-ordering constraints in the parton shower. 
These rules have not yet been worked out for any of the QCD generators.
The input that electroweak generators can give here is therefore
not so meaningful. The main thrust in this area should be an improved 
matching between the matrix-element and the parton-shower strategies
already present in QCD generators. Electroweak generators (if they 
contain QCD graphs) should therefore have the option of switching
off all QCD contributions, i.e. (the amplitudes for) the graphs above.

Some further input parameter may be required to specify whether QCD
showers should be allowed also to involve the emission of photons.
At LEP~1 we have learned that the ``competition'' between photon and
gluon emission is a not unimportant aspect, that tends to reduce
the total amount of photon radiation compared to the no-QCD-radiation
scenario. Something similar is likely to hold at LEP~2. However,
the situation is far worse here, since the number of charged particles
is much larger, and the presence of intermediate charged states
($\mbox{W}^+\mbox{W}^-$) makes a subdivision of the full emission rate 
much more complicated. One could therefore consider two extremes:
\begin{QGitemize}
\item If an EW generator attempts to do the full job of photon
radiation from all charged legs, then the QCD generator should not
add further photon radiation. In fact, if anything, one may question
whether the EW generator overestimated the amount of photon radiation
off the quarks.
\item If an EW generator only claims to have initial-state photon
radiation, then the QCD generator could add final-state radiation
inside each fermion-antifermion pair (also leptons, if implemented).
This would still not be the full answer, but likely to be better than
having no final-state radiation at all. (Since there is no unique,
gauge-independent definition of final-state $\gamma$ radiation
in four-fermion processes, the usefulness of such an approach should 
be checked from case to case.)
\end{QGitemize}

Traditionally, QCD generators are not good at handling the 
polarization of $\tau$'s in the decay treatment. This is better done
by dedicated $\tau$ decay packages. Therefore EW generators that do
provide the spin of outgoing $\tau$'s should give this information
for the $\tau$ entries in\\[2mm]  
\hspace*{1cm}{\tt COMMON/HEPSPN/SHEP(4,NMXHEP)}\\[2mm]
using the standard conventions \cite{QGIKhepevt}. A flag could be set by 
the EW generator, and used by the QCD generator to inhibit it from
decaying the $\tau$'s. 

We remind the reader that the production vertices at the femtometer 
level may be 
of interest for physics such as colour reconnection and Bose-Einstein 
effects. If any generator should provide such output,
the {\tt VHEP} part of {\tt HEPEVT} can be used to define 
vertices. The original objective was for vertices at the scale of mm, 
but also numbers of order $10^{-12}$ mm could be stored with maintained
precision so long as the primary event vertex is designed to be at the
origin. 

\subsubsection{Existing codes:}
Several interfaces now exist that are based on the philosophy outlined
above.
\begin{QGitemize}
\item Output from {\sc Excalibur}, with amplitude information for the 
different colour singlets. Can be obtained at\\
{\tt  http://wwwcn.cern.ch/$\sim$charlton/excalibur/excalibur.html}.
\item Input into {\small HERWIG}. Can be obtained at \\
{\tt http://surya11.cern.ch/users/seymour/herwig/}.
\item Input into {\sc Jetset}. Can be obtained at\\
{\tt http://thep.lu.se/tf2/staff/torbjorn/test/main07.f}.
\item Input into {\sc Ariadne}. Is part of the standard {\sc Ariadne}
distribution.
\end{QGitemize}
Further information is available in the respective files.

\subsection{Systematic errors}
At LEP~2, several physics issues will involve hadronized quarks, both for
QCD studies and for Electroweak measurements or searches. The following
can be envisaged as case studies:
\begin{QGitemize}
\item
Establish the running of $\QGalphas$ from the Z pole to LEP~2 energies.
\item
Measurement of hadronic cross-sections, e.g. e$^+\mbox{e}^-\rightarrow
\mbox{Z}/\gamma\rightarrow \mbox{q}\QGqbar$ or e$^+\mbox{e}^-\rightarrow
\mbox{W}^+\mbox{W}^-\rightarrow \mbox{q}_1\QGqbar_2\mbox{q}_3\QGqbar_4$,
and discrimination between the two.
\item
Reconstruction of jet-jet invariant masses in the above two processes,
or even in e$^+\mbox{e}^-\rightarrow\mbox{ZH}\rightarrow\mbox{q}\QGqbar
\mbox{b}\bar{\mbox{b}}$.
\end{QGitemize}

\noindent
{\em The first item} seems the easiest case. $\QGalphas$ has been measured
using a great variety of observables at the Z peak, with nearly infinite
statistics. The {\em variation} with $\sqrt{s}$ of well defined quantities
such as energy-energy correlations or their asymmetry, or jet rates for a
given $y_{\QGindx{cut}}$ should be much less prone to systematic errors than 
their relationship to $\QGalphas$ itself.

Two difficulties can be expected here. First, the flavour composition of
the sample will be different at LEP~2. In particular, the rate of
b$\bar{\mbox{b}}$ production will decrease from 22\% down to less than
10\%. The fact that the specific fragmentation parameters for b quarks
\cite{QGQCDbfrag} have been measured at LEP~1 should be of great help. In
order to extrapolate to higher energies, these results have to be
incorporated in the simulation in one way or another. The three-jet rate
has been used at the Z peak to test the universality of the strong coupling
\cite{QGQCDbalphas} with an accuracy of about 0.005. The argument can be
turned around as, the sensitivity of a determination of $\QGalphas$ to
flavour composition, leading to a rough uncertainty estimate of about
0.0005 on the difference in $\QGalphas$ from the Z peak to LEP~2.

The second difficulty will arise when one tries to go from establishing
the running to more quantitative estimates of it. The running will be
compared to the expectation from the QCD fragmentation models. Given that
the most popular generators are presently based on $\cal{O}(\QGalphas)$
exponentiated showers, one can rightfully challenge their capability to
predict the $\sqrt{s}$ evolution, because of missing higher orders. The
solution to this issue will probably have to come from a better mapping
of the shower models to second order matrix elements.

\noindent
{\em The impact on acceptance corrections} was limited at the Z peak by
two positive factors: large statistics and limited initial state radiation.
A simple event rotation technique~\cite{QGALEPHew94} was sufficient to reduce
the uncertainty on event selection down to $10^{-3}$ or better. The
precision required at LEP~2 for such studies is less stringent, statistical
errors being at the level of 1\%, so that the same method applied to high
energy annihilation events should be adequate. One difficulty will arise
from initial state radiation (ISR): the optimum sensitivity for electroweak
effects is obtained by removing the radiative return to the Z peak using an
$s'$ cut. Most of the ISR photons being emitted at small angles, the
invariant mass of the hadronic system has to be used to implement this cut.
The issue here is to understand how accurately one can reconstruct an
invariant mass from a system of boosted jets. An important experimental
constraint can presumably be placed by using e$^+\mbox{e}^-\rightarrow
\mbox{Z}+\gamma\rightarrow\mbox{q}\QGqbar+\gamma$ events. However the issue
of flavour dependence will come up again here, as the mass of the b quark
and missing energy from neutrinos are expected to have sizeable effects on
the jet angles and energies after a boost. A similar problem will be
encountered when reconstructing W$\rightarrow\mbox{q}\QGqbar$ invariant
mass, where the difference in flavour composition is even more drastic.

Finally one last but important issue is the discrimination between e$^+
\mbox{e}^-\rightarrow\mbox{Z}/\gamma\rightarrow\mbox{q}\QGqbar$ and e$^+
\mbox{e}^-\rightarrow\mbox{W}^+\mbox{W}^-\rightarrow\mbox{q}_1\QGqbar_2
\mbox{q}_3\QGqbar_4$ events for the determination of the W$^+\mbox{W}^-$
cross-section at threshold~\cite{QGWWreport}. There is a finite probability
that a four-jet event from the first process with two hard QCD-radiated
partons will mimic the second process. In the present state of QCD
generators with only $\cal{O}(\QGalphas)$ exponentiated showers, it is not
obvious that the Monte Carlo gives the right answer. One way to obtain
direct experimental information is to see how often a hadronic Z decay
can be reconstructed as two heavy systems of 45 GeV mass and compare with
the predictions of the fragmentation model. The extrapolation to the
appropriate center-of-momentum energies and invariant mass requires a
fragmentation Model.

In most of these problems, an experimental constraint can be found in e.g.
Z decays. However every time fragmentation event generators are needed to
perform the necessary extrapolations. Evaluating the corresponding
systematic errors has been performed traditionally by either i) varying
some (well chosen) input parameters within ``reasonable limits'' or ii)
comparing the results obtained when using two different models. Recently,
a more complex situation has emerged for the analysis of the jet charge
asymmetries in Z decays~\cite{QGSTDENIS}. This is a clear
example of an electroweak measurement performed using jets. The jet charge
separations are ultimately obtained from a fragmentation model, upon which
many constraints are imposed: measured production spectra for pions, kaons
and baryons (p and $\Lambda$), resonances such as $\rho,\;\mbox{K}^*$
and $\eta$, average jet charge measured from opposite hemisphere charge
correlation, {\it etc.} Imposing these constraints immediately leads to
extremely strong correlations among fragmentation parameters. In {\sc
Jetset}, it is possible to find enough parameters to describe very
completely the production of each particle species. The weak points
remain the transverse momentum distributions and the baryon spectra. In
{\small HERWIG} fewer parameters are available and the $\chi^2$ is worse.
Nevertheless the value of the electroweak asymmetry can be extracted for
both models, with systematic errors related to the goodness of fit. A
consistency check is supplied by the agreement of the values obtained
from the two models within the systematics pertaining to each model.
Similar procedures can be envisaged for measurements of electroweak
quantities at LEP~2.

To conclude, there is no unique prescription for evaluating systematic
errors. In each problem specific sources of errors and the corresponding
fragmentation model parameters have to be found. Incorporating
experimental constraints generally leads to very strong correlations
among parameters, but this can be solved by for example using a combined
linearized fit. The most general problem in extrapolating results obtained
at the Z pole to LEP~2 will be the change in flavour composition. A mapping
of the parton shower models to the second order matrix elements would be 
most useful.

\clearpage
\section{Summary and Recommendations}
\label{QGsectsummary}

It is useful to remember the last words of the LEP~1 QCD generators 
report \cite{QGLEPone}: {\em Due to the large uncertainty present in
any realistic Monte Carlo, physics studies must be based on the use of 
at least two complete and independent programs}. Nothing has been
changed in this regard; QCD is still not solved and the need for models 
is as large as ever. 

The QCD generators of today may be considered more mature than the 
pre-LEP ones, in that they have successfully survived a number of
experimental tests. However, there is always the danger that
``incorrect'' models do not just fade away --- they are only modified and
retuned for agreement. The increased energy lever arm provided by 
LEP~2 could give additional discrimination power, or at lease 
necessitate further fine tuning of programs.

Furthermore, in comparing with the LEP~1 data, we see that no generator 
is perfect. Depending on the physics area studied, it is therefore 
important to beware of generators with known shortcomings in that
area. These shortcomings may indicate basic problems in the models,
but could also come from further effects (e.g. higher-order 
matrix-element corrections) that authors never claimed to include. 
Generator authors are encouranged to sort out known problems in
the light of LEP~1 experience, and in particular those with implications
for LEP~2 studies. For some areas, such as colour reconnection and 
Bose-Einstein effects, the modelling is only in its infancy, 
and further efforts obviously are required. 

The World Wide Web offers new opportunities to make programs accessible, 
including manuals, update notes, sample main programs and so on.
To the extent authors did not yet adapt their distribution practices
to the new opportunities, they are encouraged to do so. A common
practice of having a ``home page'' for each generator will allow
the construction of useful generator directories. 

Standardization is as important as ever, in order to avoid confusion
among experimentalists required to run a multitude of different codes.
We have here proposed modifications to the {\tt /HEPEVT/} standard,
extensions and a few corrections to the PDG particle code,
a standardized decay table and an interface between electroweak
four-fermion generators and QCD generators. A continued dialogue
about possible standards would be very useful.

\end{document}